\documentclass[a4paper,11pt]{article}
\pdfoutput=1
\usepackage{graphicx}
\usepackage{graphics}
\usepackage{dcolumn}
\usepackage{bm}
\usepackage{epstopdf}
\usepackage{mathrsfs}
\usepackage{amssymb}
\usepackage{amsmath}
\usepackage{multirow,cite}

\newcommand{\beq}{\begin{eqnarray}}
\newcommand{\eeq}{\end{eqnarray}}
\newcommand{\be}{\begin{equation}}
\newcommand{\ee}{\end{equation}}

\newcommand{\bea}{\begin{eqnarray}}
\newcommand{\eea}{\end{eqnarray}}

\newcommand{\la}{\left\langle}
\newcommand{\ra}{\right\rangle}
\newcommand{\lc}{\left[}
\newcommand{\rc}{\right]}
\newcommand{\lp}{\left(}
\newcommand{\rp}{\right)}

\usepackage{url}
\usepackage{hyperref}

\begin{document}

\vspace{-2.0cm}
\begin{flushright}
OUTP-15-14P
\end{flushright}

\begin{center}
  {\Large \bf Boosting Higgs Pair Production in the $b\bar{b}b\bar{b}$ Final State\\[0.2cm]
    with
  Multivariate Techniques}
\vspace{.7cm}

J. Katharina Behr, Daniela Bortoletto, James A. Frost,
  Nathan P. Hartland, \\Cigdem Issever and Juan Rojo

\vspace{.3cm}
{\it Physics Department, 1 Keble Road, University of Oxford, United Kingdom }

\vspace{1cm}

{\bf \large Abstract}\\
\end{center}
\vspace{0.1cm}
The measurement of Higgs pair production will be a cornerstone of
  the LHC program in the coming years.
  Double Higgs production provides a crucial window upon the mechanism of electroweak
  symmetry breaking and has a unique sensitivity
  to the Higgs trilinear coupling.
  We study the feasibility of a measurement of
  Higgs pair production in the
  $b\bar{b}b\bar{b}$ final state at the LHC.
  Our analysis is based on a combination of traditional cut-based
  methods with state-of-the-art
  multivariate techniques.
  We account for all relevant
  backgrounds, including the contributions from light and charm
  jet mis-identification, which are ultimately comparable in size 
  to the irreducible $4b$ QCD background.
  We demonstrate the robustness of our analysis strategy in
  a high pileup environment.
  For an integrated luminosity of $\mathcal{L}=3$ ab$^{-1}$,
  a signal significance of  $S/\sqrt{B}\simeq 3$ is obtained,
  indicating
  that the $b\bar{b}b\bar{b}$ final state
alone could allow for the observation of double Higgs production
at the High Luminosity LHC.

\clearpage

\section{Introduction}

The measurement of double Higgs production will be one of the central
physics goals of the LHC program in its recently started high-energy
phase, as well as for its future high-luminosity upgrade (HL-LHC)
which aims to accumulate a total integrated
luminosity of 3 ab$^{-1}$~\cite{ATLAS:2013hta,CMS:2013xfa}.
Higgs pair production~\cite{baglio} is directly sensitive to the
Higgs trilinear coupling $\lambda$ and 
provides crucial
information on the electroweak symmetry breaking mechanism.
It also probes the underlying strength of the Higgs interactions
at high energies, and can be used to test the composite nature of the 
Higgs boson~\cite{Giudice:2007fh,Contino:2010mh}.
While Standard Model (SM) cross-sections are small,
many Beyond the SM (BSM)
scenarios predict enhanced rates for double Higgs production, therefore searches have already been performed by ATLAS and CMS with Run I data~\cite{Aad:2015xja,Aad:2015uka,Aad:2014yja,Khachatryan:2015yea,Chatrchyan:2011wt}
and will continue at Run II.
The study of Higgs pair production will also be relevant to
any future high-energy 
collider, either at a 100 TeV circular machine~\cite{Arkani-Hamed:2015vfh,Barr:2014sga,Papaefstathiou:2015iba,Azatov:2015oxa} or at
a linear or circular electron-positron collider~\cite{Contino:2013gna}.

Analogously to single Higgs production~\cite{Dittmaier:2012vm}, 
in the SM the dominant mechanism for the production of a pair of
Higgs bosons at the LHC is 
gluon fusion (see~\cite{baglio,Frederix:2014hta} and
references therein).
For a center-of-mass energy of $\sqrt{s} = 14\,$TeV, the
next-to-next-to-leading order (NNLO)
total cross section is approximately $40\,$fb~\cite{deFlorian:2013jea},
which is increased by a further few percent once
next-to-next-to-leading logarithmic
(NNLL) corrections
are accounted for~\cite{deFlorian:2015moa}.
Feasibility studies in the case of a SM-like Higgs boson
in the gluon-fusion channel
at the LHC have been performed for different final states, including
$b\bar b\gamma\gamma$~\cite{Baur:2003gp,Barger:2013jfa,Lu:2015jza},
$b\bar{b}\tau^+\tau^-$~\cite{Baur:2003gpa,Barr:2013tda,Dolan:2012rv,Dolan:2013rja},
$b\bar{b}W^+W^-$~\cite{Dolan:2012rv,Papaefstathiou:2012qe} and
$b\bar{b}b\bar{b}$~\cite{Baur:2003gpa,Dolan:2012rv,Wardrope:2014kya,deLima:2014dta,Barger:2013jfa}.
While these studies differ in their quantitative conclusions,
a consistent picture emerges 
that the ultimate precision in the determination of the Higgs trilinear
coupling $\lambda$ requires the full integrated luminosity
of the HL-LHC, $\mathcal{L}=3$ ab$^{-1}$,
and should rely on the combination of different final states.
The interplay between  kinematic
distributions for the
extraction of $\lambda$ from the measured
cross-sections, and the role of the associated theoretical
uncertainties, have been intensely scrutinized
recently~\cite{Slawinska:2014vpa,Chen:2014xra,Goertz:2013kp,
  Frederix:2014hta,Dawson:2015oha,Maltoni:2014eza,Maierhofer:2013sha,Grigo:2013rya,Grigo:2014jma}.

In addition to the gluon-fusion channel, Higgs pairs
can also be produced in the vector-boson fusion
channel $hhjj$~\cite{Contino:2010mh,Dolan:2013rja,Dolan:2015zja,
  Brooijmans:2014eja},
the associated production modes
$hhW$ and $hhZ$~\cite{Barger:1988jk,baglio,Cao:2015oxx}
(also known as Higgs-Strahlung),
and also in association
with top quark pairs $hht\bar{t}$~\cite{Englert:2014uqa}.
All these channels are challenging due to the small production
rates: at 14 TeV, the inclusive total cross-sections are
2.0 fb for VBF $hhjj$~\cite{Liu-Sheng:2014gxa},
0.5 fb for $W(Z)hh$~\cite{baglio}
and 1.0 for $hht\bar{t}$~\cite{Englert:2014uqa}.

While the SM production rates for Higgs
pairs are small, they are substantially
enhanced in a variety of BSM scenarios.
Feasibility studies of Higgs pair production in New Physics
models have been performed in a number of different frameworks,
including Effective Field
Theories (EFTs) with higher-dimensional
operators and anomalous 
Higgs couplings~\cite{Nishiwaki:2013cma,Dall'Osso:2015aia,Azatov:2015oxa,Liu:2014rba,Goertz:2014qta,He:2015spf,Grober:2015cwa,Cao:2015oaa}, resonant production
in models such as extra dimensions~\cite{Gouzevitch:2013qca,Cooper:2013kia,No:2013wsa,Wen-Juan:2015gqg}, and Supersymmetry and
Two Higgs Doublet models (2HDMs)~\cite{Belyaev:1999kk,Han:2013sga,Hespel:2014sla,Wu:2015nba,Cao:2014kya,Ellwanger:2013ova,Cao:2013si}.
Since BSM dynamics modify
the kinematic distributions of the Higgs decay products, for
instance boosting the di-Higgs system,
different analysis strategies  might be required for BSM
Higgs pair searches as compared to SM measurements.

Searches for the production of Higgs pairs
have already been performed with 8 TeV Run I data
by ATLAS in the $b\bar{b}b\bar{b}$~\cite{Aad:2015uka}
and $b\bar{b}\gamma\gamma$~\cite{Aad:2014yja} final states,
and by
CMS in the same $b\bar{b}b\bar{b}$~\cite{Khachatryan:2015yea}
and $b\bar{b}\gamma\gamma$~\cite{Chatrchyan:2011wt} final
states.
In addition, ATLAS has presented~\cite{Aad:2015xja} a combination
of its di-Higgs searches in the $bb\tau\tau,$
$\gamma\gamma WW^*$, $\gamma\gamma bb$, and $bbbb$ final states.
Many other exotic searches involve Higgs pairs in the final
state, such as the recent
search for heavy Higgs bosons $H$~\cite{Khachatryan:2015tha}.

In the context of SM production,
the main advantage of the $b\bar{b}b\bar{b}$ final state is the
enhancement of the signal yield
from the large branching fraction of Higgs bosons into
$b\bar{b}$
pairs, ${\rm BR}\lp H\to b\bar{b}\rp\simeq 0.57$~\cite{Dittmaier:2012vm}.
However a measurement in this channel
needs to deal with an overwhelming QCD multi-jet background.
Recent studies of Higgs pair production in this
final state~\cite{Wardrope:2014kya,deLima:2014dta}
estimate that, for an integrated
luminosity of
$\mathcal{L}=3$ ab$^{-1}$,
a signal significance of around $S/\sqrt{B}\simeq 2.0$ can be obtained.
In these analysis, irreducible backgrounds such as $4b$ and
$t\bar{t}$ are included, however the
reducible components, in particular $bbjj$ and
$jjjj$, are neglected.
These can contribute to the signal yield when 
light and charm jets are mis-identified as $b$-jets.
Indeed, due to both
selection effects and $b$-quark radiation in the
parton shower, the
contribution of the $2b2j$ process is as significant as
the irreducible $4b$ component.

In this work, we revisit the feasibility of SM Higgs pair production by
gluon-fusion
in the $b\bar{b}b\bar{b}$ final state at the LHC.
 Our strategy is based upon a combination of traditional cut-based
 methods and multivariate analysis (MVA).
  We account for  all relevant
  backgrounds, including the contribution from mis-identified
  light and charm jets.
  We also assess the robustness of our analysis strategy in
  an environment with high pileup (PU).
 Our results indicate that 
  the $b\bar{b}b\bar{b}$ 
final state
alone should allow for the observation of double Higgs production
  at the HL-LHC.

The structure of this paper proceeds as follows.
In Sect.~\ref{mcgeneration} we present the modeling of the signal
and background processes with Monte Carlo event generators.
In Sect.~\ref{sec:analysis}
we introduce our analysis strategy, in particular
the classification of individual events into
different categories according to their topology.
Results of the cut-based analysis
are then presented in Sect.~\ref{sec:results}.
In Sect.~\ref{sec:mva} we illustrate the enhancement of signal
significance using multivariate techniques, and
assess the robustness of our results against the effects of PU.
In Sect.~\ref{sec:conclusions} we conclude and outline
future studies to estimate the accuracy
in the determination of the trilinear coupling $\lambda$ and
to provide
 constraints in
BSM scenarios.


\section{Modeling of signal and background processes}
\label{mcgeneration}

In this section we discuss the Monte Carlo generation of the signal and background
process samples used in this analysis.
We shall also discuss the modelling of detector
resolution effects.

\subsection{Higgs pair production in gluon-fusion}

Higgs pair production is simulated at leading order (LO) using
{\tt MadGraph5\_aMC@NLO}~\cite{Alwall:2014hca}.
We use a tailored  model~\cite{Maltoni:2014eza}
for gluon-fusion Higgs boson pair production
which includes mass effects
from the
exact form factors for the top-quark triangle and box
loops~\cite{Plehn:1996wb}.
Equivalent results can be obtained using
the recently available functionalities
for the calculation of loop-induced processes~\cite{Hirschi:2015iia}
in {\tt MadGraph5\_aMC@NLO}.
The calculation is performed in the
$n_f$=4 scheme,  accounting for  $b$-quark mass effects. 
The renormalization and factorization
scales are taken to be $\mu_F=\mu_R=H_T/2$,
with
\be
H_T\equiv \sum_i \sqrt{p_{T,i}^2+m_i^2} \, ,
\ee
the scalar sum of the
transverse masses of all final state particles.
For the input parton distribution functions (PDFs) we 
adopt the NNPDF 3.0 $n_f=4$ LO set~\cite{Ball:2014uwa} with
$\alpha_s(m_Z^2)=0.118$,
interfaced via {\tt LHAPDF6}~\cite{Buckley:2014ana}.
The Higgs boson couplings
and branching ratios are set to their SM values,
and its mass is taken to be
$m_h=125$ GeV~\cite{Aad:2014aba,Khachatryan:2014jba,Aad:2015zhl}.
In the SM, the Higgs trilinear coupling
is given by $\lambda=m_h^2/2v^2$, with
$v\simeq 246$ GeV the Higgs vacuum expectation
value.
%

\begin{figure}[t]
\begin{center}
  \includegraphics[width=0.96\textwidth]{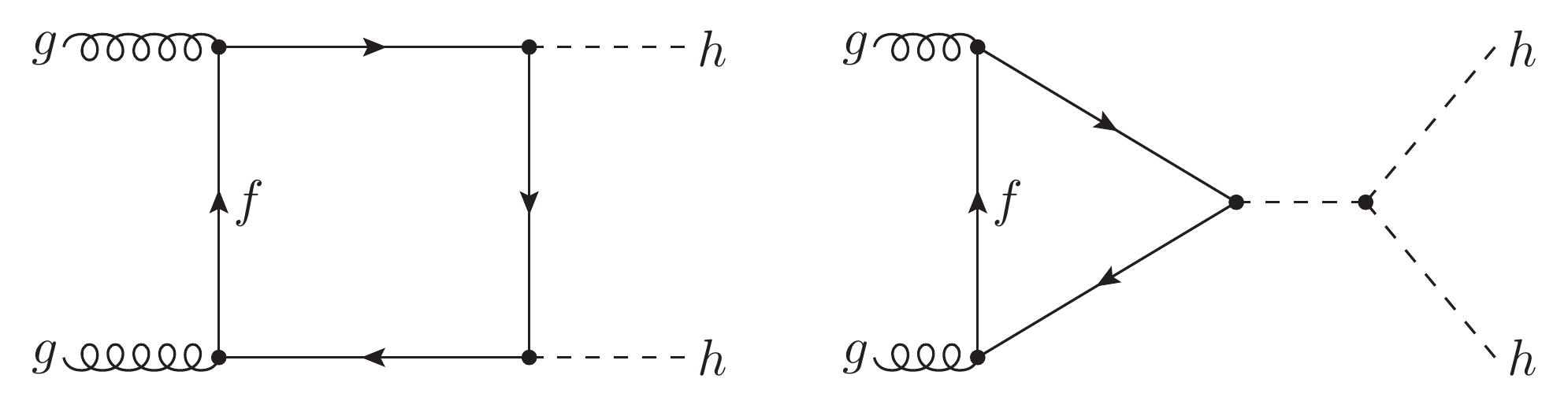}
  \caption{\small Representative Feynman diagrams
    for Higgs pair production in gluon fusion at
    leading order.
    Only the fermion triangle loop diagram (right) is
    directly sensitive to the Higgs trilinear coupling
    $\lambda$.
    In the SM, the fermion loops are dominated by the
    contribution from the top quark.
}
\label{fig:hhFeyn}
\end{center}
\end{figure}

In Fig.~\ref{fig:hhFeyn} we show representative Feynman diagrams
    for LO Higgs pair production in gluon fusion.
    The non-trivial interplay between the heavy quark box and the triangle loop diagrams
    can lead to either constructive or destructive interference
    and complicates the extraction of
    the trilinear coupling
    $\lambda$ from the measurement of the Higgs pair
    production cross-section.
    Higher-order corrections~\cite{deFlorian:2013jea,Frederix:2014hta}
    are dominated by gluon radiation
    from either the initial state gluons or from the heavy quark loops.

    The total inclusive cross-section for this processes is
    known up to NNLO~\cite{deFlorian:2013jea}.
    Resummed NNLO+NNLL calculations for Higgs pair production are
    also available~\cite{deFlorian:2015moa},
leading to a moderate enhancement of the order of
few percent as compared to the fixed-order NNLO calculation.
To achieve the correct higher-order value of the
integrated cross-section, we rescale our LO signal sample to match the
NNLO+NNLL
inclusive calculation.
This corresponds to
a $K$-factor $\sigma_{\rm NNLO+NNLL}/\sigma_{\rm LO}=2.4$, as indicated
in Table~\ref{tab:samples}.

Parton level signal events are then showered with the {\tt Pythia8} Monte
Carlo~\cite{Sjostrand:2007gs,Sjostrand:2014zea}, version {\tt v8.201}.
We use the default settings for the modeling
of the underlying event (UE), multiple parton
interactions (MPI), and PU, by means
of the Monash 2013 tune~\cite{Skands:2014pea},
based on the NNPDF2.3LO PDF set~\cite{Ball:2012cx,Ball:2013hta}.

\subsection{Backgrounds}

Background samples are generated at leading order
with {\tt SHERPA}~\cite{Gleisberg:2008ta} {\tt v2.1.1}.
As in the case of the signal generation,
the NNPDF 3.0 $n_f = 4$ LO set with strong coupling
$\alpha_s(m_Z^2)=0.118$ is used for all samples, and
we use as
factorisation and renormalisation scales $\mu_F=\mu_R=H_T/2$.
We account for all relevant background
processes that can mimic the
 $hh\to 4b$ signal process.
This includes  QCD $4b$ multi-jet production, as well as
QCD $2b2j$ and $4j$ production, and top quark pair
production.
The latter is restricted to the fully hadronic final state,
since 
leptonic decays of top quarks can be removed by requiring
a lepton veto.
Single Higgs production processes such as $Z(\to b\bar{b})h(\to b\bar{b})$
and $t\bar{t}h(\to b\bar{b})$ (see Appendix~\ref{app:singlehiggs})
along with electroweak backgrounds {\it e.g} $Z(\to b\bar{b})b\bar{b}$,
are much smaller than the
QCD backgrounds~\cite{Wardrope:2014kya,deLima:2014dta}
and are therefore not included in the present analysis.

The LO cross-sections for
the background samples have been rescaled so that the integrated
distributions reproduce known higher-order QCD results.
For the $4j$ sample, we rescale the LO cross-section
using the {\tt BLACKHAT}~\cite{Bern:2011ep}
calculation, resulting in
an NLO/LO $K$-factor of 0.6.
For the $4b$ and $2b2j$ samples NLO/LO $K$-factors of 1.6 and 1.3
respectively have been determined
using {\tt MadGraph5\_aMC@NLO}~\cite{Alwall:2014hca}.
Finally, the LO cross-section for $t\bar{t}$ production has been rescaled
to match the NNLO+NNLL calculation of Ref.~\cite{Czakon:2013goa}, leading
to a $K$-factor of 1.4.
The $K$-factors that we use to rescale
the signal and background samples are summarised in
Table~\ref{tab:samples}.

\begin{table}[h]
  \small
\begin{center}
\begin{tabular}{|c|c|c|c|c|c|}
\hline
Process &  Generator & $N_{\mathrm{evt}}$ & $\sigma_{\mathrm{LO}}$ (pb)  & $K$-factor \\
\hline
\hline
$pp \to hh\to 4b$ &  {\tt MadGraph5\_aMC@NLO} & 1M & $6.2\cdot10^{-3}$  &  2.4  (NNLO+NNLL~\cite{deFlorian:2013jea,deFlorian:2015moa}) \\
\hline
\hline
$pp \to b\bar{b}b\bar{b}$ &  {\tt SHERPA} & 3M &$1.1 \cdot10^3$  & 1.6 (NLO~\cite{Alwall:2014hca}) \\
$pp \to b\bar{b}jj$ &  {\tt SHERPA} & 3M & $2.7 \cdot 10^5$ & 1.3 (NLO~\cite{Alwall:2014hca}) \\
$pp \to jjjj$ &  {\tt SHERPA} & 3M  & $9.7\cdot 10^6$ &  0.6 (NLO~\cite{Bern:2011ep})\\
$pp \to t\bar{t}\to b\bar{b}jjjj$ &  {\tt SHERPA} & 3M & $2.5\cdot 10^3$   & 1.4 (NNLO+NNLL~\cite{Czakon:2013goa})\\
\hline
\end{tabular}
\caption{\small Details of the signal and background Monte
  Carlo samples used in this work.
  Also provided are the inclusive $K$-factors
  which are applied to reproduce the known
  higher-order results. \label{tab:samples}
} 
\end{center}
\end{table}%

At the generation level, the following loose selection 
cuts are applied to
background events.
Each final-state particle in the hard process must have $p_T \ge 20$ GeV, and be located
in the central  rapidity
region with
$| \eta | \le 3.0$.
At the matrix-element level
all final-state particles must also be separated by a minimum $\Delta R_{\mathrm{min}} =0.1$.
We have checked that these generator-level cuts are loose enough to have
no influence over the analysis cuts.
From Table~\ref{tab:samples}
we see that the $t\bar{t}$ and QCD $4b$ cross-sections are of
the same order of magnitude. However the former can be efficiently
reduced by using top quark reconstruction criteria.
The $bbjj$ cross-section is more than two orders
of magnitude larger than the $4b$ result, but will be suppressed
by the light and charm jet mis-identification rates,
required to contribute to the $4b$ final state.

As a cross-check of the {\tt SHERPA}
background cross-sections reported in Table~\ref{tab:samples}, we have produced leading-order
multi-jet samples
using {\tt MadGraph5\_aMC@NLO},
benchmarked with the results for the same processes reported in
Ref.~\cite{Alwall:2014hca}.
Using common settings, we find
agreement, within scale uncertainties,  between the
{\tt MadGraph5\_aMC@NLO} and {\tt SHERPA} calculations of
the multi-jet backgrounds.

\subsection{Modelling of detector resolution}
\label{sec:detectormodeling}

While it is beyond the scope of this work to perform a full
detector simulation, it is important to include an estimate of detector
effects in the analysis, particularly for the finite energy
and angular resolutions which directly
degrade the reconstruction of important kinematic variables, such as
the invariant mass of the Higgs candidates.
Here we simulate the finite energy resolution of the ATLAS and CMS
hadronic calorimeters by applying a Gaussian smearing of the transverse
momentum $p_T$ with mean zero and standard deviation $\sigma_E$ for all
final-state particles before jet clustering, that is,
\be
\label{eq:smearing}
p_T^{(i)} \, \to \, p_T^{(i)\prime}= \lp 1+ r_i\cdot\sigma_E \rp\, p_T^{(i)} \, , \quad
i=1,\ldots,N_{\rm part} \, ,
\ee
with $r_i$ a univariate Gaussian random number, different for each
of the $N_{\rm part}$ particles in the event.
We take as a baseline value for the transverse momentum smearing a
factor of $\sigma_E=5\%$.

To account for the finite angular resolution of the calorimeter,
the $\lp \eta,\phi\rp$ plane is divided into regions of
$\Delta \eta \times \Delta \phi=0.1\times 0.1$,
and each final state particle
which falls in each of these cells is set to the same $\eta$
and $\phi$ values of the center of the
corresponding cell.
Finally, the energy
of each final-state particle
is recalculated from the smeared $p_T^\prime$,
$\eta^\prime$ and $\phi^\prime$ values to ensure that the resulting
four-momentum is that of a light-like particle, since we neglect all
jet constituent masses in this analysis.

Our modelling of detector simulation has been tuned
to lead to a mass resolution of
the reconstructed Higgs candidates consistent
with the hadronic mass resolutions of the
ATLAS and CMS detectors~\cite{Aad:2012gxa,Chatrchyan:2013zna,Aad:2014xzb},
as discussed in Sect.~\ref{sec:pileup}.

\section{Analysis strategy}
\label{sec:analysis}

In this section we describe our analysis strategy.
First of all we discuss the settings
for jet clustering and the strategy for jet $b$-tagging.
Following this we discuss the categorisation of events into different
topologies, and how the different topologies may be prioritised.
We motivate our choice of analysis cuts by comparing signal and background
distributions for representative kinematic variables.
Finally, we describe the simulation of PU and validate
the PU subtraction strategy.

\subsection{Jet reconstruction}

After the parton shower, final state particles
are clustered using the
jet reconstruction algorithms
of
{\tt FastJet}~\cite{Cacciari:2011ma,Cacciari:2005hq},
{\tt v3.1.0}.
Here we use the following jet definitions:
\begin{itemize}
\item {\it Small-$R$ jets}.

  These are jets  reconstructed with the
  anti-$k_T$ clustering algorithm~\cite{Cacciari:2008gp} with $R=0.4$ radius.
  These small-$R$ jets are required
  to have transverse momentum $p_T \ge 40$~GeV
  and pseudo-rapidity $|\eta|<2.5$, within the central 
  acceptance of ATLAS and CMS, and therefore within the region
  where $b$-tagging is possible.

\item {\it Large-$R$ jets}.

  These jets are also constructed with the
  anti-$k_T$ clustering algorithm, now using a $R=1.0$ radius.
  Large-$R$ jets are required to have
  $p_T \ge 200$~GeV and lie in a pseudo-rapidity region of
  $|\eta|<2.0$.
  The more restrictive range  in pseudo-rapidity
  as compared to the small-$R$ jets
  is motivated by mimicking the  experimental requirements
  in ATLAS and CMS
  related to the track-jet based calibration~\cite{Aad:2014bia,ATLAS:2012kla}.

  In addition to the basic $p_T$ and $\eta$
  acceptance requirements, large-$R$ jets should also
  satisfy the  BDRS mass-drop tagger (MDT)~\cite{Butterworth:2008iy}
  conditions, where the {\tt FastJet} default
  parameters of  $\mu_{\rm mdt} = 0.67$ and $y_{\rm mdt}=0.09$ are used.
  Before applying the BDRS tagger, the large-$R$ jet
  constituents are reclustered with the Cambridge/Aachen (C/A)
  algorithm~\cite{Dokshitzer:1997in,Wobisch:1998wt}
  with $R=1.0$.

  In the case of the analysis including PU, a trimming
  algorithm~\cite{Krohn:2009th}
  is applied to all large-$R$ jets to mitigate the effects of PU,
  especially on the jet mass.
  For further details see Sect.~\ref{sec:pileup}.
  
\item {\it Small-$R$ subjets}.

 All final-state particles are clustered using the
 anti-$k_T$ algorithm, but this time with
 a smaller radius parameter, namely $R=0.3$.
 The resulting anti-$k_T$ $R=0.3$ (AKT03) jets
 are then ghost-associated to each large-$R$ jets
 in order to define its subjets~\cite{Aad:2015uka}.

 These AKT03 subjets
  are required to satisfy
  $p_T > 50$~GeV and $|\eta|<2.5$, and
   will be the main input for
  $b$-tagging in the boosted category.
\end{itemize}

For the   boosted and intermediate categories,
which involve the use of large-$R$ jets,
we use jet substructure variables~\cite{Salam:2009jx,Aad:2013gja} to
improve the significance of the discrimination between signal and background
events in the MVA.
In particular we  consider the following
substructure variables:
\begin{itemize}
\item The $k_T$-splitting scale~\cite{Butterworth:2002tt,Butterworth:2008iy}.

  This variable is obtained by reclustering the constituents of a jet with the
  $k_t$ algorithm~\cite{Ellis:1993tq},
  which usually clusters last the harder constituents, and then
  taking the $k_t$ distance measure between the two subjets at the final stage of the recombination
  procedure,
  \be
  \label{eq:ktsplitting}
\sqrt{d_{12}} \equiv {\rm min}\lp p_{T,1},p_{T,2}\rp \cdot \Delta R_{12} \, .
\ee
with $p_{T,1}$ and $p_{T,2}$ the transverse momenta of the two subjets merged
in the final step of the clustering, and $\Delta R_{12}$ the corresponding
angular separation.
  
\item The ratio of 2-to-1 subjettiness $\tau_{21}$~\cite{Thaler:2010tr,Thaler:2011gf}.

  The $N$-subjettiness variables $\tau_N$ are defined by clustering the constituents
  of a jet with the exclusive $k_t$ algorithm~\cite{Catani:1993hr}
  and requiring that $N$ subjets are found,
  \be
  \tau_N \equiv \frac{1}{d_0} \sum_k p_{T,k}\cdot {\rm min}\lp \delta R_{1k}, \ldots,
  \delta R_{Nk}\rp \, , \qquad d_0\equiv \sum_k p_{T,k}\cdot R \, ,
  \ee
  where $p_{T,k}$ is the $p_T$ of the constituent particle $k$ and $\delta R_{ik}$ the distance from
  subjet $i$ to constituent $k$.
  In this work we use as input to the MVA the ratio of 2-subjettiness to 1-subjettiness, namely
  \be
  \label{eq:tau21}
\tau_{21} \equiv \frac{\tau_2}{\tau_1} \, ,
  \ee
  which provides good discrimination 
  between QCD jets and jets arising from the decay of
  a heavy resonance.
  
\item The ratios of energy correlation functions (ECFs)  $C^{(\beta)}_2$~\cite{Larkoski:2013eya} and
  $D_2^{(\beta)}$~\cite{Larkoski:2014gra}.

  The ratio of energy correlation functions $C_2^{(\beta)}$ is defined as
  \be
  \label{eq:c2}
C_2^{(\beta)} \equiv \frac{ {\rm ECF}(3,\beta) {\rm ECF}(1,\beta)}{\lc {\rm ECF}(2,\beta)\rc ^2} \, ,
\ee
while $D_2^{(\beta)}$ is instead defined as a double ratio of ECFs, that is,
\be
e_3^{(\beta)}\equiv \frac{ {\rm ECF}(3,\beta)}{\lc {\rm ECF}(1,\beta)\rc^3} \, , \quad
  e_2^{(\beta)}\equiv \frac{ {\rm ECF}(2,\beta)}{\lc {\rm ECF}(1,\beta)\rc^2} \, , \quad
  \label{eq:d2}
D_2^{(\beta)} \equiv \frac{ e_3^{(\beta)})}{\lp e_2^{(\beta)} \rp^3} \, .
\ee
The energy correlation functions ${\rm ECF}(N,\beta)$ are defined
  in~\cite{Larkoski:2013eya} with the motivation that $(N+1)$-point correlators
  are sensitive to $N$-prong substructure.
  The free parameter $\beta$ is set to a value of $\beta=2$,
  as recommended by Refs.~\cite{Larkoski:2013eya,Larkoski:2014gra}.
\end{itemize}

\subsection{Tagging of $b$-jets}
\label{sec:btagging}

In this analysis we adopt
a $b$-tagging strategy along the lines
of current ATLAS performance~\cite{Aad:2013gja,Aad:2015ydr},
though differences with respect to
the corresponding CMS
settings~\cite{Khachatryan:2011wq,Chatrchyan:2012jua}
do not modify qualitatively our results.
For each jet definition described above, a different
$b$-tagging strategy is adopted:

\begin{itemize}

\item {\it Small-$R$ jets}.

  If a small-$R$ jet has at least one $b$-quark among their constituents,
  it will be tagged as a $b$-jet with probability $f_b$.
  In order to be considered in the $b$-tagging algorithm,
  $b$-quarks inside the small-$R$ jet
  should satisfy $p_T \ge 15$ GeV~\cite{Aad:2015ydr}.
  The probability of tagging a jet is not modified
  if more than one $b$-quark is found among the jet constituents.

  If no $b$-quarks are found among the constituents
  of this jet, it can be still be tagged as a $b$-jet with
  a mistag rate of $f_l$, unless a charm quark is present instead,
  and in this case the mistag rate is $f_c$.
  Only jets that contain at least one (light or charm)
  constituent
  with $p_T \ge 15$ GeV can induce a fake $b$-tag.

  We attempt to $b$-tag only the four (two) hardest small-$R$ jets
  in the resolved (intermediate) category.
  Attempting to $b$-tag all of the
  small-$R$ jets that satisfy the acceptance cuts worsens the
  overall performance as the probability of fake $b$-tags increases
  substantially.

  \item {\it Large-$R$ jets}.

    Large-$R$ jets are $b$-tagged by
    ghost-associating anti-$k_T$ $R=0.3$ (AKT03)
    subjets to the original large-$R$
    jets~\cite{Cacciari:2007fd,Aad:2013gja,
      ATLAS-CONF-2014-004,Aad:2015uka}.
    A large-$R$ jet is considered $b$-tagged if both
    the leading and subleading AKT03 subjets, where the ordering
    is done in the subjet $p_T$, are both individually $b$-tagged,
    with the same criteria as the small-$R$ jets.
     Therefore, a large-$R$ jet where the two leading
    subjets have at least one $b$-quark will be tagged
    with probability $f_b^2$.
    
    As in the case
    of small-$R$ jets, we only attempt to $b$-tag the two leading subjets,
    else one finds a degradation of the
    signal significance.
    The treatment of the $b$-jet mis-identification
    from light and charm jets
    is the same as for the small-$R$ jets.
  
\end{itemize}

For the $b$-tagging probability $f_b$, along with
the $b$-mistag probability of light ($f_l$) and charm ($f_c$) jets,
we use the values $f_b=0.8$, $f_l=0.01$
and  $f_c=0.1$.

\subsection{Event categorisation}
\label{sec:categorisation}

The present analysis follows a strategy similar to the
scale-invariant resonance tagging of Ref.~\cite{Gouzevitch:2013qca}.
Rather than restricting ourselves to a specific event topology,
we aim to consistently combine the information from
the three possible topologies: boosted, intermediate and
resolved, with the optimal cuts for each category being determined
separately.
This approach is robust
under variations of
the underlying production model of Higgs pairs,
for instance in the case of
BSM dynamics, which can substantially increase
the degree of boost in the final state.

The three categories are defined as follows:
\begin{itemize}
\item {\it Boosted category}.

  An event which
  contains at least two large-$R$ jets, with the two leading jets
 being $b$-tagged.
 Each of these two  $b$-tagged, large-$R$ jets are 
 therefore candidates
 to contain the decay products of a Higgs boson.

\item {\it Intermediate category}.

  An event with exactly one  $b$-tagged, large-$R$ jet, which
  is assigned to be the leading Higgs candidate.
  In addition, we require at least two $b$-tagged, small-$R$ jets,
  which must be separated with respect to the large-$R$ jet
  by an angular distance of $\Delta R\ge 1.2$.

  The subleading Higgs boson candidate is reconstructed
  by selecting the two $b$-tagged small-$R$ jets that minimize the difference
  between the invariant mass of the large-$R$ jet
  with that of the dijet obtained
  from the sum of the two small-$R$ jets.
  
\item {\it Resolved category}.

 An event with at least
  four $b$-tagged small-$R$ jets.
  The two Higgs candidates are reconstructed out of the
  leading four small-$R$ jets in the event
  by considering all possible combinations of forming two pairs of jets
  and then choosing the configuration that minimizes the relative difference of
  dijet masses.

\end{itemize}

Once a Higgs boson candidate has been identified,
its invariant mass is required to lie within a fixed window
of width $80~{\rm GeV}$ around the nominal Higgs boson mass of $m_h= 125$
GeV.
Specifically we require the condition
\be
\label{higgsmasswindow}
|m_{h,j} - 125~{\rm GeV}| < 40~{\rm GeV} \, ,\, j=1,2 \, ,
\ee
where $m_{h,j}$ is the invariant mass of each of the two reconstructed  Higgs candidates.
This cut is substantially looser than the corresponding
cut used in the typical ATLAS and CMS $h\to b\bar{b}$
analyses~\cite{Aad:2012gxa,Chatrchyan:2013zna}.
The motivation
for such a loose cut 
is that further improvements of the
signal significance will be obtained using an MVA.
Only events where the two Higgs candidates satisfy
Eq.~(\ref{higgsmasswindow}) are classified as signal events.

These three categories are not exclusive:
a given event can be assigned to more than one category, for
example, satisfying the requirements of both the intermediate
and resolved
categories at the same time.
The exception is the boosted and intermediate categories, which have
conflicting jet selection requirements.

This is achieved as follows.
First of all we perform an inclusive analysis, and optimise the
signal significance
$S/\sqrt{B}$ in each of the three categories separately, including
the MVA.
We find that the category with highest significance is
the boosted one,
followed by the intermediate and the resolved topologies, the latter two
with similar significance.
Therefore, when ascertaining in
which category an event is to be exclusively placed:
if the event satisfies the boosted requirements, it is assigned to
this category, else we check if it suits the intermediate
requirements.
If the event also fails the intermediate category
requirements, we
then check if it passes the resolved selection criteria.
The resulting exclusive event samples are then separately processed
through the MVA, allowing for a consistent combination
of the significance of the three event categories.

\subsection{Motivation for basic kinematic cuts}

We now motivate the
kinematic cuts applied to the different categories, 
comparing representative kinematic distributions between
signal and background events.
Firstly we present results without PU, and then
discuss
the impact of PU
on the description of the kinematic
distributions.
In the following, all
distributions are normalized to their total integral.

\begin{figure}[t]
\begin{center}
 \includegraphics[width=0.48\textwidth]{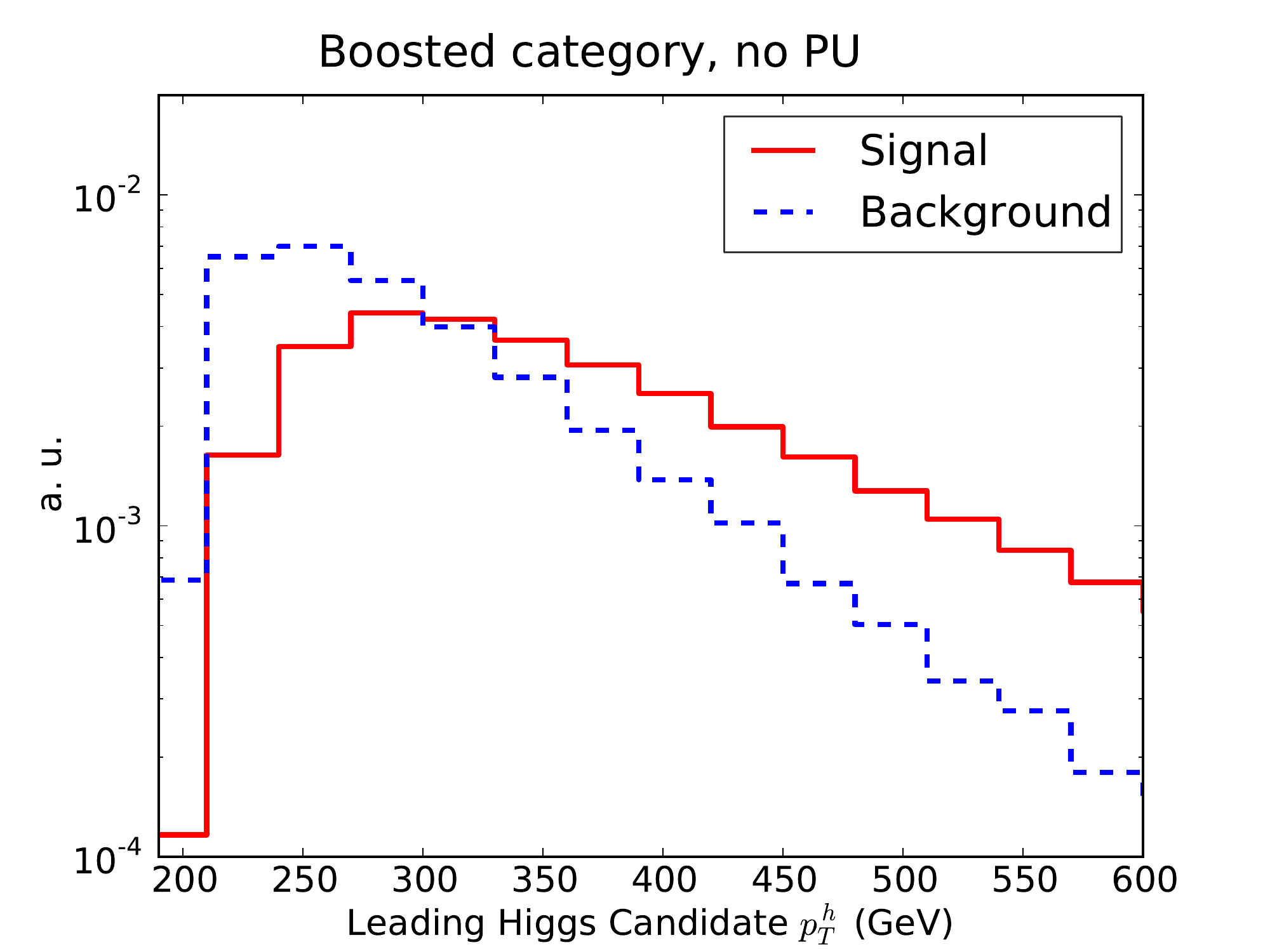}
 \includegraphics[width=0.48\textwidth]{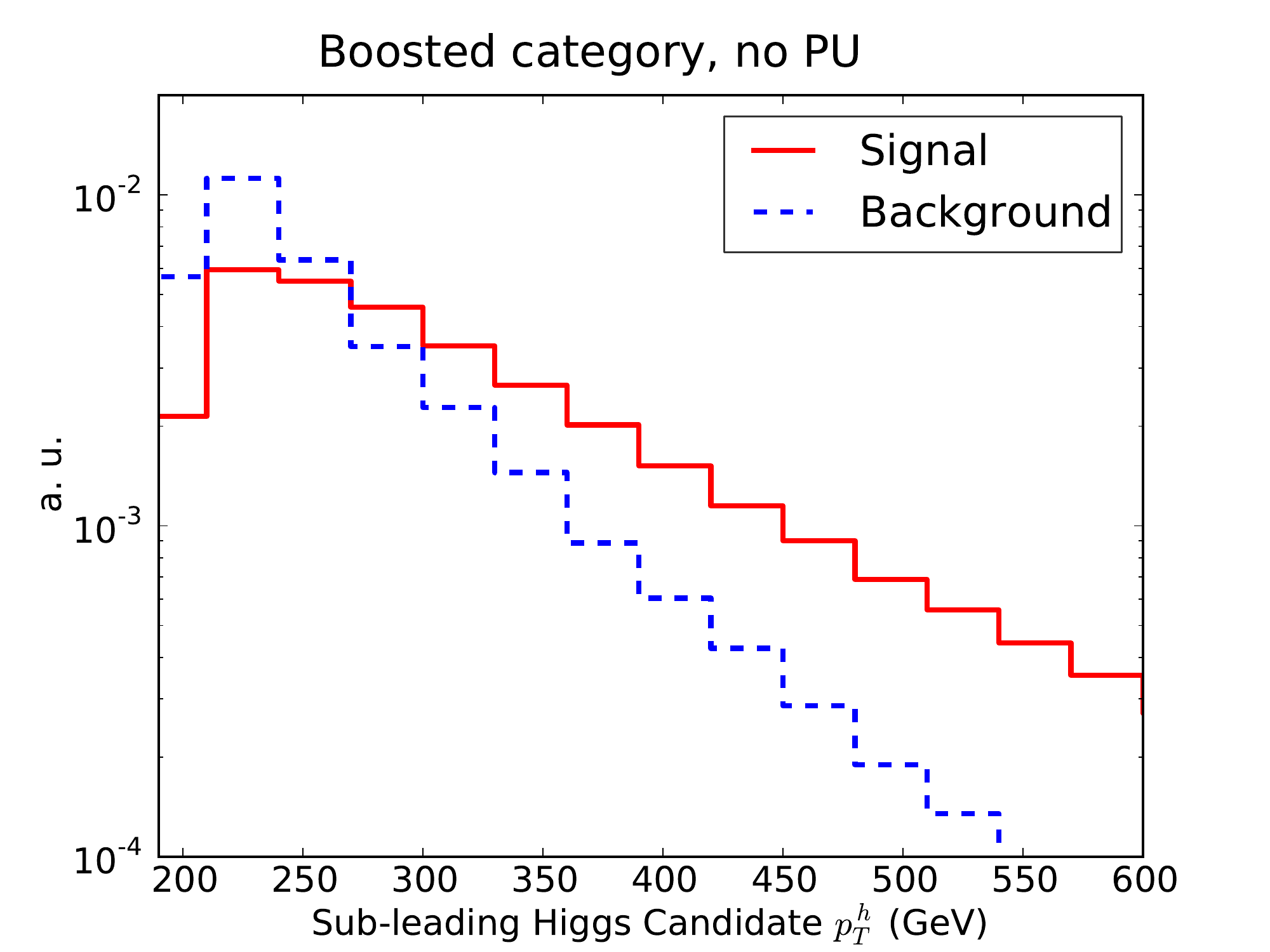}
\caption{\small  Comparison of the $p_T$ distributions of the
  leading (left) and
  subleading (right) large-$R$ jets in the boosted category,
  for signal and background events.
  Distributions have been normalized to unity.
  The total background is the sum of all components
  listed in Table~\ref{tab:samples}.
}
\label{fig:cutplots1}
\end{center}
\end{figure}

In Fig.~\ref{fig:cutplots1} we show
the $p_T$ distributions
of the
  leading and subleading large-$R$ jets in the boosted category.
  We observe that the background distribution
falls off more rapidly as a function of $p_T$ than the di-Higgs signal.
  On the other hand, the cut in $p_T$ cannot be too strong to avoid
  a substantial degradation of signal selection efficiency,
  specially taking into account the subleading large-$R$ jet.
  This comparison justifies the cut of $p_T \ge 200$ GeV
  for the large-$R$ jets that we impose in the boosted category.

\begin{figure}[t]
\begin{center}
 \includegraphics[width=0.48\textwidth]{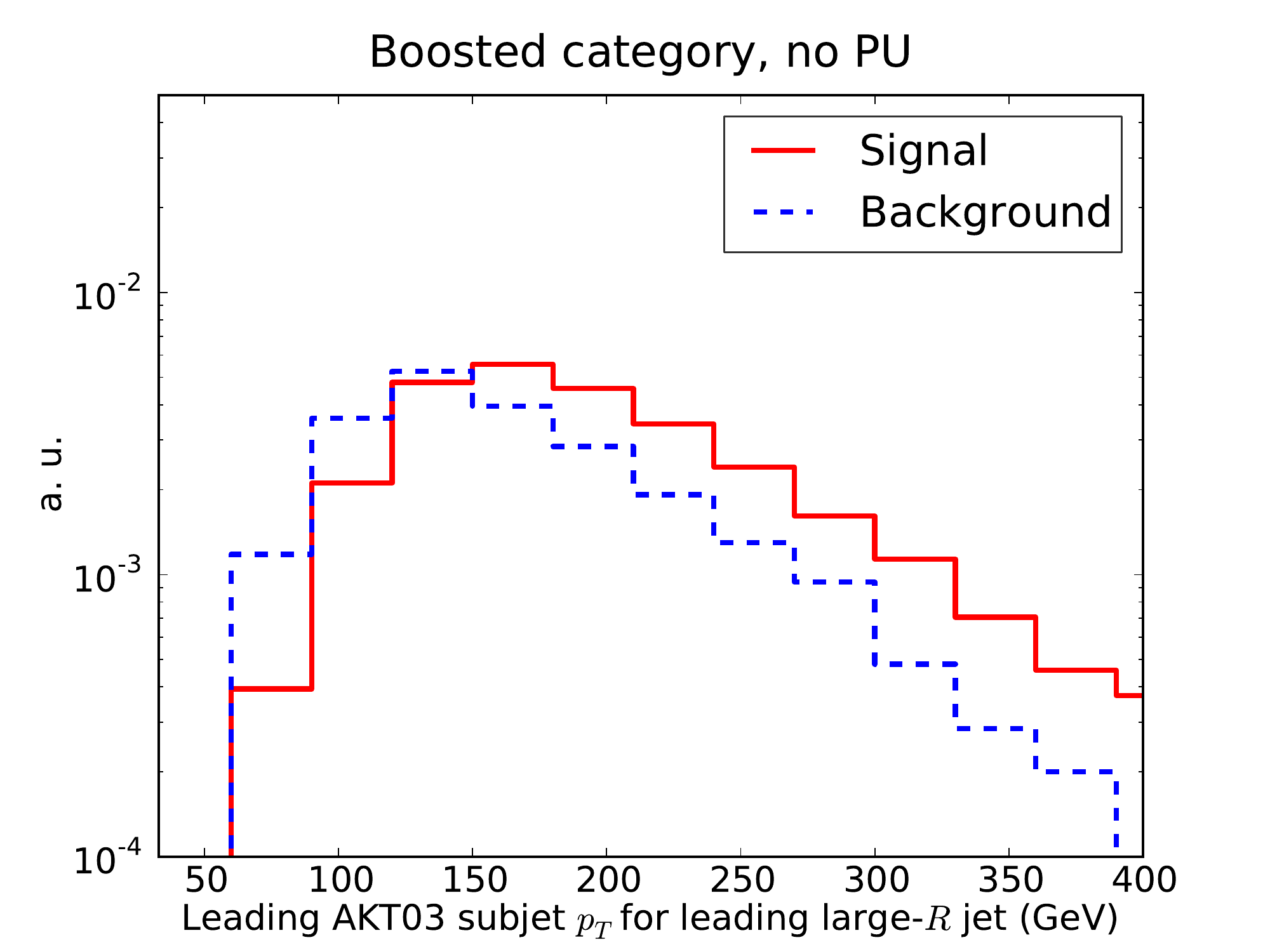}
 \includegraphics[width=0.48\textwidth]{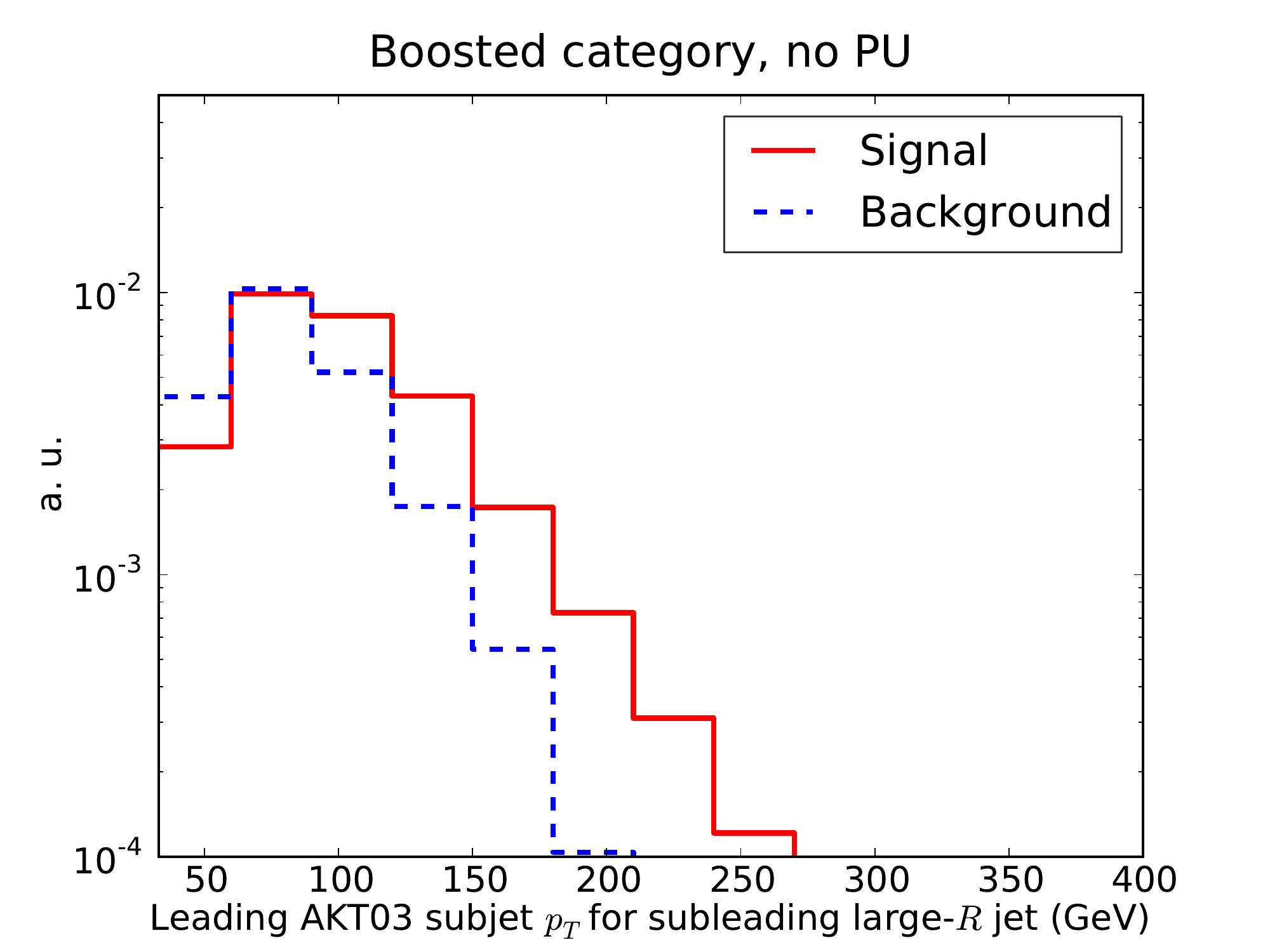}
 \caption{\small  Same as Fig.~\ref{fig:cutplots1} for the leading (left)
   and subleading (right) AKT03
   subjets in the subleading Higgs candidate large-$R$ jet.
}
\label{fig:cutplots22}
\end{center}
\end{figure}

Another selection requirement for the boosted category is that the two
leading AKT03 subjets of the large-$R$ jet
should satisfy $p_T \ge $ 50 GeV.
To motivate this cut, in Fig.~\ref{fig:cutplots22}
we show the distribution in $p_T$ of the leading
and subleading AKT03 subjets in the subleading large-$R$ jet in events
corresponding to the boosted category.
It is clear from the comparison that the subjet $p_T$ spectrum is
relatively harder in the signal with respect to the background.
On the other hand, considering the subleading AKT03 subjet,
this cut in $p_T$
cannot be too harsh, to maintain a high signal selection
efficiency.
Therefore,
as for the previous distribution, the chosen cut value is
a compromise between suppressing backgrounds but keeping a large fraction of
signal events is crucial.

\begin{figure}[t]
\begin{center}
 \includegraphics[width=0.48\textwidth]{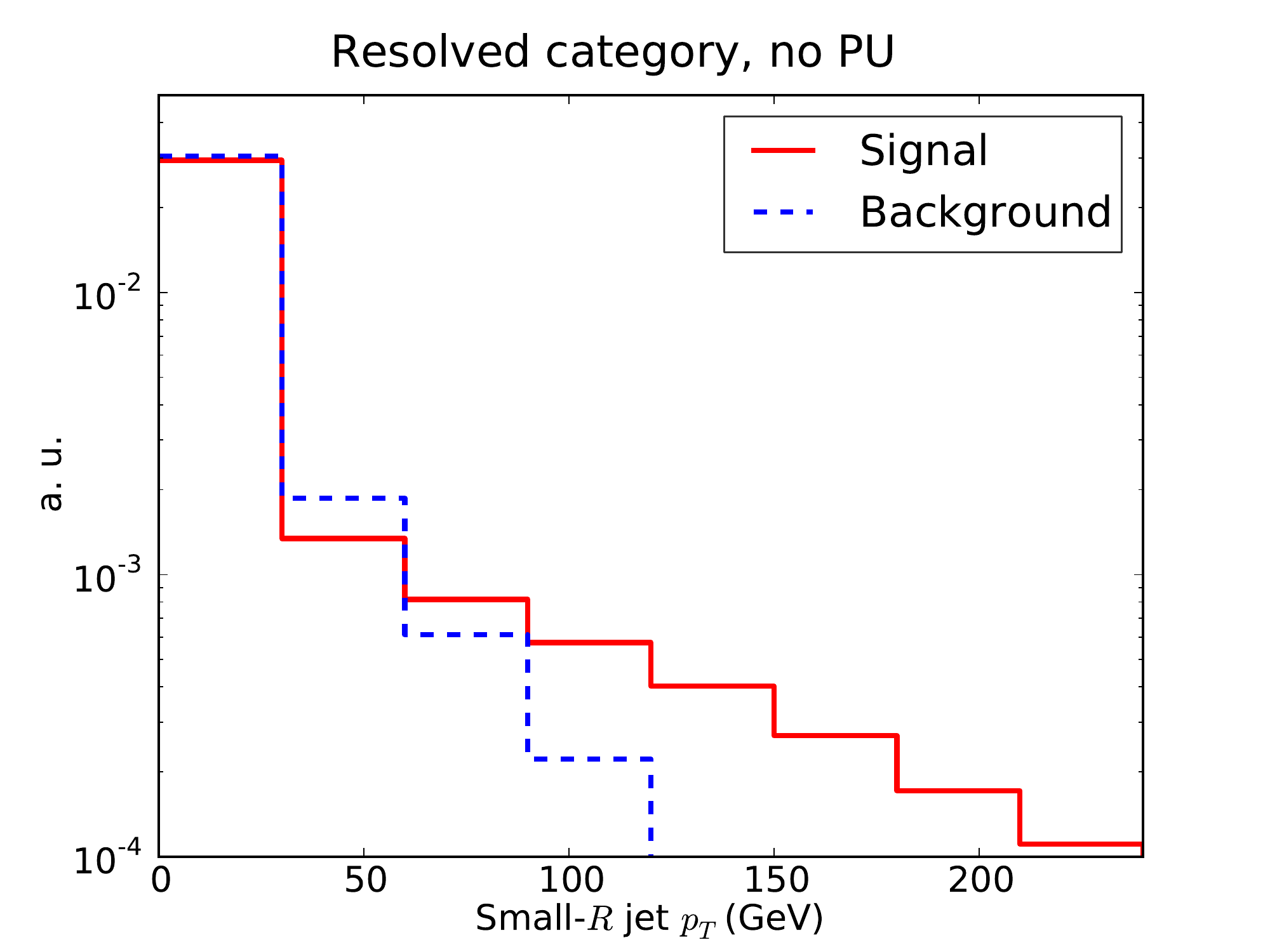}
 \includegraphics[width=0.48\textwidth]{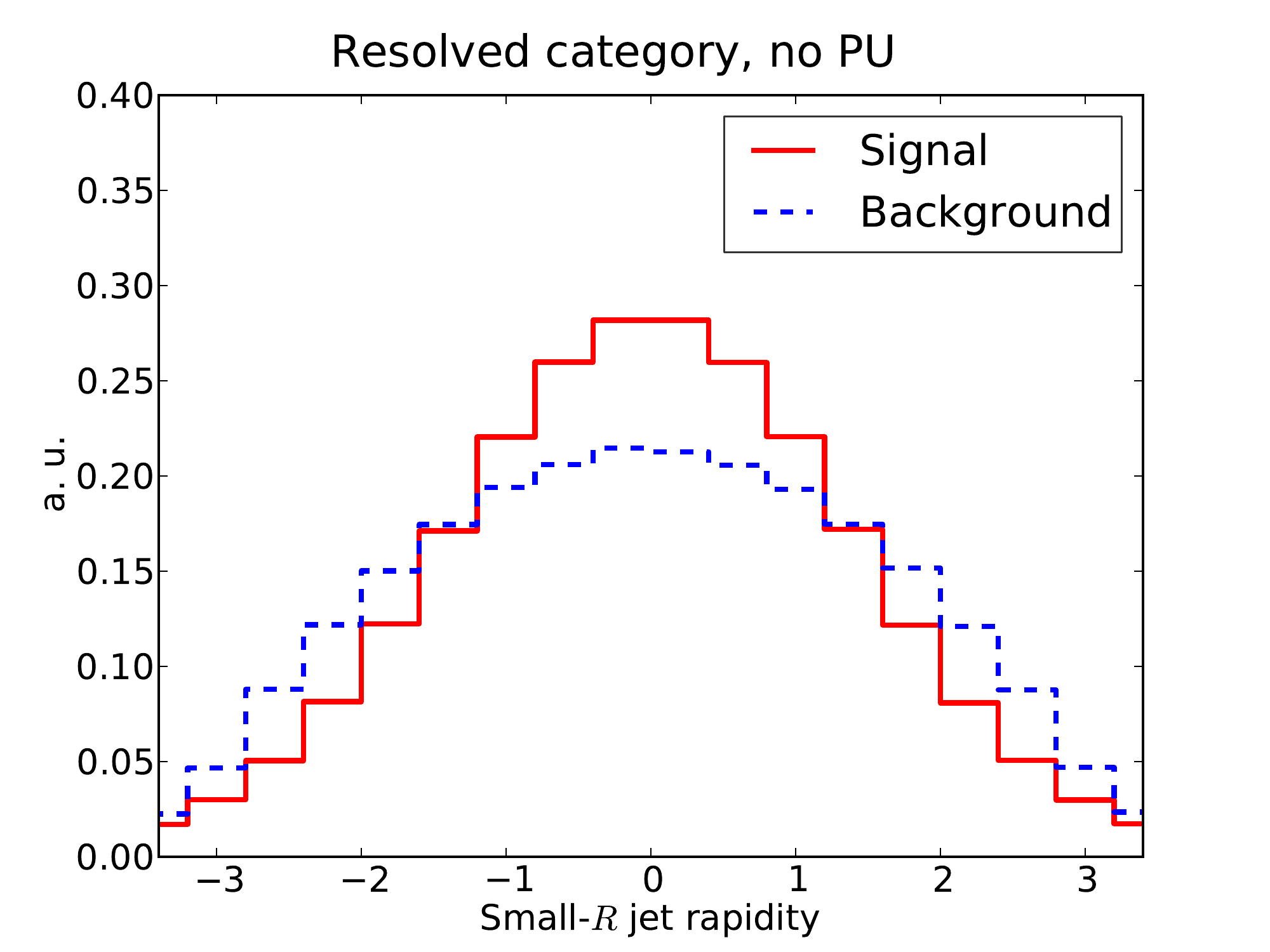}
 \caption{\small Same as Fig.~\ref{fig:cutplots1}, now for the
   $p_T$ and rapidity distributions of the small-$R$
   jets corresponding to the resolved selection.
}
\label{fig:cutplots23}
\end{center}
\end{figure}

Turning to the resolved category, an important aspect to account for
in the  selection
cuts is the fact that the $p_T$ distribution
of the four leading small-$R$ jets of the event can be relatively soft,
specially for the subleading jets.
As noted in~\cite{deLima:2014dta}, this is due to the fact
that the boost from the Higgs decay is moderate,
therefore the $p_T$ selection cuts for the small-$R$ jets cannot be too large.
In Fig.~\ref{fig:cutplots23}
we show the distribution in $p_T$ of the four leading
small-$R$ jets in signal and background events: we observe that both
distributions peak at $p_T \le 50$ GeV, with the signal distribution
falling off less steeply at large $p_T$.
The feasibility of triggering on four small-$R$ jets with a relatively
soft $p_T$ distribution is one of the experimental challenges for
exploiting the resolved category in this final state,
and hence the requirement that $p_T \ge 40$ GeV for
the small-$R$ jets.
In  Fig.~\ref{fig:cutplots23} we also show the
rapidity distribution of the the small-$R$
jets in the resolved category.
As expected, the production
is mostly central, more so in the case
of signal events, since backgrounds are dominated by
QCD $t$-channel exchange, therefore the
selection criteria on the jet rapidity are very efficient.

\begin{figure}[t]
\begin{center}
  \includegraphics[width=0.48\textwidth]{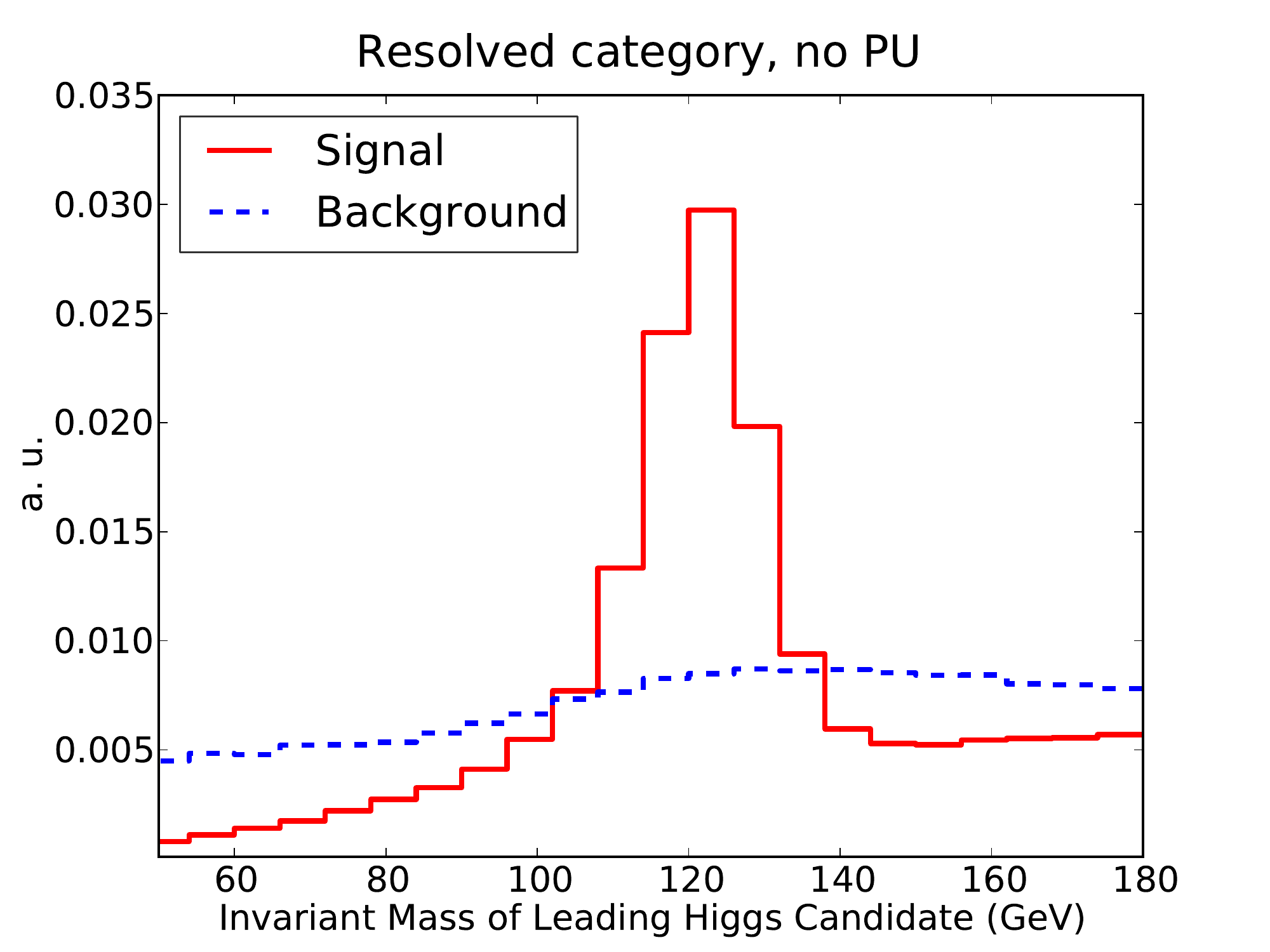}
  \includegraphics[width=0.48\textwidth]{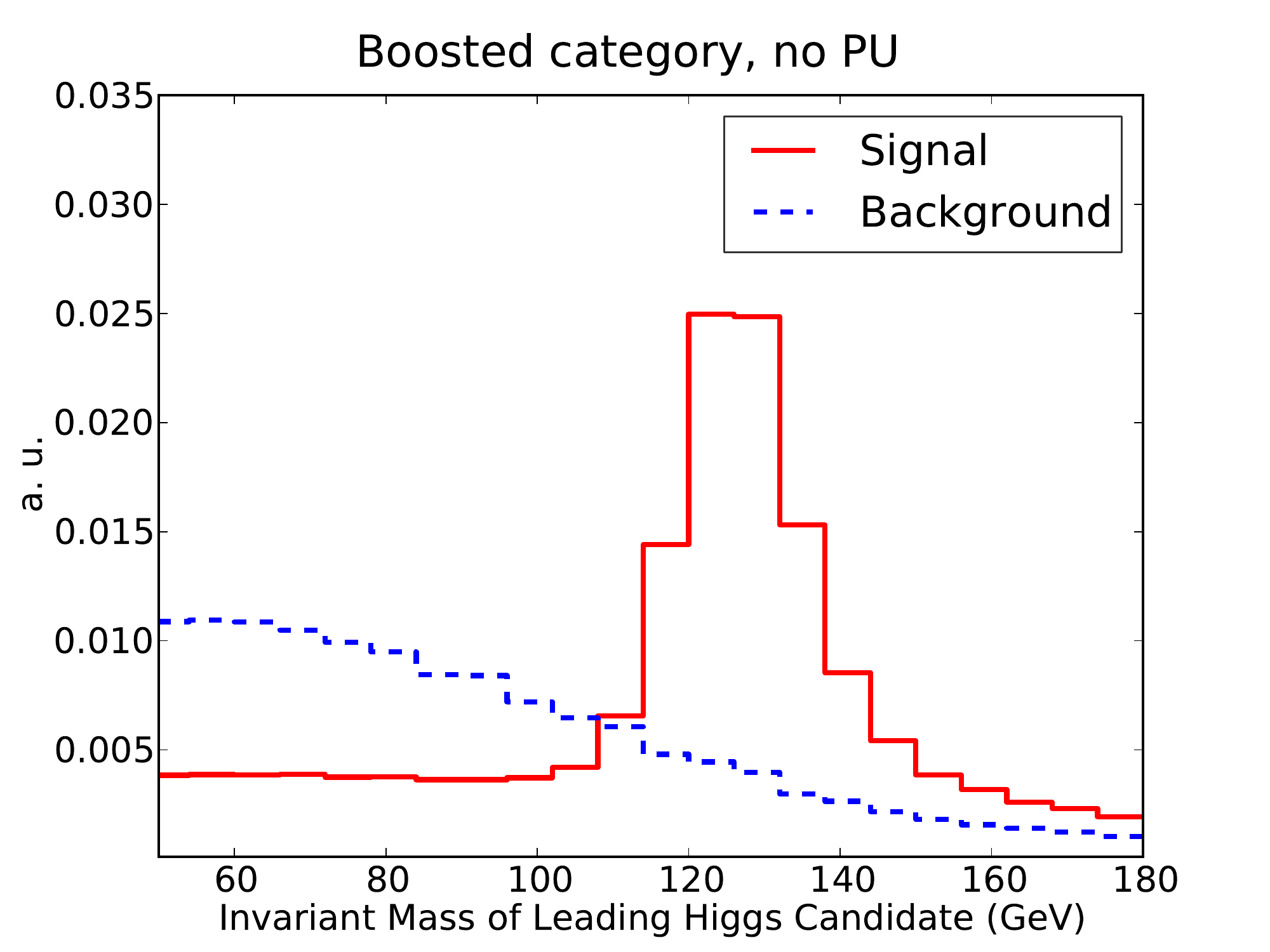}
  \caption{\small Same as   Fig.~\ref{fig:cutplots1} for the invariant
    mass distribution of the leading Higgs candidates in the resolved
    (left) and boosted (right) selections.
}
\label{fig:mHHinv}
\end{center}
\end{figure}

One of the most discriminating selection cuts is the requirement
that the invariant mass of the Higgs candidate (di)jets must lie within a window
around the nominal Higgs value, Eq.~(\ref{higgsmasswindow}).
In Fig.~\ref{fig:mHHinv} we show the invariant mass
of the leading reconstructed Higgs candidates, before the Higgs mass window
selection
  is applied, for the resolved and boosted categories.
While the signal distribution naturally peaks at the
nominal Higgs mass, the background distributions
show no particular
structure.
The
width of the Higgs mass peak is driven both from QCD effects,
such as initial-state radiation (ISR)
and out-of-cone radiation, as well
as from the four-momentum smearing applied to final state particles
as part of our minimal detector simulation.
%

\begin{figure}[t]
\begin{center}
  \includegraphics[width=0.48\textwidth]{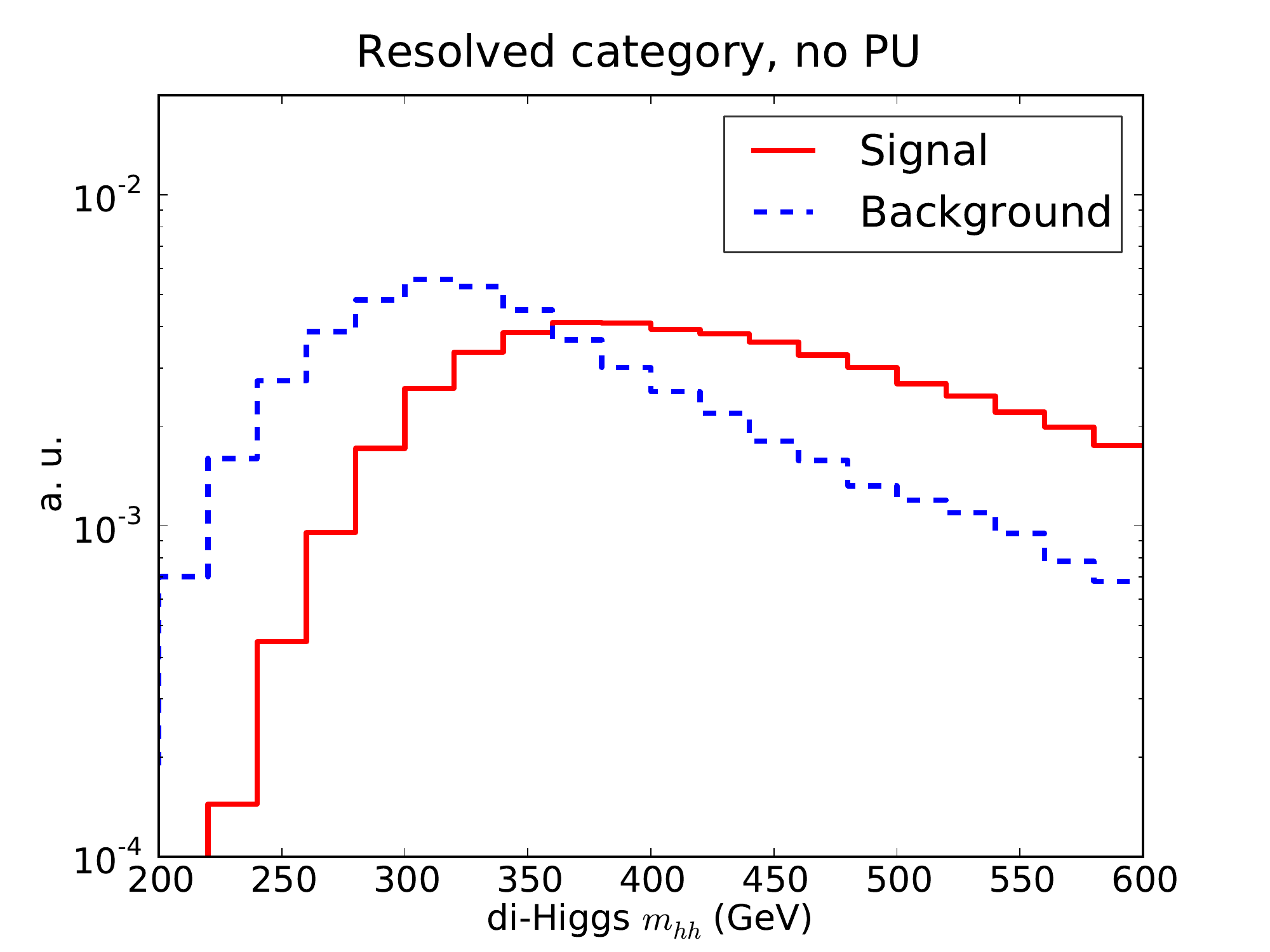}
  \includegraphics[width=0.48\textwidth]{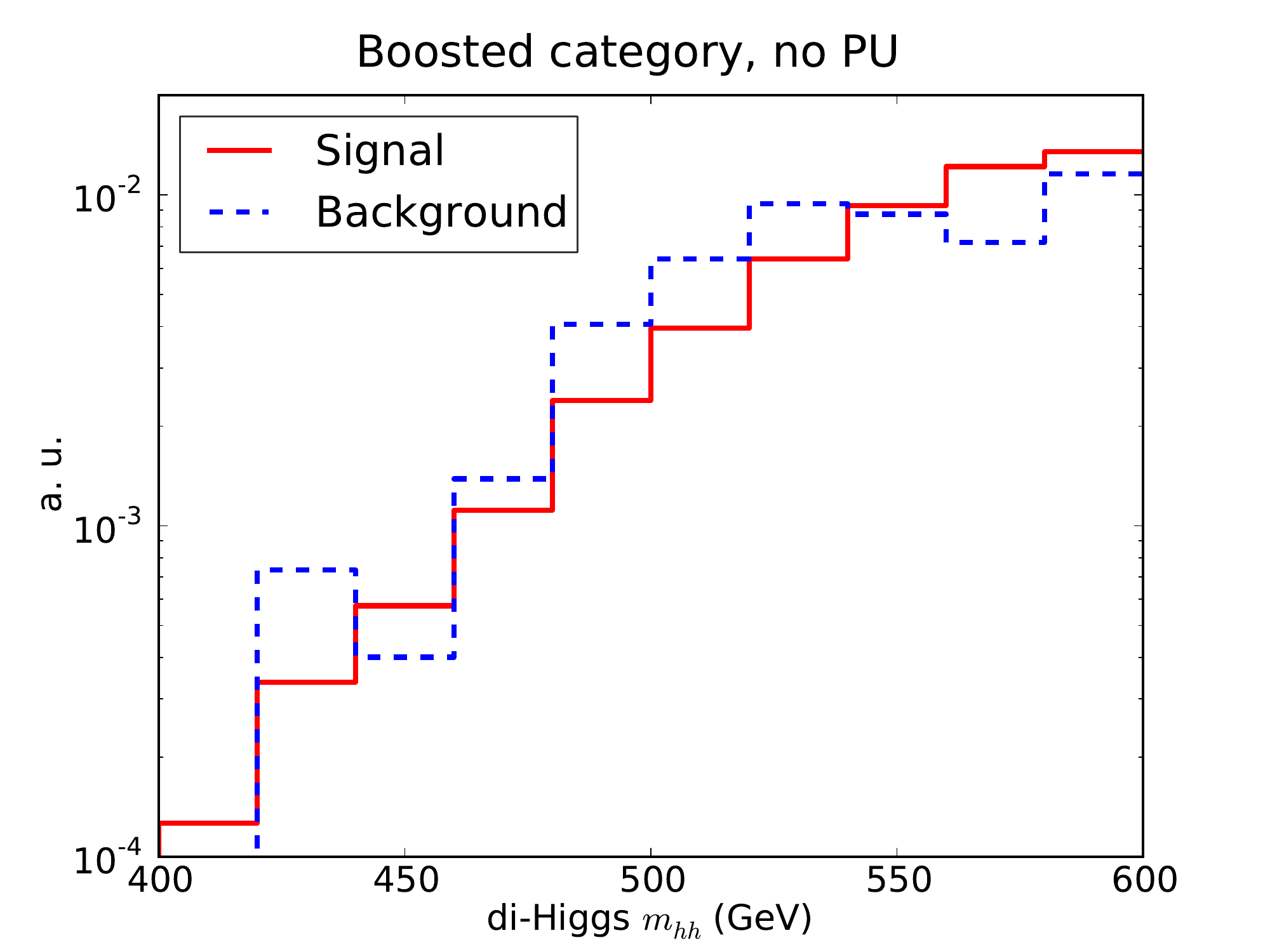}
  \caption{\small
Same as   Fig.~\ref{fig:cutplots1} for the invariant
mass distribution of the di-Higgs system $m_{hh}$, in
the resolved (left) and boosted (right) categories.
}
\label{fig:mhh}
\end{center}
\end{figure}

The invariant mass of the di-Higgs system is another important
kinematic distribution for this process.
The di-Higgs invariant mass is a direct measure of the boost of the system,
which in  BSM scenarios can be substantially
enhanced, for instance due to
specific $d=6$ EFT operators~\cite{Azatov:2015oxa}.
One important advantage of the $b\bar{b}b\bar{b}$ final state for
di-Higgs production is that it significantly increases the reach
in $m_{hh}$ as compared to other channels with smaller branching
ratios,
such as $2b2\gamma$ or $2b2\tau$.
In Fig.~\ref{fig:mhh} we show the invariant mass distribution of the
reconstructed Higgs pairs,
comparing the resolved and the boosted categories.

In the resolved case, we see that the distribution
in $m_{hh}$ is rather harder for the signal as compared
to the background,
and therefore one expects that cutting in $m_{hh}$ would help signal
discrimination.
For the boosted category the overall trend of the $m_{hh}$ distribution
is different because of the selection criteria, and the
distribution now peaks at higher values of the invariant mass.
In this case, signal and background distributions are not significantly
differentiated.
Note that at parton-level the $m_{hh}$ distribution
for signal events has a kinematic
cut-off at $m_{hh}^{\rm min}=250$ GeV, which is smeared due
to parton shower and detector resolution effects.

\begin{figure}[t]
\begin{center}
  \includegraphics[width=0.48\textwidth]{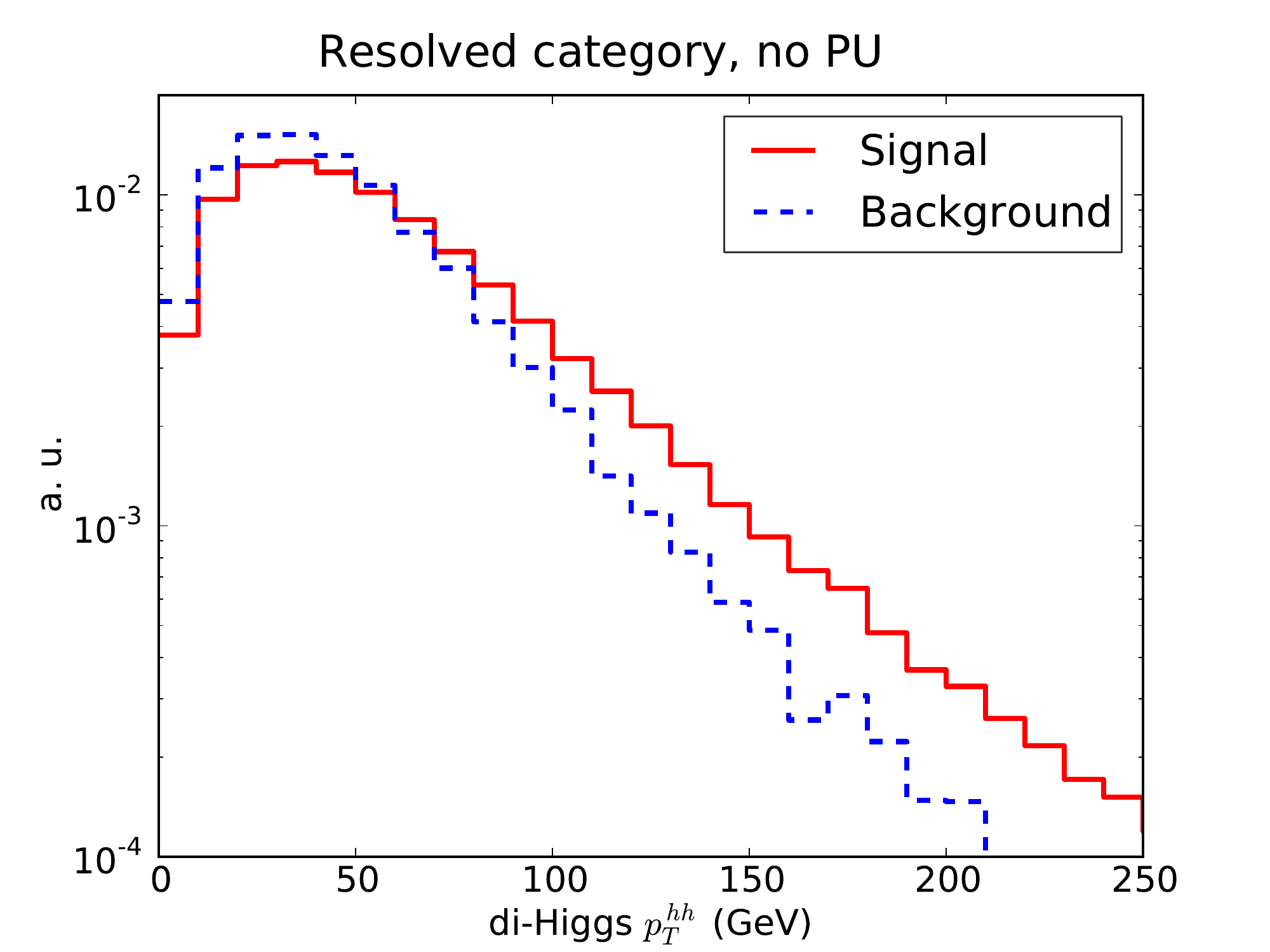}
  \includegraphics[width=0.48\textwidth]{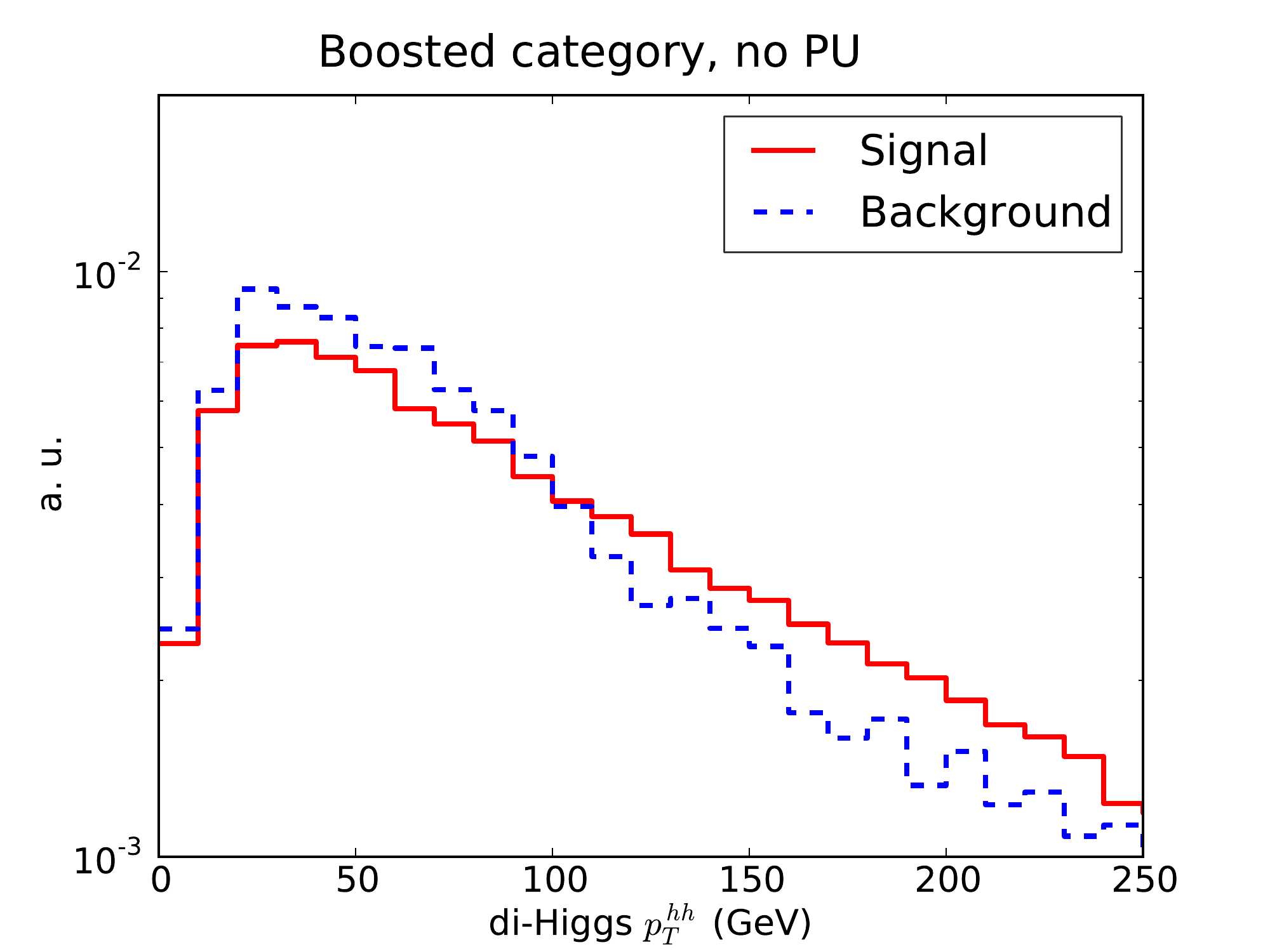}
  \caption{\small Same as Fig.~\ref{fig:cutplots1} for the transverse momentum
    distribution of the di-Higgs system $p_T^{hh}$.
}
\label{fig:pthh}
\end{center}
\end{figure}

In Fig.~\ref{fig:pthh} we show the transverse momentum of
the di-Higgs system, $p_T^{hh}$,
for the resolved and boosted categories.
Once more we see that the background has a steeper fall-off in $p_T^{hh}$
than the signal, in both categories, therefore this variable
should provide additional discrimination power, motivating its inclusion
as one of the inputs for the MVA.
In our LO simulation the $p_T^{hh}$ distribution is generated
by the parton shower, an improved theoretical
description would require
merging higher-multiplicity
matrix elements~\cite{Maierhofer:2013sha} or matching to
the NLO calculation~\cite{Frederix:2014hta},
%

\begin{figure}[t]
  \begin{center}
    \includegraphics[width=0.48\textwidth]{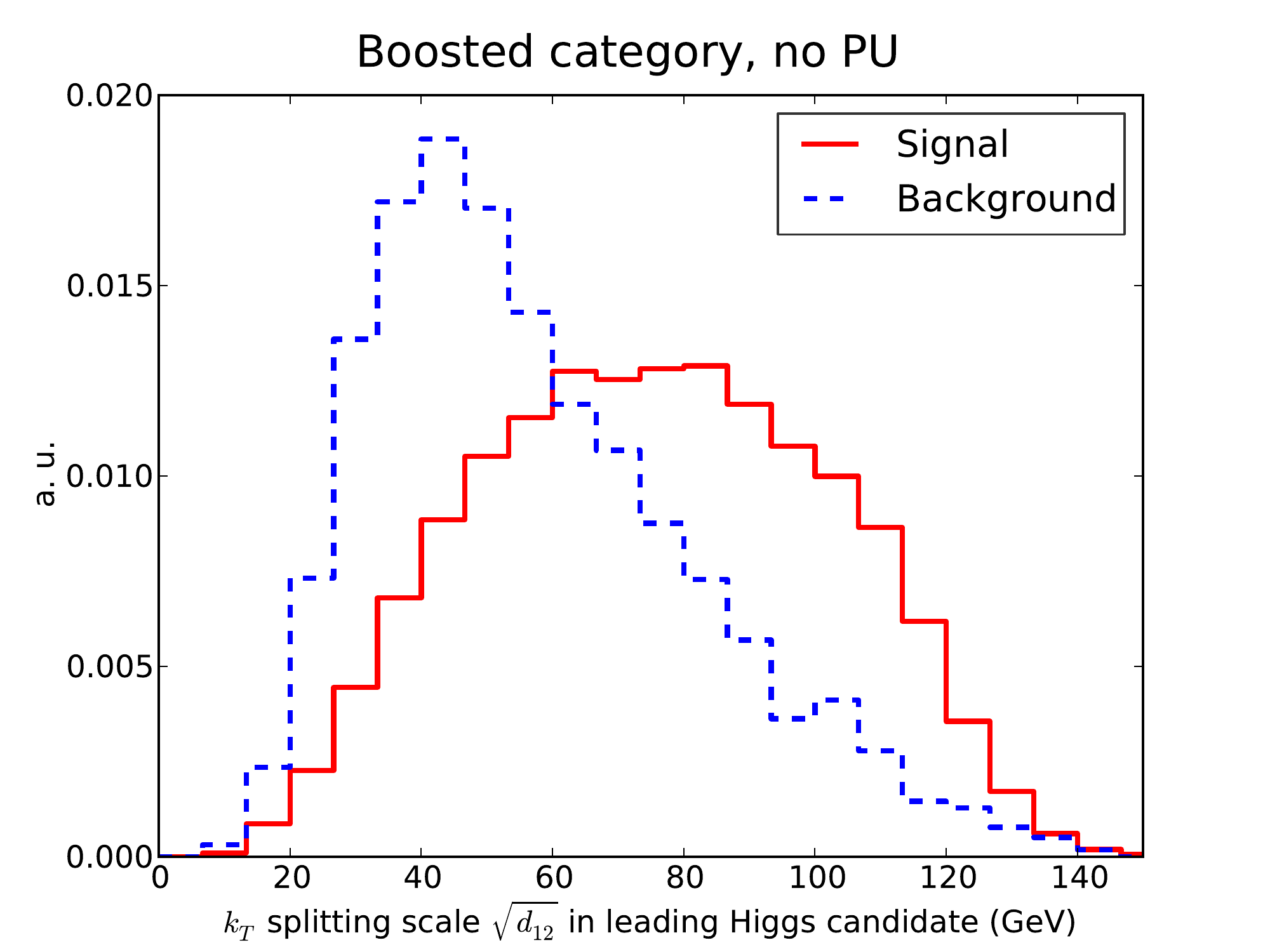} 
  \includegraphics[width=0.48\textwidth]{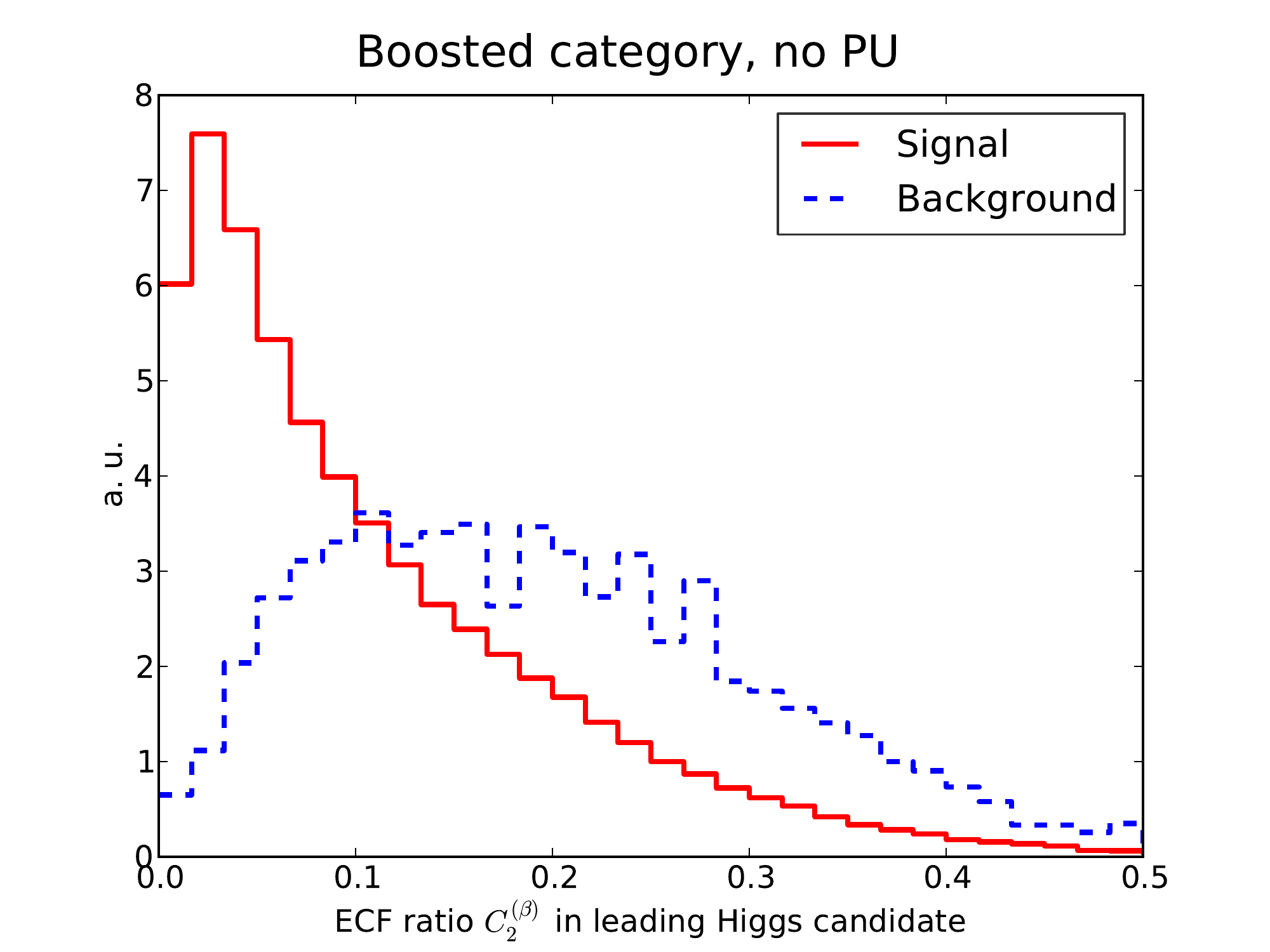}
  \includegraphics[width=0.48\textwidth]{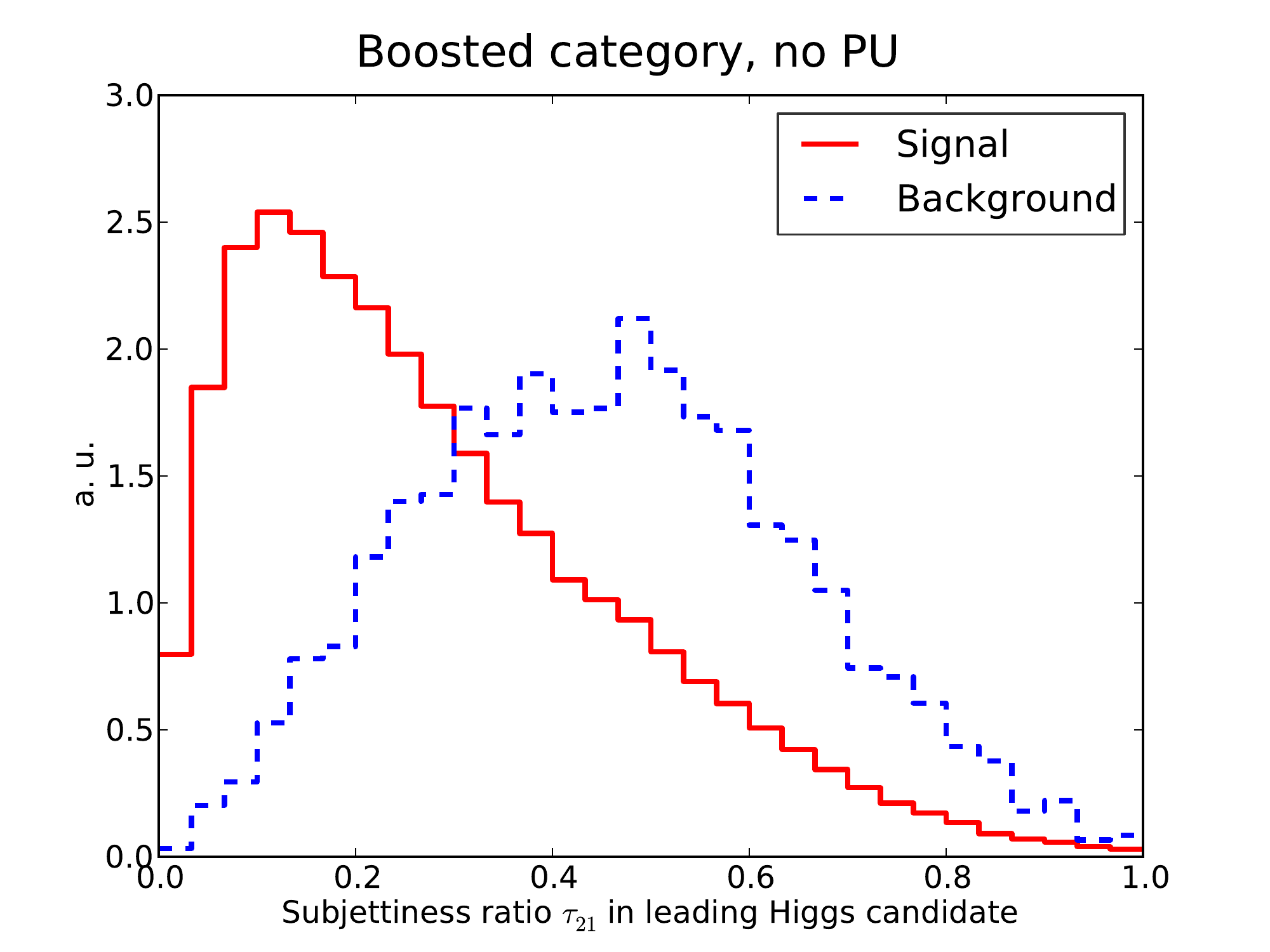}
  \includegraphics[width=0.48\textwidth]{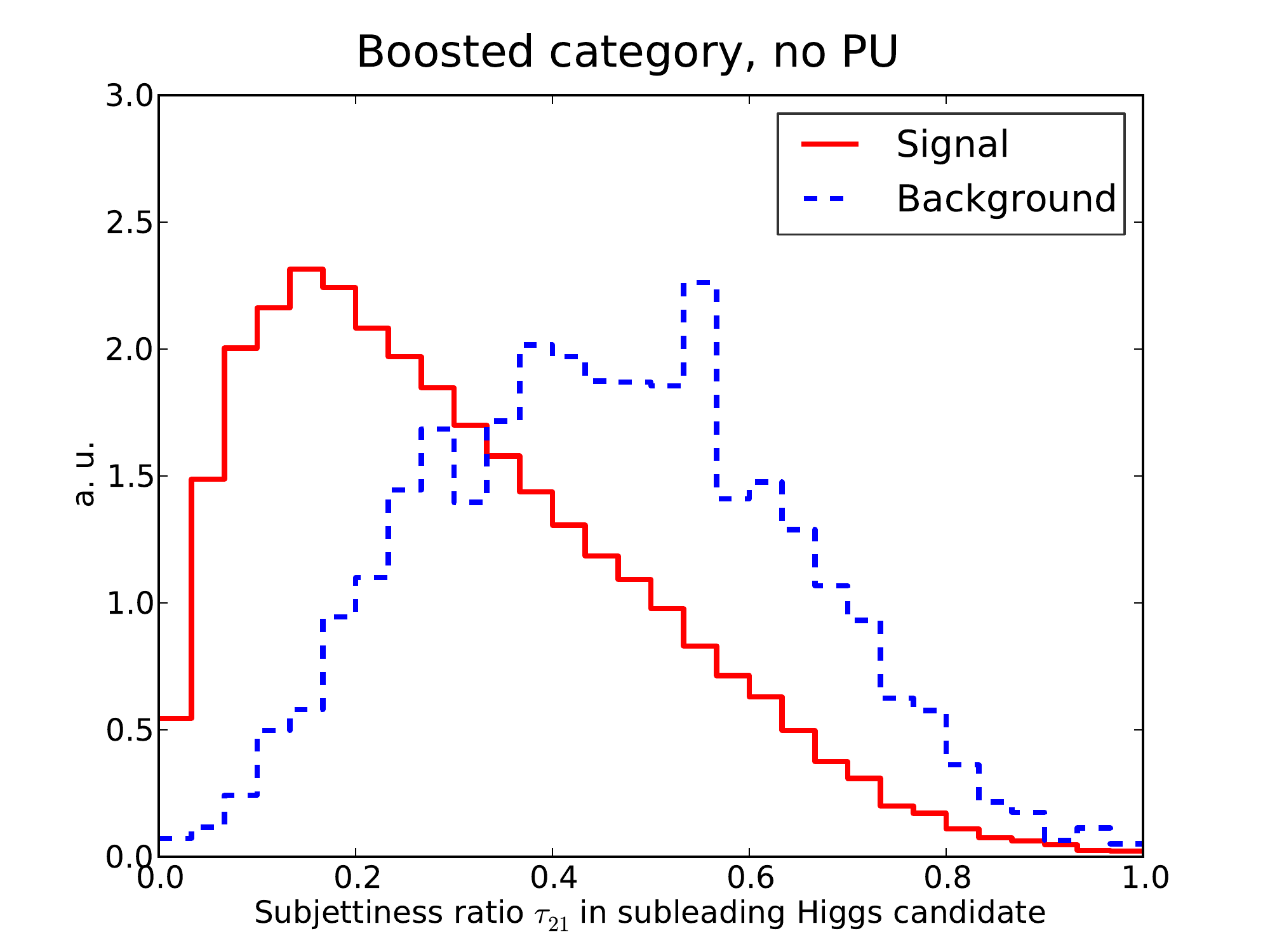}
  \caption{\small Distribution of representative substructure variables
    in the boosted category at the end of the cut-based
    analysis, to be used as input to the MVA.
    From top to bottom and from left to right we show  the
    $k_t$ splitting scale $\sqrt{d_{12}}$,
    the energy correlation ratio $C_2^{(\beta)}$
    and the  subjettiness ratio $\tau_{21}$ for the leading
    Higgs.
    In the case of 
    $\tau_{21}$  the distributions for the subleading Higgs are also given.
 }
\label{fig:mva_substructure_1}
\end{center}
\end{figure}

We shall now investigate the discrimination power
provided by jet substructure
quantities.
In Fig.~\ref{fig:mva_substructure_1}
we show the distributions of representative
 substructure variables for the boosted category: the
$k_t$ splitting scale $\sqrt{d_{12}}$, Eq.~(\ref{eq:ktsplitting}), 
the ECF ratio $C_2^{(\beta)}$,
Eq.~(\ref{eq:c2}), and
the 2--to--1 subjettiness ratio $\tau_{21}$, Eq.~(\ref{eq:tau21}),
all for the leading
Higgs candidates, and also $\tau_{21}$ for the subleading
Higgs candidates.

From Fig.~\ref{fig:mva_substructure_1}
we observe how for these substructure variables the shapes of the signal
and background distributions reflect
the inherent differences in the internal structure of
QCD jets and jets originating from Higgs decays.
Signal and background distributions peak
in rather
different regions. For example, the $k_t$ splitting scale $\sqrt{d_{12}}$
peaks around 80 GeV (40 GeV) for signal (background) events, while
the distribution of the
ECF ratio $C_2^{(\beta)}$ is concentrated at small values
for signal and is much broader for background events.
From Fig.~\ref{fig:mva_substructure_1} we also see
the distributions of the subjettiness ratio $\tau_{21}$ are
reasonably similar
for both the leading and the subleading jets.

\subsection{Impact of pileup}
\label{sec:pileup}

Now we turn to discuss how the description of kinematic
distributions for signal
and background processes are
modified in the presence of pileup.
To study the impact of PU,
Minimum Bias events have been generated
with {\tt Pythia8}, and then
superimposed to the signal
and background samples described in Sect.~\ref{mcgeneration}.
We have explored two scenarios,
one with a number of
PU vertices per bunch crossing of $n_{\rm PU}=80$,
and another
with $n_{\rm PU}=150$.
In the following we adopt $n_{\rm PU}=80$ as our baseline,
and denote this scenario by PU80.
We have verified that the combined signal significance is
similar if $n_{\rm PU}=150$ is adopted instead.

In order to subtract PU in hadronic collisions, a number of techniques
are available~\cite{Cacciari:2009dp,TheATLAScollaboration:2013pia,Butterworth:2008iy,Cacciari:2007fd,Krohn:2009th,Krohn:2013lba,Cacciari:2008gd,Ellis:2009me,Bertolini:2014bba,Cacciari:2014gra,Cacciari:2014jta,Berta:2014eza,Larkoski:2014wba}.\footnote{
These techniques have also important applications in the subtraction
of the UE/MPI contamination for jet reconstruction
in heavy ion collisions~\cite{Cacciari:2010te}.
}
In this work, PU  is subtracted
with the {\tt SoftKiller} (SK)
method~\cite{Cacciari:2014gra}, as implemented in {\tt FastJet},
whose performance has been shown to
improve the commonly used area-based subtraction~\cite{Cacciari:2009dp}.
The idea underlying {\tt SoftKiller} consists of eliminating particles
below a given cut-off in their transverse momentum, $p_T^{\rm (cut)}$, whose
value is dynamically determined so that the event-wide
transverse-momentum flow density $\rho$ vanishes, where $\rho$ is
defined as
\be
\rho\equiv{\rm median}_i \Bigg\{ \frac{p_{Ti}}{A_i}\Bigg\} \, ,
\ee
and where the median is computed over all the regions $i$ with area
$A_i$ and transverse momentum $p_{Ti}$ in which the $\lp \eta,\phi\rp$ plane
is partitioned.

From its definition in terms of the median,
it follows that the value of $p_T^{(\rm cut)}$
will be dynamically raised until half of the regions have
$p_{Ti}=0$.
The size and number of these regions is a free parameter of the algorithm -
here we will use square regions with length $a=0.4$.
We restrict ourselves to the central rapidity region,
$|\eta| \le 2.5$, for the estimation of the
$p_T$ flow density $\rho$.
The {\tt SoftKiller} subtraction is then
applied to particles at the end of the parton shower, before
jet clustering.

In addition, jet trimming~\cite{Krohn:2009th}, as implemented in {\tt FastJet}, is applied to large-$R$ jets.
The trimming parameters are chosen such that the constituents of a given jet are reclustered into $k_T$ subjets with $R_{\textrm{sub}} = 0.2$.
Subjets with transverse momentum less than 5\% of the total
transverse momentum of the large-$R$ jet are then removed.
The use of trimming in addition to PU removal with {\tt SoftKiller} is necessary to correct the jet mass in the boosted category,
which is particularly susceptible
to soft, wide-angle contaminations.
No trimming is applied to the small-$R$ jets and
to the case without PU.

\begin{figure}[t]
  \begin{center}
    \includegraphics[width=0.49\textwidth]{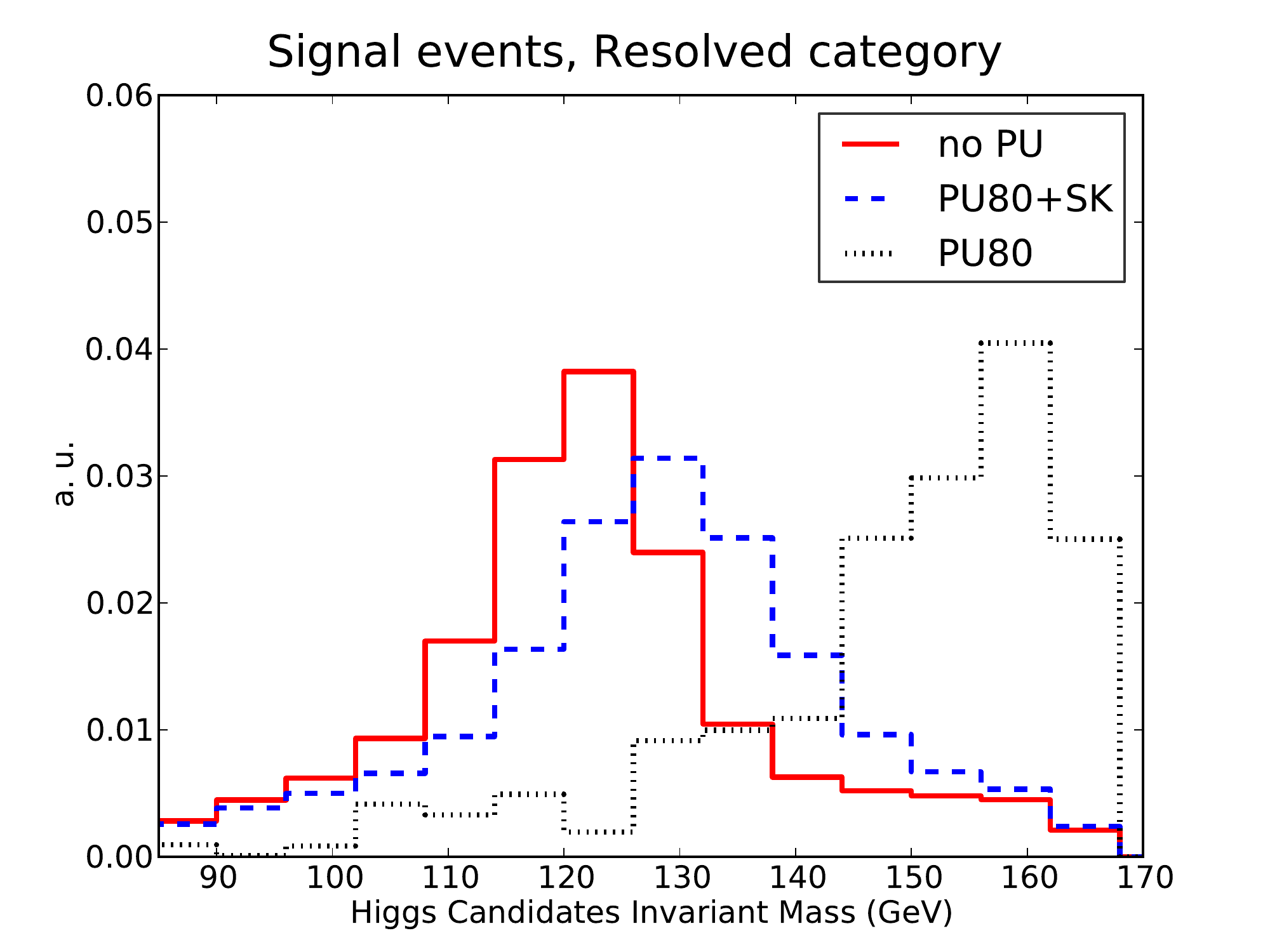}
    \includegraphics[width=0.49\textwidth]{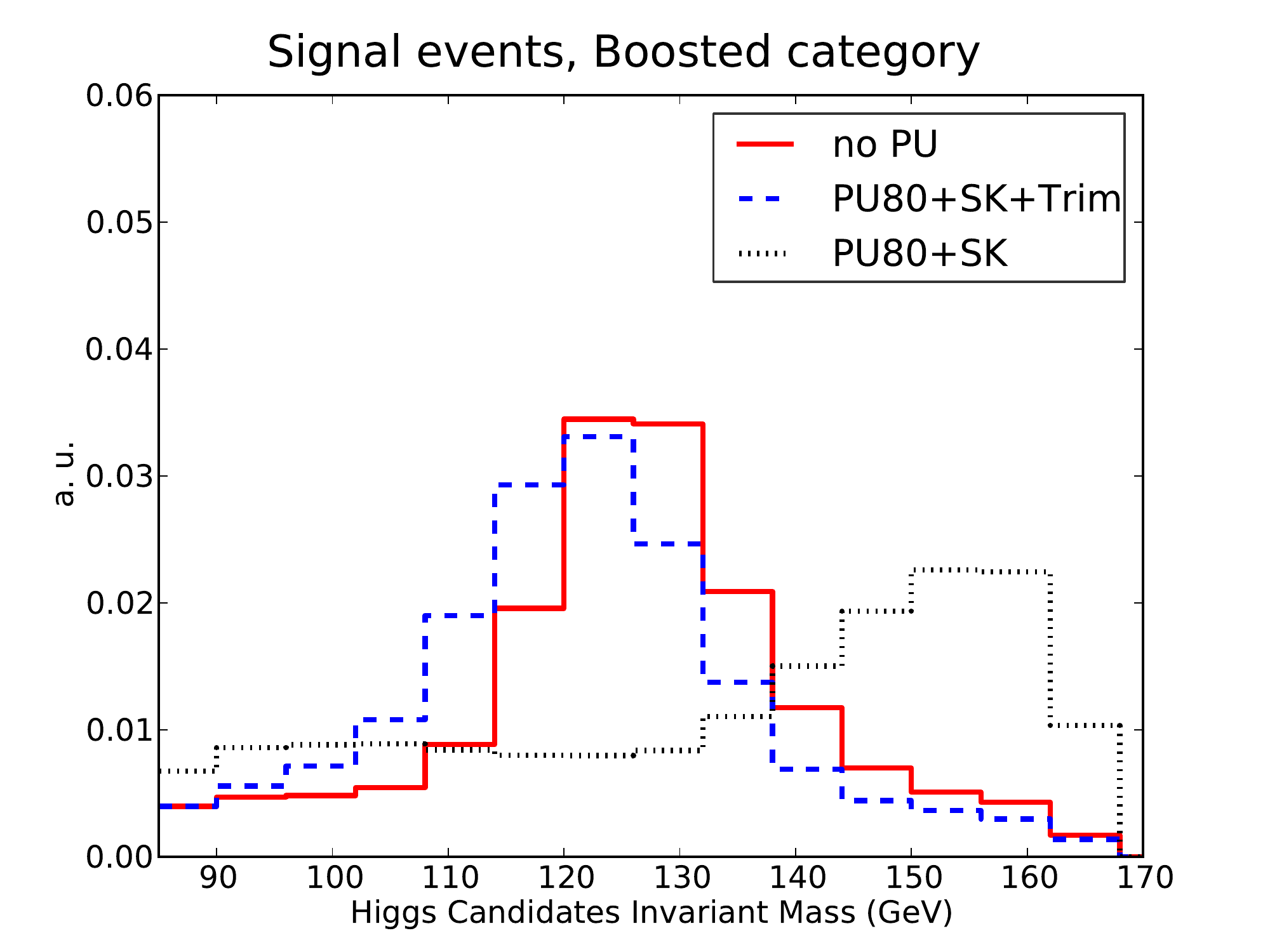}
    \caption{\small
    The invariant mass distributions of Higgs candidates in signal
    events in the resolved (left) and boosted
    (right) categories.
    In the resolved category,
    we compare  the results without PU
    with those with PU80
    with and without SK subtraction.
    In the boosted case, the comparison is performed between no PU,
    PU with only SK subtraction,
    and PU with both SK and trimming.
}
\label{fig:PUvalidation}
\end{center}
\end{figure}

%
In Fig.~\ref{fig:PUvalidation} we show the
invariant mass distributions of the
Higgs candidates for signal
events in the resolved and boosted categories.
 In the resolved category,
    we compare  the results without PU
    with those with PU80, with and without SK subtraction.
    If PU is not subtracted, there is a large shift in the Higgs mass
peak, by more than 30 GeV.
Once SK subtraction is performed, we recover a distribution much closer
to the no PU case, with only a small shift of a few GeV
and a broadening of the mass
distribution.
    In the boosted case, the comparison is performed between no PU,
    PU with only SK subtraction,
    and PU with both SK and trimming.
We find that
the mass distribution for jets to which no trimming
is applied peaks at around 160~GeV, even
after PU subtraction with {\tt SoftKiller}.
When trimming is applied in addition to {\tt SoftKiller}, 
the distribution peaks close to the nominal Higgs mass, as in the case
of the resolved category.

\begin{figure}[t]
  \begin{center}
  \includegraphics[width=0.49\textwidth]{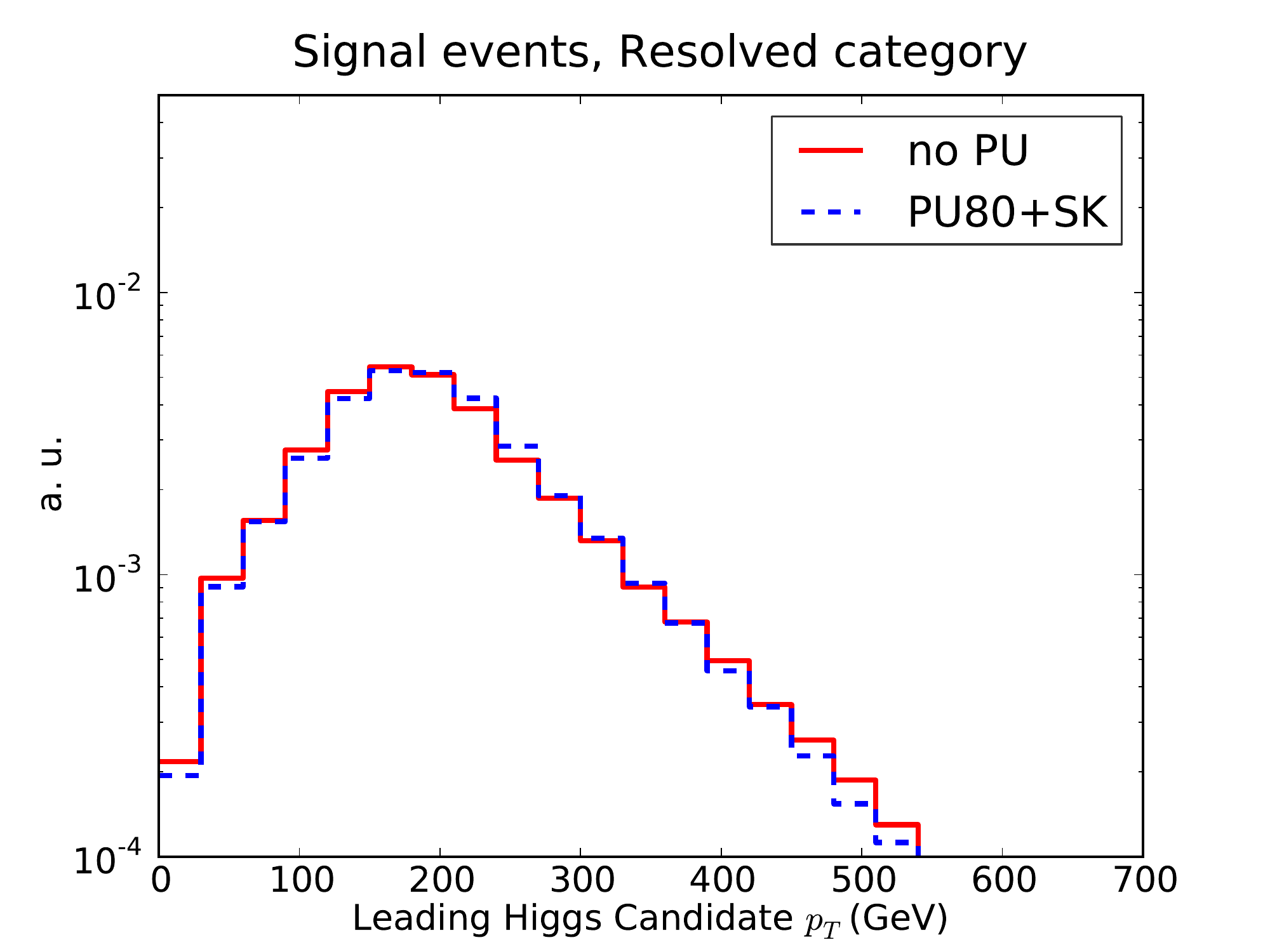}
  \includegraphics[width=0.49\textwidth]{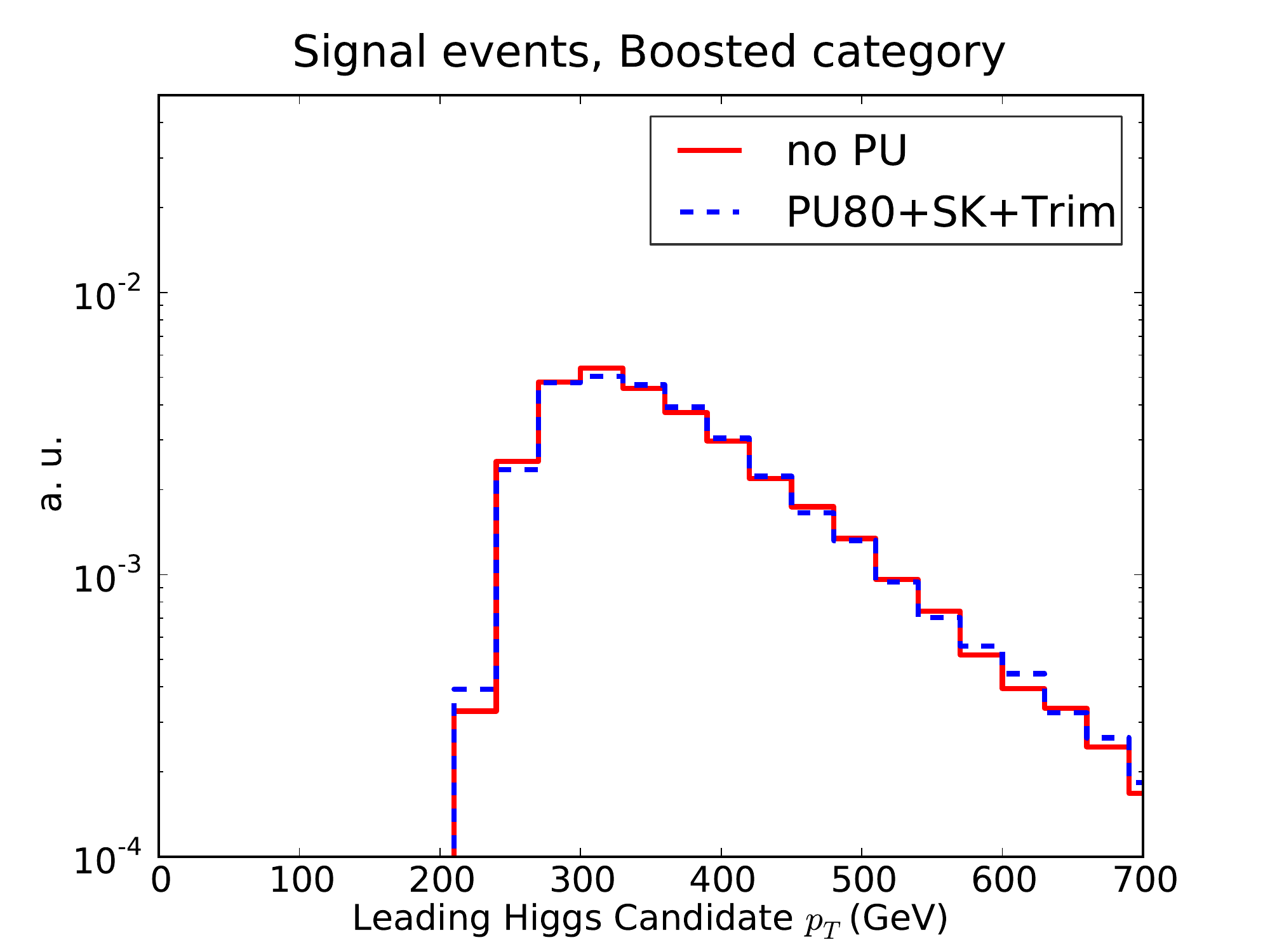}
  \includegraphics[width=0.49\textwidth]{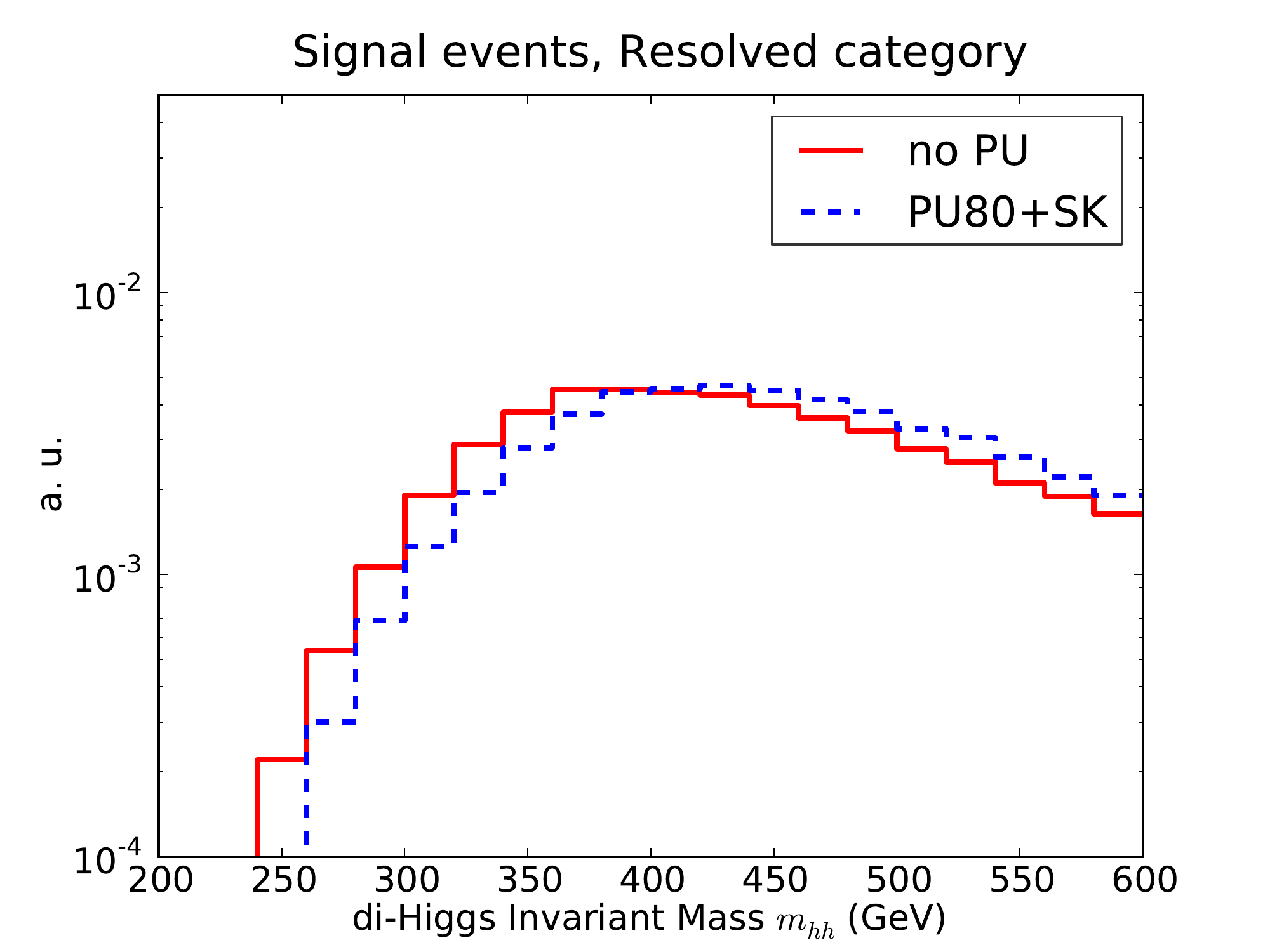}
  \includegraphics[width=0.49\textwidth]{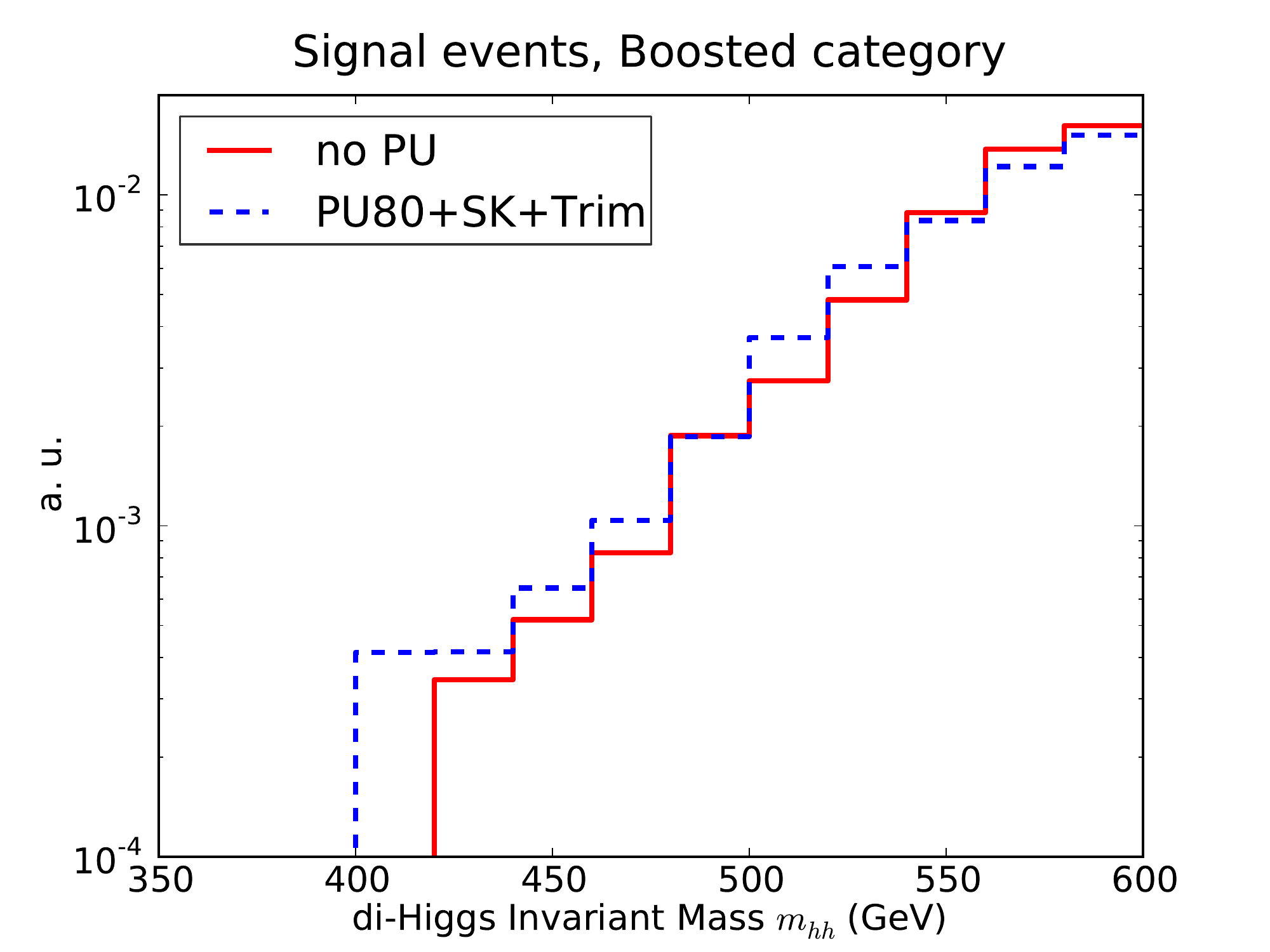}
  \caption{\small
   The
   transverse momentum $p_T^h$ of the leading
    Higgs candidate (upper plots) and of the invariant mass $m_{hh}$
    of the di-Higgs system (lower plots) in the resolved
    (left) and boosted (right) categories.
    We compare the results without PU with those with PU80
    and SK+Trim subtraction,
    as explained in the text.
}
\label{fig:mHH_PU}
\end{center}
\end{figure}

In Fig.~\ref{fig:mHH_PU}
we compare the transverse momentum of the leading Higgs
candidate, $p_t^{h}$ and the invariant mass of the di-Higgs system
$m_{hh}$, in both the boosted and resolved categories,
between the no PU and the PU+SK+Trim cases.
In the case of the $p_T^{h}$ distribution, the differences between the selection
criteria for the resolved
and boosted categories is reflected in the rightward shift of the latter.
After subtraction,
the effects of PU are small in the two categories.
A similar behaviour is observed in the di-Higgs invariant mass distribution.

\begin{figure}[t]
  \begin{center}
    \includegraphics[width=0.49\textwidth]{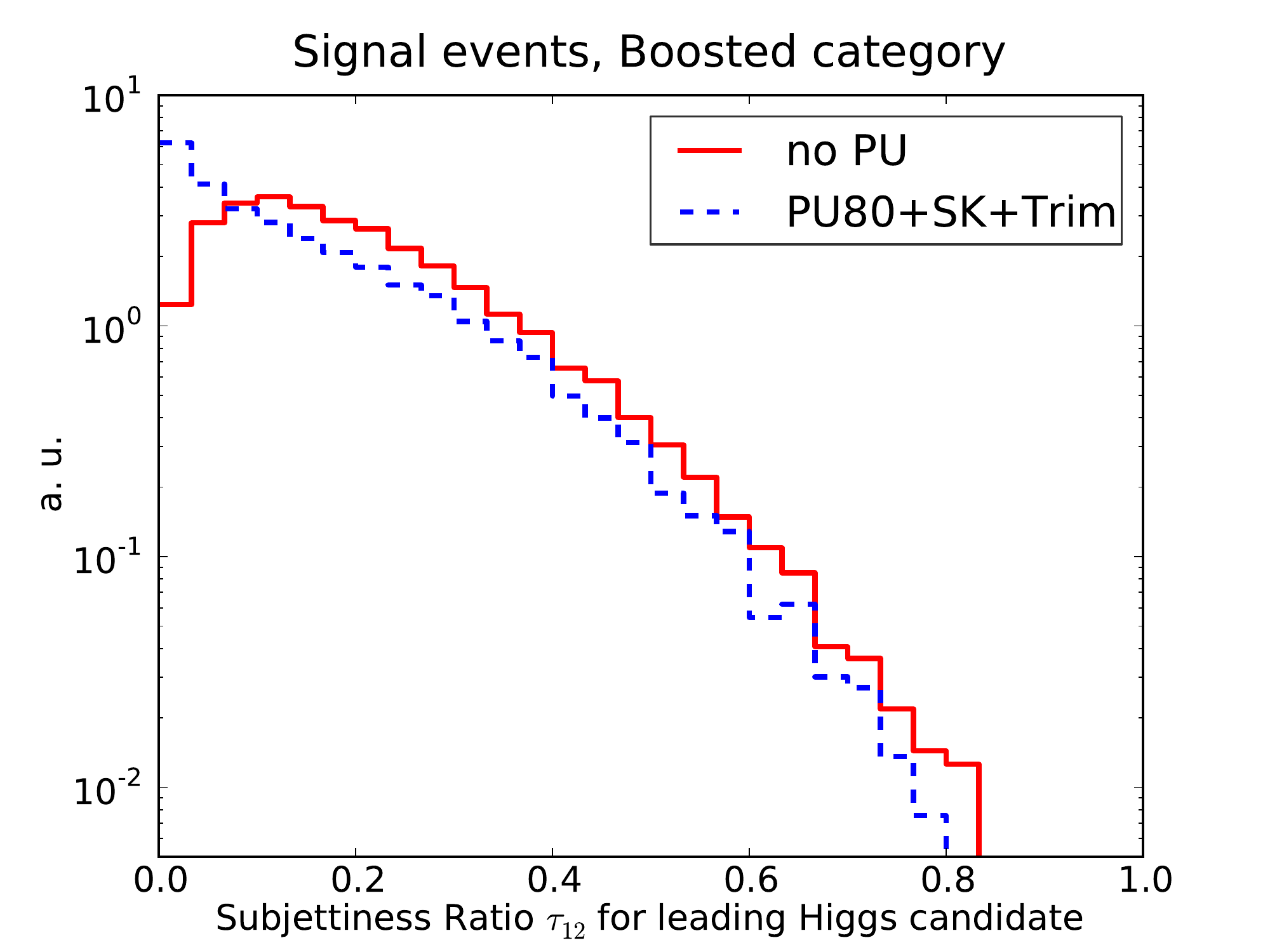}
  \includegraphics[width=0.49\textwidth]{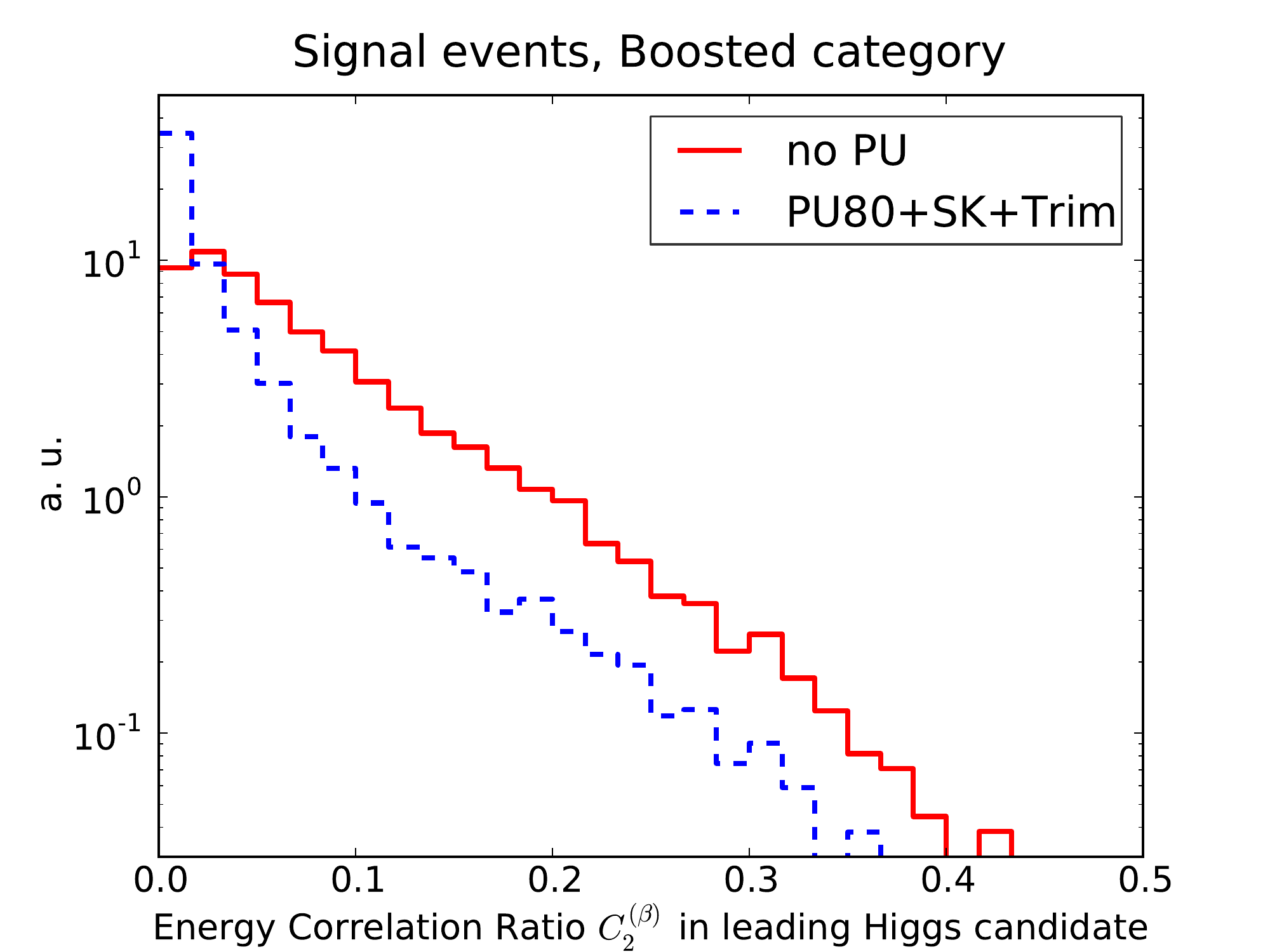}
    \caption{\small
    Same as Fig.~\ref{fig:mHH_PU} for the
    substructure variables $\tau_{21}$  (left)
     and  $C_2^{(\beta)}$ (right)
     for the leading Higgs candidates in the boosted category.
}
\label{fig:Substructure_PU}
\end{center}
\end{figure}

We can also assess the impact of PU on the
substructure variables that will be 
used as input to the MVA in the boosted
and intermediate categories.
In Fig.~\ref{fig:Substructure_PU} we show the 2-to-1 subjettiness ratio
$\tau_{21}$, Eq.~(\ref{eq:tau21}), and the ratio
of energy correlation functions $C_2^{(\beta)}$,
Eq.~(\ref{eq:c2}), for the leading Higgs candidate.
We observe that
the shapes of both substructure variables
are reasonably robust in a environment including significant PU.
Therefore we can consider the PU subtraction strategy
as validated for the purposes of this study, although
further optimisation should still be possible, both in terms of
the {\tt SoftKiller} and of the trimming
input settings.

\begin{figure}[t]
  \begin{center}
    \includegraphics[width=0.49\textwidth]{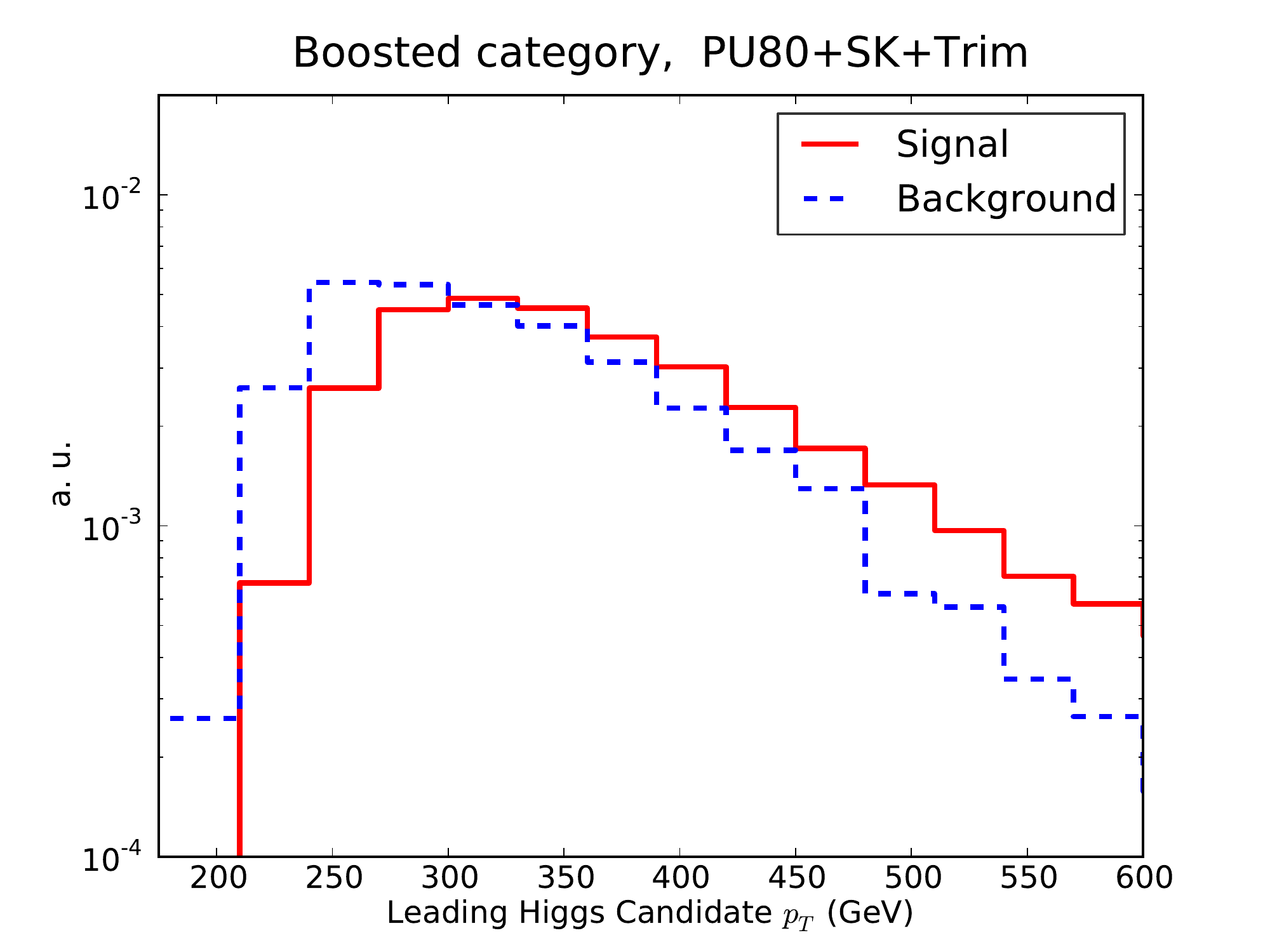}
    \includegraphics[width=0.49\textwidth]{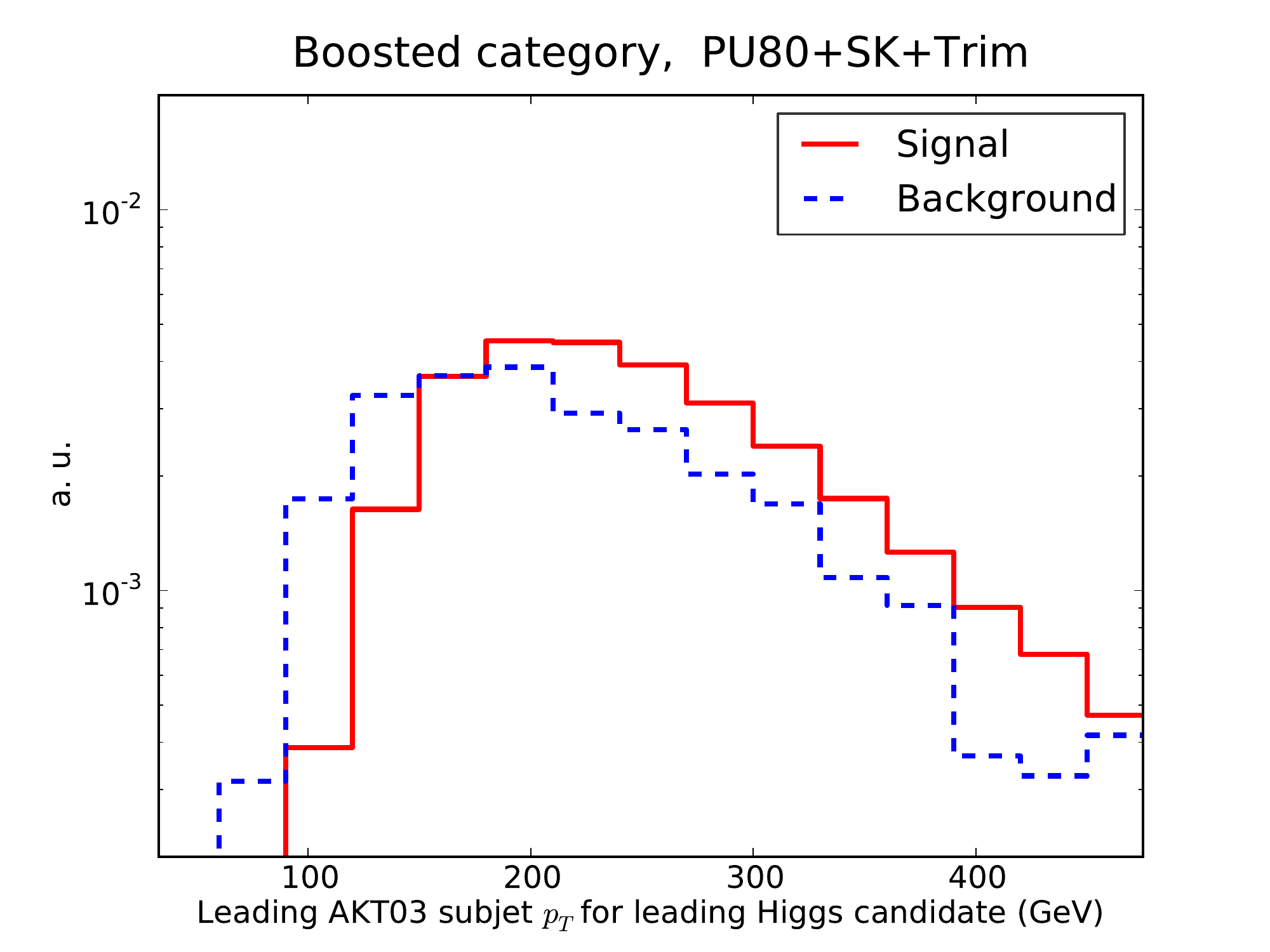}
   \includegraphics[width=0.49\textwidth]{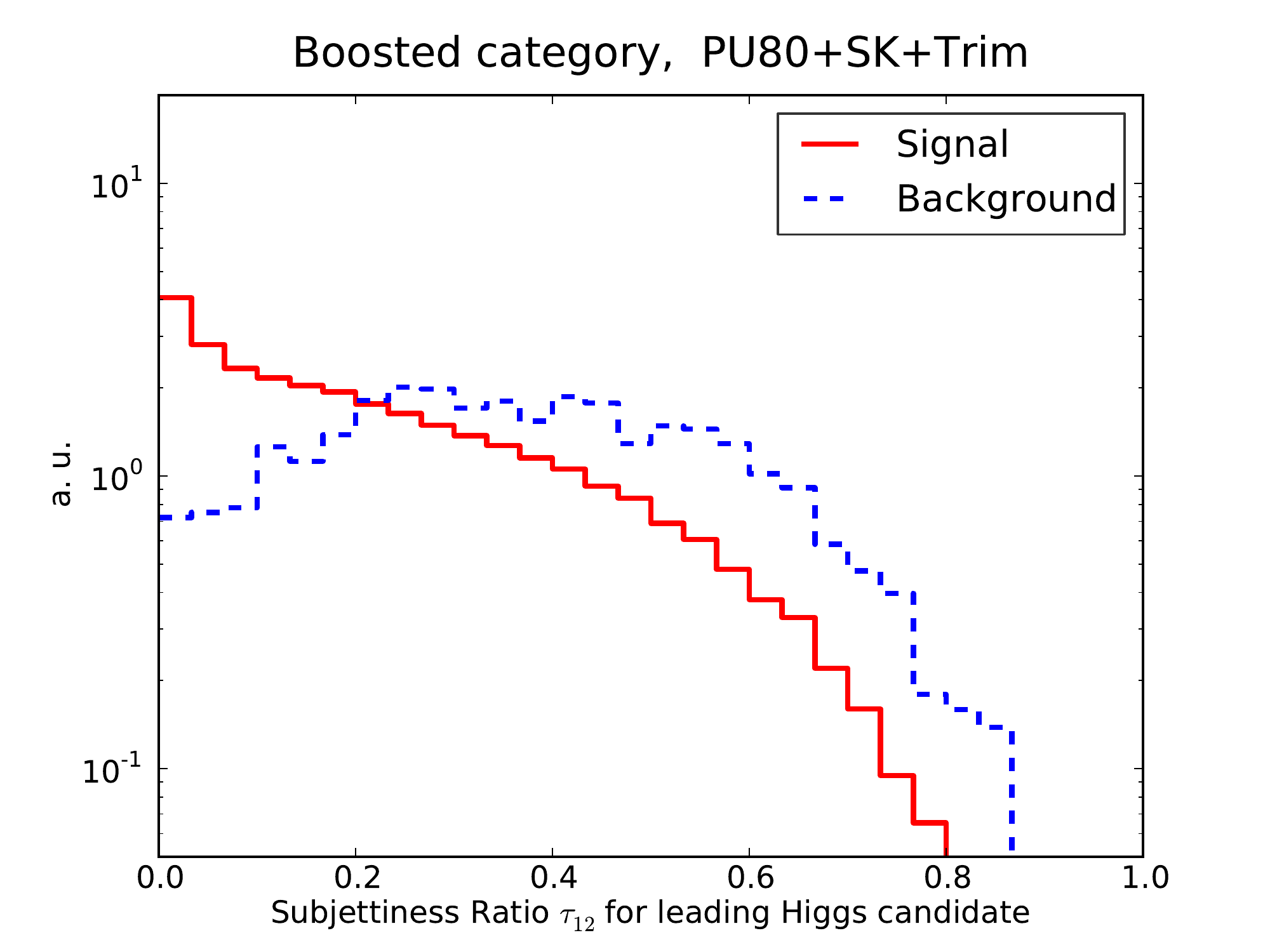}
   \includegraphics[width=0.49\textwidth]{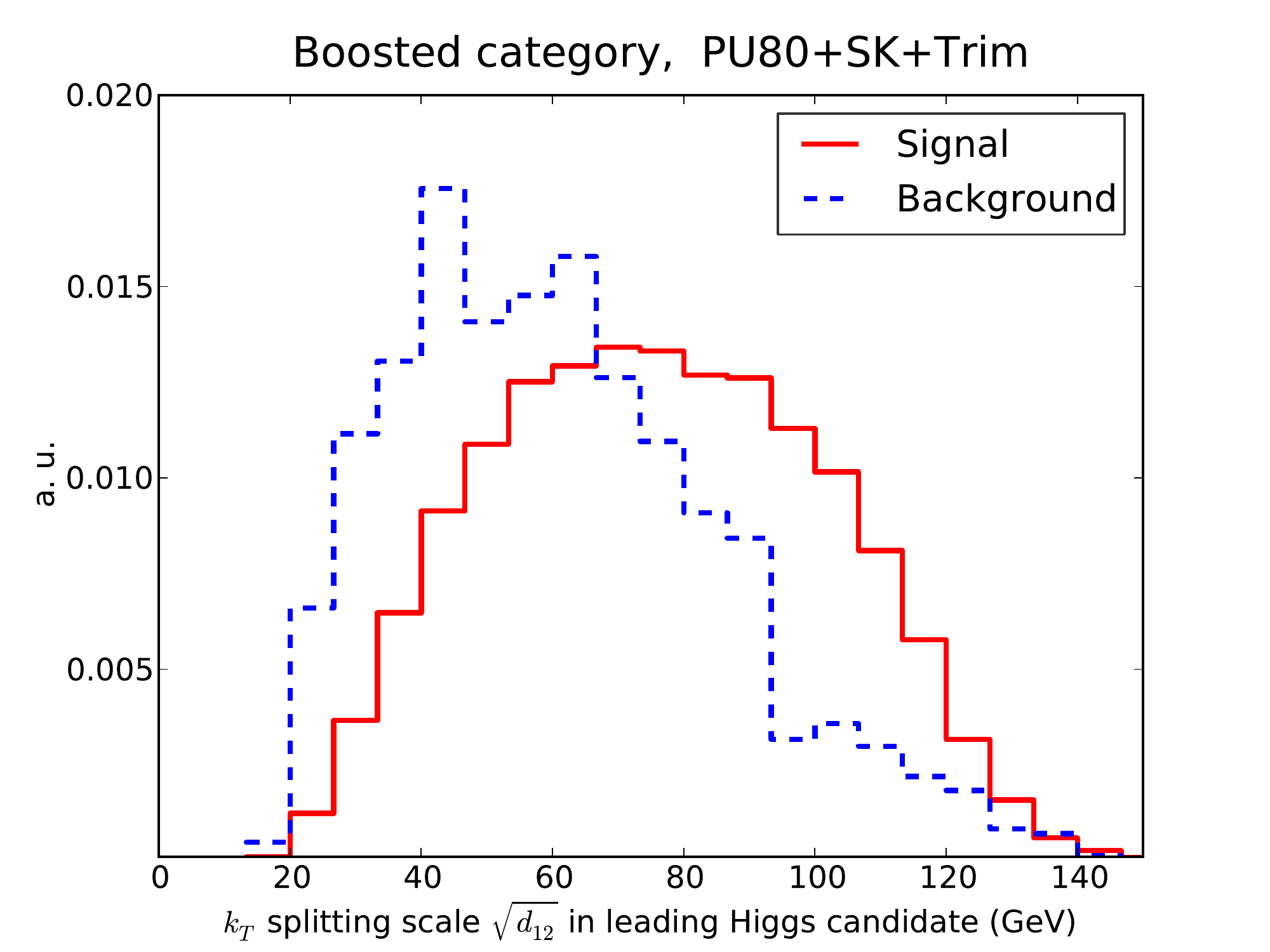}
    \caption{\small
      Comparison of kinematic distributions for the leading
      Higgs candidate, in
     the boosted category, for signal and background events
    in the case of PU subtracted with SK+Trim:
    its transverse momentum  $p_T$,
    the $p_T$ of its leading AKT03 subjet,
    and the substructure variables $\tau_{21}$ and $\sqrt{d_{12}}$.
  }
\label{fig:signal-vs-back-boosted}
\end{center}
\end{figure}

It is also interesting to quantify how
the relative differences between
signal over background distributions are modified by the inclusion of PU.
Considering the boosted category initially,
in Fig.~\ref{fig:signal-vs-back-boosted} we compare
various kinematic distributions for signal and background events,
with and without PU for the leading Higgs candidate: the transverse
momentum distribution $p_T$,
the $p_T$ of the leading AKT03 subjet,
     the 2--to--1 subjettiness ratio $\tau_{21}$, and 
     the $k_T$ splitting scale $\sqrt{d_{12}}$.
      We verify that the relevant
      qualitative differences between signal
      and background distributions are maintained  in the presence of PU.
      This is especially noticeable for the substructure variables, which
      exhibit a similar discriminatory power both with and without
      PU.

\begin{figure}[t]
  \begin{center}
      \includegraphics[width=0.49\textwidth]{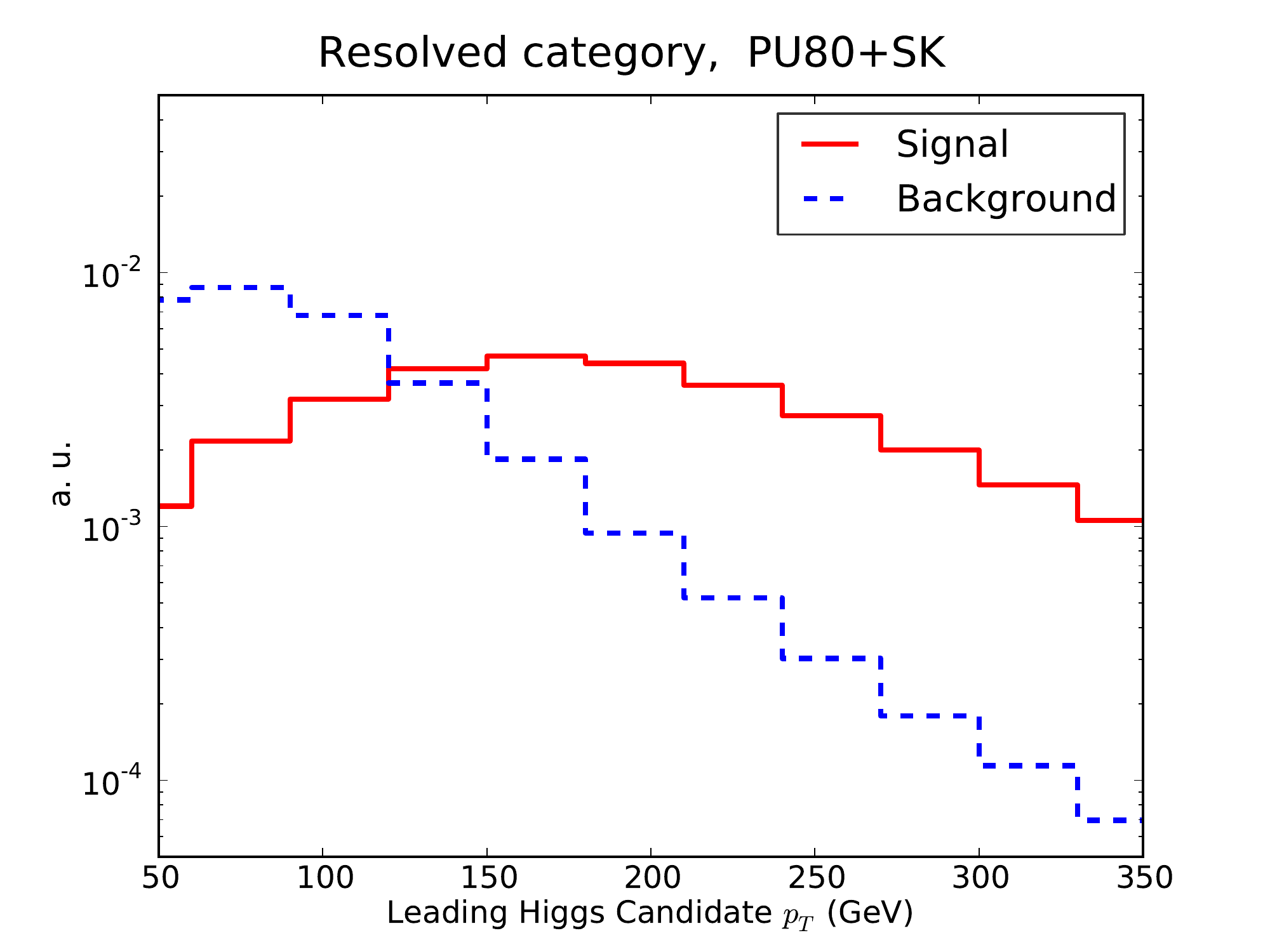}
   \includegraphics[width=0.49\textwidth]{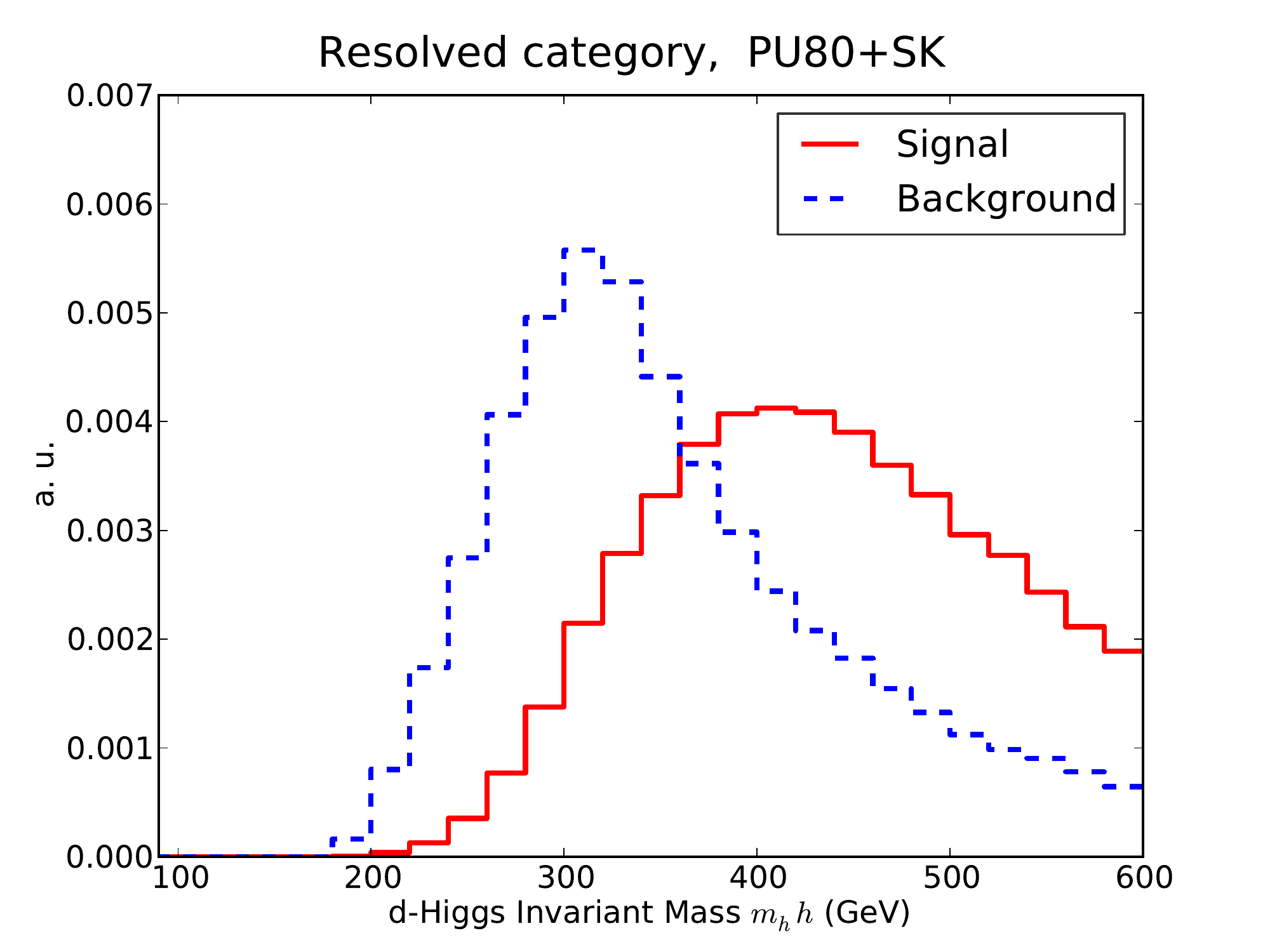}
     \caption{\small
       Same as Fig.~\ref{fig:signal-vs-back-boosted} for the resolved category.
}
\label{fig:signal-vs-back-resolved}
\end{center}
\end{figure}

We can also perform a similar comparison for
the resolved category.
In Fig.~\ref{fig:signal-vs-back-resolved} we compare
the kinematic distributions for signal and background events,
with and without PU, for the invariant mass and the
transverse momentum of the leading
     Higgs candidate.
     Again, the PU-subtracted background distributions
     appear reasonably close
     to their counterparts without PU, and thus
     the distinctive features between signal and background
     are maintained after PU subtraction.

    \begin{table}[h]
      \centering
      \begin{tabular}{|c|c|c|c|}
        \hline
        \multicolumn{4}{|c|}{Resolved category}\\
        \hline
        \hline
        &   &   $\la m_h^{\rm reco}\ra-m_h$  &  $\sigma_{m_h}$  \\
              \hline
        \multirow{2}{*}{no PU}  & leading $h$  &  -3.8 GeV   & $\lp 8.5\pm 0.2\rp$ GeV   \\
          & subleading $h$   & -5.8 GeV  &  $\lp 9.1\pm 0.3\rp$ GeV \\
        \hline
          \multirow{2}{*}{PU80}  & leading $h$  &  +33 GeV   & $\lp 8.8\pm 1.5\rp$ GeV   \\
          & subleading $h$   & +31 GeV  &  $\lp 11.7\pm 3.3\rp$ GeV \\
          \hline
            \multirow{2}{*}{PU80+SK}  & leading $h$  &  +3.9 GeV   & $\lp 10.7\pm 0.3\rp$ GeV   \\
          & subleading $h$   & +2.1 GeV  &  $\lp 10.5\pm 0.3\rp$ GeV \\
            \hline
            \multicolumn{4}{c}{}\\
             \hline
        \multicolumn{4}{|c|}{Boosted category}\\
        \hline
        \hline
        &   &   $\la m_h^{\rm reco}\ra-m_h$  &  $\sigma_{m_h}$  \\
              \hline
        \multirow{2}{*}{no PU}  & leading $h$  &  +2.0 GeV   & $\lp 8.2\pm 0.5\rp$ GeV   \\
          & subleading $h$   & +1.0 GeV  &  $\lp 8.8\pm 0.5\rp$ GeV \\
        \hline
              \multirow{2}{*}{PU80+SK+Trim}  & leading $h$  &  -2.2 GeV   & $\lp 8.7\pm 0.7\rp$ GeV   \\
          & subleading $h$   & -4.9 GeV  &  $\lp 9.0\pm 0.8 \rp$ GeV \\
        \hline
        \end{tabular}
      \caption{\label{tab:massresolution}
        Resolution of the invariant mass distribution of
        reconstructed Higgs candidates in the resolved 
        and boosted  categories.
        We show three cases: no PU, with PU80
        without subtraction (only for resolved),
        and the same with SK+Trim subtraction.
        We indicate the shift of the fitted invariant
        mass peak $\la m_h^{\rm reco}\ra$ for
        the Higgs candidates as compared
        to the nominal Higgs mass $m_h$, as well as the
        the fitted  Gaussian  width  $\sigma_{m_h}$.
        }
    \end{table}

It is illustrative to determine
the mass resolution obtained for the
reconstructed Higgs candidates in the various cases considered in
the present study.
In Table~\ref{tab:massresolution}
we indicate the shift of the fitted invariant mass peak as compared
to the nominal Higgs mass, $\la m_h^{\rm reco}\ra-m_h$,
and the corresponding width of the distribution, $\sigma_{m_h}$,
obtained from fitting a Gaussian to the mass distributions
of leading and subleading Higgs candidates in
the resolved and boosted categories.
        We show results for three cases: without PU, with PU80
        but without subtraction (only for the
        resolved category), and the same with SK+Trim subtraction.

        In both categories,
        we find a mass resolution of around 9 GeV in the case
        without PU.
        In the case of PU
        with SK+Trim subtraction,
        in the resolved category the mass resolution
        worsens only slightly
        to around 11 GeV, while in the boosted category we find
        the same resolution as in the no PU case.
        We also note that after SK+Trim subtraction, the peak of
        the invariant mass distributions of Higgs candidates
        coincides with the nominal values of $m_h$ within a few GeV
        for the two categories.

\section{Pre-MVA loose cut-based analysis}

\label{sec:results}

In this section we present the results of the pre-MVA
loose cut-based analysis described in the previous section, and provide 
cut-flows for the different analysis steps.
We study how the signal significance
is affected if only the $4b$ component of the
QCD multi-jet background is taken into account.
This section presents the results in an environment
without pileup; the following one contains those
obtained including significant PU.

\subsection{Cut-flow and signal significance}

Here we compare the cross-sections for
signal and background events at various
stages of the analysis.
We consider all relevant backgrounds (see Sect.~\ref{mcgeneration}),
and discuss how results are modified in the case where only the $4b$
background is considered.
In Table~\ref{tab:cutflowdetails}
the different
steps of the cut-flow in the present analysis are summarised,
separated into the boosted, intermediate,
    and resolved topologies.
    The different analysis steps proceed as follows:
   
\begin{table}[t]
  \centering
  \begin{tabular}{|c|c|c|c|}
\hline
&  Boosted  &   Intermediate &  Resolved  \\
\hline
\hline
\multirow{2}{*}{{\bf C1a}} & $N_{\rm jets}^{R10}\ge 2$ & $N_{\rm jets}^{R04}\ge 2$, $N_{\rm jets}^{R10}=1$  &
$N_{\rm jets}^{R04}\ge 4$ \\
 & \multicolumn{3}{c|}{+$p_T$ cuts and rapidity cuts} \\
\hline
\multirow{2}{*}{{\bf C1b}} & +$N_{\rm MDT}\ge 2$ & +$N_{\rm jets}^{R10}=1$ with MDT  &
 +Higgs reconstruction \\
  &  &
 +Higgs reconstruction  & \\
 \hline
{\bf C1c} & \multicolumn{3}{c|}{ +$m_h$ window cut} \\
\hline
{\bf C2} & \multicolumn{3}{c|}{+$b$-tagging}    \\
\hline
  \end{tabular}
  \caption{\small Definition of the cuts imposed successively for the three selections.
      \label{tab:cutflowdetails}
  }
\end{table}

\begin{itemize}
    \item {\bf C1a}:  check that we have at least
      two large-$R$ jets (in the boosted case),
      one large-$R$ jet and at least 2 small-$R$ jets (in the intermediate
      case) and at least four small-$R$ jets (in the resolved case).

      In addition,
  require that these jets 
      satisfy the corresponding $p_T$ thresholds;
      $p_T \ge 200$ GeV for large-$R$ jets and
      $p_T \ge 40$ GeV for small-$R$ jets, as well as
      the associated
      rapidity acceptance constraints.
    \item {\bf C1b}: the two leading large-$R$ jets must
      be mass-drop tagged in the boosted category.
      In the intermediate category, the large-$R$ jet
      must also be mass-drop tagged.
    \item {\bf C1c}: after the two Higgs candidates  have been reconstructed,
      their invariant masses are required to lie within a window around $m_H$,
      in particular between 85 and 165 GeV, Eq.~(\ref{higgsmasswindow}).
          \item {\bf C2}: the
            $b$-tagging conditions are
            imposed (see
            Sect.~\ref{sec:btagging}), and the event is categorised exclusively
            into one of the three topologies, according
            to the hierarchy determined in Sect.~\ref{sec:categorisation}.
      \end{itemize}
    Signal and background events satisfying all the analysis cuts up to the
    C2 level
    are then used as input for the MVA training, to be described next
    in Sect.~\ref{sec:mva}.
    
    In Table~\ref{tab:cutflow_noPU_1} we collect
    the values for the signal and background cross-sections
    at the different analysis steps.
    Results are divided into the resolved, intermediate and boosted categories,
    and are inclusive up to the C2 level, where exclusivity is imposed.
In Table~\ref{tab:cutflow_noPU_1} we also  provide the signal over
      background ratio, $S/B$, and the signal
      significance, $S/\sqrt{B}$, corresponding to an integrated
      luminosity of $\mathcal{L}=3$ ab$^{-1}$.
      These are computed either
      taking into account all the background components or
      the $4b$ QCD background only.
      We find that after $b$-tagging, the  $2b2j$ component is
      of the same order of magnitude as the $4b$ component in all categories.
      This implies that the signal significance at the end of the cut-based
      analysis is degraded due to the contribution
      of light and charm jets being mis-identified as $b$-jets.
    
\begin{table}[t]
  \centering
  \scriptsize
  \begin{tabular}{|l|cc|cccc|cc|cc|}
  \hline
\multicolumn{11}{|c|}{HL-LHC, Resolved category, no PU}\\
\hline
&  \multicolumn{6}{c|}{Cross-section [fb]} &  \multicolumn{2}{c|}{$S/B$}  &  \multicolumn{2}{c|}{$S/\sqrt{B}$}  \\
   &  $hh4b$ &  total bkg  &   $4b$    &  $2b2j$   &   $4j$    &
$t\bar{t}$ &
tot & $4b$ & tot & $4b$ \\
  \hline
  \hline
 C1a   & $9$  &   $2.2\cdot 10^8$   & $6.9\cdot 10^4$ & $1.5\cdot 10^7$ & $2.0\cdot 10^8$ & $2.1\cdot 10^5$ &  $4.0\cdot 10^{-8}$  &  $1.3\cdot 10^{-4}$  &  0.03   & 1.9         \\
 C1b   & $9$  &   $2.2\cdot 10^8$   & $6.9\cdot 10^4$ & $1.5\cdot 10^7$ & $2.0\cdot 10^8$ & $2.1\cdot 10^5$ &  $4.0\cdot 10^{-8}$  & $1.3\cdot 10^{-4}$   &  0.03   & 1.9         \\
 C1c   & $2.6$  &   $4.4\cdot 10^7$   & $1.6\cdot 10^4$ & $3.2\cdot 10^6$ & $4.1\cdot 10^7$ & $8.8\cdot 10^4$ &   $6.1\cdot 10^{-8}$  &  $1.6\cdot 10^{-4}$   &   0.02   & 1.1         \\
 C2    & $0.5$  &   $4.9\cdot 10^3$   & $1.7\cdot 10^3$ & $2.9\cdot 10^3$ & $2.1\cdot 10^2$ & $47$ &            $ 1.1\cdot 10^{-4}$   & $2.9\cdot 10^{-4}$   &   0.4   & 0.6       \\
\hline
\end{tabular}

  $\,$ \\
  \vspace{0.5cm}
  \begin{tabular}{|l|cc|cccc|cc|cc|}
  \hline
\multicolumn{11}{|c|}{HL-LHC, Intermediate category, no PU}\\
\hline
&  \multicolumn{6}{c|}{Cross-section [fb]} &  \multicolumn{2}{c|}{$S/B$}  &  \multicolumn{2}{c|}{$S/\sqrt{B}$}  \\
   &  $hh4b$ &  total bkg  &   $4b$    &  $2b2j$   &   $4j$    &
$t\bar{t}$ &
tot & $4b$ & tot & $4b$ \\
  \hline
  \hline
C1a     & 2.8  &   $8.4\cdot 10^7$   & $2.1\cdot 10^4$ & $5.3\cdot 10^6$ & $7.9\cdot 10^7$ & $3.3\cdot 10^4$ &  $3.4\cdot 10^{-8}$   & $1.3\cdot 10^{-4}$  &   0.02   & 1.1 \\
 C1b     & 2.6  &   $5.8\cdot 10^7$   & $1.4\cdot 10^4$ & $3.6\cdot 10^6$ & $5.5\cdot 10^7$ & $3.0\cdot 10^4$ &  $4.5\cdot 10^{-8}$   & $1.9\cdot 10^{-4}$  &   0.02  & 1.2\\
 C1c     & 0.5  &   $3.5\cdot 10^6$   & $8.7\cdot 10^2$ & $2.1\cdot 10^5$ & $4.3\cdot 10^7$ & $8.8\cdot 10^3$ &  $1.6\cdot 10^{-7}$   & $6.1\cdot 10^{-4}$  &   0.02   & 1.0\\
 C2      & 0.09  &  $1.8\cdot 10^2$   & $56$ & $96$ & $22$ & 3.1             & $5.3\cdot 10^{-4}$    & $1.6\cdot 10^{-3}$  &   0.4   & 0.6 \\
\hline
\end{tabular}

  $\,$ \\
  \vspace{0.5cm}
    \begin{tabular}{|l|cc|cccc|cc|cc|}
  \hline
\multicolumn{11}{|c|}{HL-LHC, Boosted category, no PU}\\
\hline
&  \multicolumn{6}{c|}{Cross-section [fb]} &  \multicolumn{2}{c|}{$S/B$}  &  \multicolumn{2}{c|}{$S/\sqrt{B}$}  \\
   &  $hh4b$ &  total bkg  &   $4b$    &  $2b2j$   &   $4j$    &
$t\bar{t}$ &
tot & $4b$ & tot & $4b$ \\
  \hline
  \hline
 C1a     & 3.9  &   $4.6\cdot 10^7$   & $1.1\cdot 10^4$ & $2.9\cdot 10^6$ & $4.3\cdot 10^7$ & $2.4\cdot 10^4$   &   $8.2\cdot 10^{-8}$   & $3.4\cdot 10^{-4}$ &  0.03   & 2.0   \\
 C1b     & 2.7  &   $3.7\cdot 10^7$   & $7.5\cdot 10^3$ & $2.1\cdot 10^6$ & $3.5\cdot 10^7$ & $2.2\cdot 10^4$   &   $7.4\cdot 10^{-8}$   & $3.7\cdot 10^{-4}$ &  0.03   & 1.7   \\
 C1c     & 1.0  &   $3.9\cdot 10^6$   & $8.0\cdot 10^2$ & $2.3\cdot 10^5$ & $3.7\cdot 10^6$ & $7.1\cdot 10^3$   &   $2.6\cdot 10^{-7}$   & $1.3\cdot 10^{-3}$ &  0.03   & 2.0   \\
 C2      & 0.16  &   $2.5\cdot 10^2$  & $53$ & $1.9\cdot 10^2$ & $13$ & 1.6               &   $5.7\cdot 10^{-4}$   & $2.7\cdot 10^{-3}$ &   0.5                 & 1.1   \\
\hline
\end{tabular}

    \caption{\small The cross-sections
      for the signal and the background
      processes at different steps of the
      analysis (see Table~\ref{tab:cutflowdetails}), for the resolved (upper),
      intermediate (middle) and boosted
      (lower table) categories, for the analysis
      without PU.
      For each step, the signal over
      background ratio $S/B$, and the signal
      significance $S/\sqrt{B}$ for
       $\mathcal{L}=3$ ab$^{-1}$ are also provided, considering either
      the total background or only the $4b$ component.
 \label{tab:cutflow_noPU_1}}
\end{table}

%
In the boosted category, at the end of the loose cut-based
analysis, we find that around 500 events
are expected
at the HL-LHC, with a large number,
$\simeq 10^6$, of background events.
This leads to a pre-MVA signal significance of
$S/\sqrt{B}=0.5$ and a signal over background
ratio of $S/B=0.06\%$.
From Table~\ref{tab:cutflow_noPU_1}
it is also possible to compute the corresponding pre-MVA
expectations for the LHC Run II with
$\mathcal{L}=300$ fb$^{-1}$: one expects in the boosted
category around
50 signal events, with signal significance dropping down to
$S/\sqrt{B}\simeq 0.16$.
Such signal
significances could have been enhanced
by applying tighter selection requirements,
but our analysis cuts have been left deliberately loose
so that such optimisation may be performed by the MVA.

The resolved category benefits from higher signal yields,
but this enhancement is compensated for by the
corresponding
increase in the QCD multi-jet background.
In both resolved and intermediate categories
the signal significance is
$S/\sqrt{B}\simeq 0.4$,
similar to that of the boosted category.
A further
drawback of the resolved case is
that $S/B$
is substantially reduced as compared to the boosted and
intermediate cases.

Combining the results
from the boosted, intermediate and resolved categories,
we obtain an overall pre-MVA 
significance for the observation of the Higgs pair production
in the $b\bar{b}b\bar{b}$ final
state at the HL-LHC 
of  $(S/\sqrt{B})_{\rm tot} \simeq 0.8$.

\subsection{The role of light  and charm jet mis-identification}

One of the main differences in the present study as compared
to previous works is the inclusion of both irreducible
and reducible background components, which allows us to
quantify
the impact of light and charm jet mis-identification. 
Two recent studies that have also studied the
feasibility of SM Higgs pair production in the $b\bar{b}b\bar{b}$
final state are from the UCL group~\cite{Wardrope:2014kya} and from
the
Durham group~\cite{deLima:2014dta}.
The UCL study is based
on requiring at least four $b$-tagged $R=0.4$ anti-$k_T$ jets
in central acceptance with $p_T \ge 40$ GeV, which are
then used to construct dijets (Higgs candidates) with
$p_T \ge 150$ GeV, $85 \le m_{\rm dijet} \le 140$ GeV
and $\Delta R \le 1.5$ between the two components
of the dijet.
In addition to the basic selection cuts, the constraints
from additional kinematic variables are included by means of a 
Boosted Decision Tree (BDT) discriminant.
The backgrounds included are the $4b$ and
$2b2c$ QCD multijets, as well as
$t\bar{t}$, $Zh$, $t\bar{t}h$ and $hb\bar{b}$.
For the HL-LHC, a signal significance of $S/\sqrt{B}\simeq 2.1$ 
is obtained.

The Durham group study~\cite{deLima:2014dta} requires events
to have two $R=1.2$ C/A jets with $p_T\ge 200$ GeV, and in
addition
two $b$-tagged subjets inside each large-$R$ jet with
$p_T \ge$ 40 GeV each.
To improve the separation between
signal and background, both the BDRS
method and the Shower Deconstruction (SD)~\cite{Soper:2011cr,Soper:2012pb}
technique are used.
The backgrounds considered are QCD $4b$ as well as $Zb\bar{b}$, $hZ$ and
$hW$.
At the HL-LHC, their best result is obtained by requiring two
SD-tagged large-$R$ jets, which leads to $S/\sqrt{B}\simeq 2.1$.
Using the BDRS tagger
results in slightly poorer performance.

 From our results in Table~\ref{tab:cutflow_noPU_1}, we observe
 that the signal significance for the boosted, intermediate,
 and resolved categories is increased to 1.1, 0.6 and 0.6, respectively,
 when only the QCD $4b$ background is included.
 Combining
 the signal significance in the three categories,
 we
 obtain $(S/\sqrt{B_{\rm 4b}})_{\rm tot}\simeq 1.4$, twice
 as large as the result found when
 all background components are included.
 Note the importance of
 the combination of the three exclusive event topologies,
 as opposed the exploitation of a single specific category.
 Taking into account the loose selection cuts, we
 see that
 our pre-MVA results including only the $4b$ background are consistent
 with those reported in previous studies.

 From Table~\ref{tab:cutflow_noPU_1} we
 can compare the interplay
 between the reducible and irreducible components of the
 QCD backgrounds.
 In all cases, the $4b$ and $2b2j$ components have comparable
 magnitudes within the uncertainties from missing higher-order
 corrections.
 On the other hand, the $4j$ component
is always substantially smaller.
  So while the $4j$ component can be safely
 neglected, the inclusion of the
 $2b2j$ component is essential to assess the feasibility
 of measuring Higgs pairs in this final state robustly,
 especially
 in the boosted category.
 This has the important
 consequence that a promising avenue to improve the prospects
 of this measurement would be to reduce, as much as possible,
 the light and charm jet mis-identification rate.

\begin{figure}[t]
\begin{center}
 \includegraphics[width=0.49\textwidth]{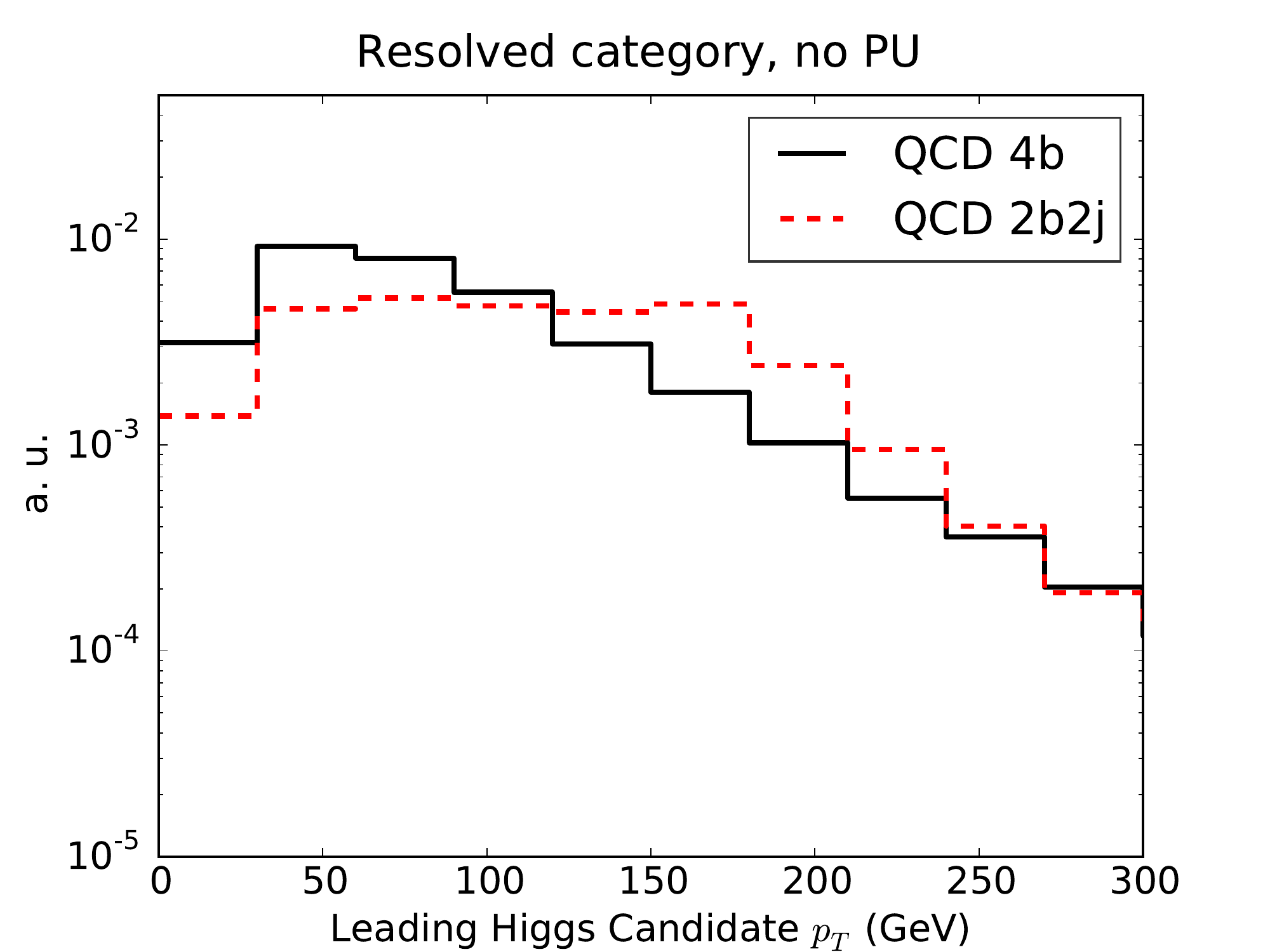}
 \includegraphics[width=0.49\textwidth]{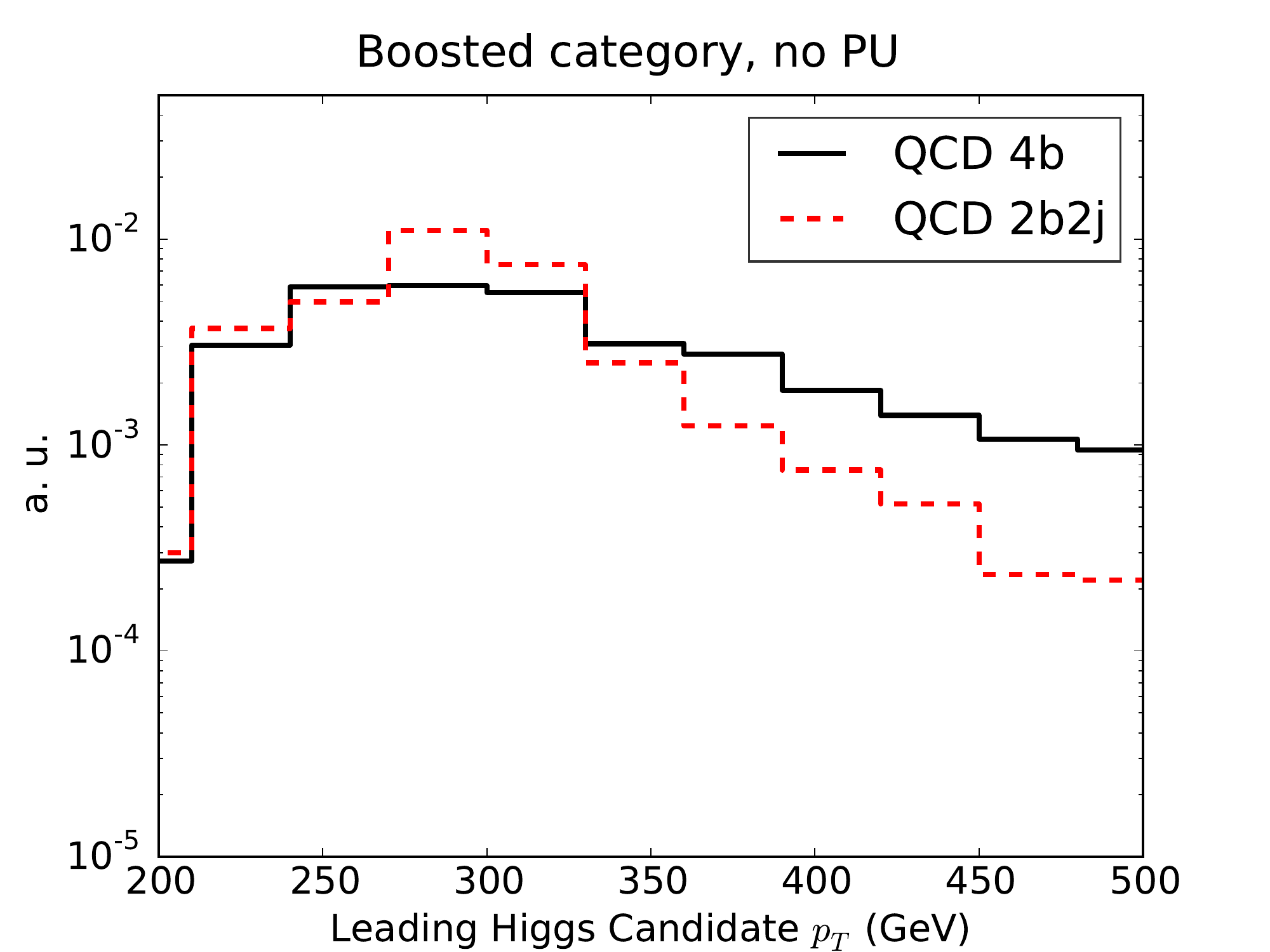}
  \includegraphics[width=0.49\textwidth]{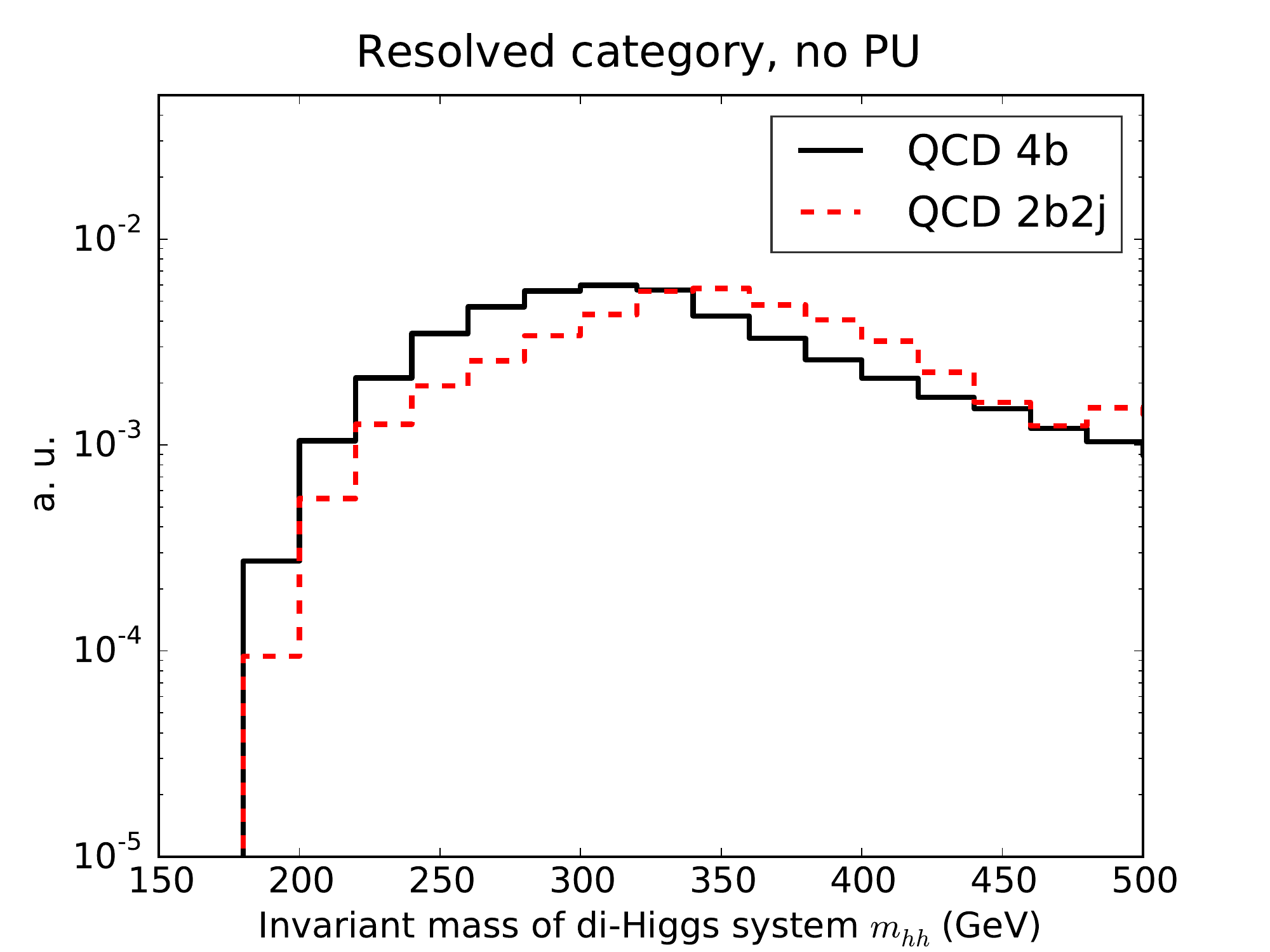}
  \includegraphics[width=0.49\textwidth]{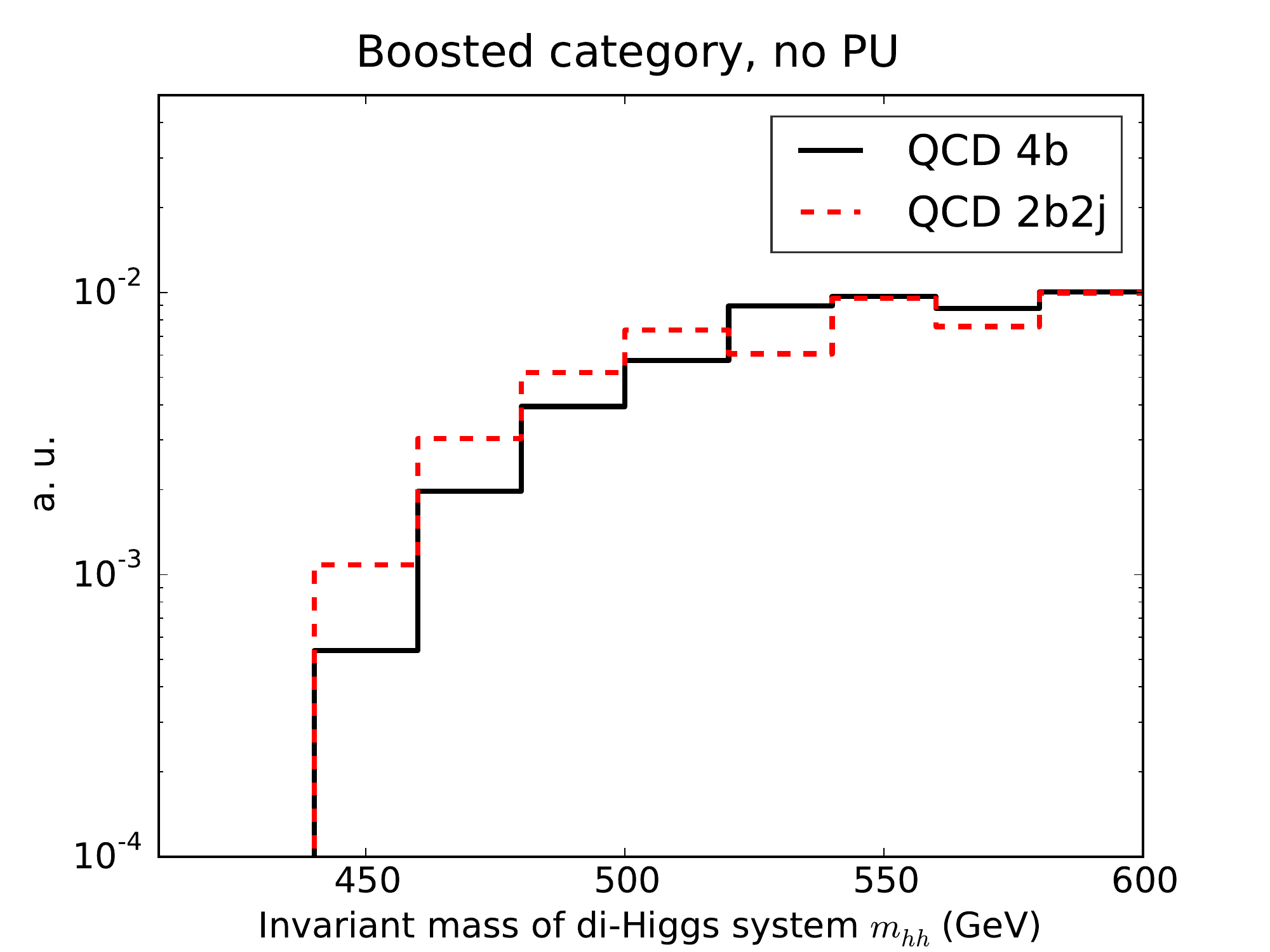}
  \caption{\small
    Upper plots: comparison
    of the shapes of the $4b$ and $2b2j$
components of the QCD background for the $p_T^h$ of the leading
Higgs candidate in the resolved
(left plot) and boosted (right plot) categories.
Lower plots:  same comparison for the invariant
mass $m_{hh}$ of the
    reconstructed di-Higgs system.
}
\label{fig:histoBack}
\end{center}
\end{figure}

 In Fig.~\ref{fig:histoBack} we show a
 comparison
    of the shapes of the $4b$ and $2b2j$
    components of the QCD background for the transverse momentum
    $p_T^h$ of the leading
Higgs candidate and for invariant
mass $m_{hh}$ of the
    reconstructed di-Higgs system in the resolved
    and boosted  categories.
    The two components possess a rather similar shape
    for the two distributions, albeit with some 
    differences.
    In the boosted
    category, the $4b$ component exhibits a less steep fall-off of
    the $p_T^h$ distribution at large $p_T$,
    while in the resolved case
    the $2b2j$ component has a slightly harder
    distribution of the  invariant
    mass $m_{hh}$.
    We also observe that the $2b2j$ distributions
    are affected by somewhat larger
    Monte Carlo fluctuations as compared to $4b$, despite the large size
of the initial sample.
%

In the resolved category, 
the cross-section  before
$b$-tagging is two orders
of magnitude larger in the $2b2j$
sample as compared to the $4b$ sample.
After $b$-tagging, a naive assessment would
suggest a suppression of the $2b2j$ cross-section by a factor $(f_l/f_b)^2 \simeq
1.5\cdot 10^{-4}$, as compared to the $4b$ component,
since a total of four $b$-tags are required to classify the
event as a Higgs candidate.
In this case the ratio of $2b2j$ over $4b$ would be
around $\simeq 3\%$, and therefore negligible.
While  we have checked that this expectation is borne
out at the parton level,
we find that  when parton shower effects
are accounted for the situation is different, due both to radiation of $b\bar{b}$ pairs
and from selection effects.
Due to these,
the
number of  $b$ quarks in the  final state is
increased substantially in the $2b2j$ component as compared
to the parton level, while at the same
time the number of events in the $4b$ sample
with 4 $b$-jets passing selection cuts is reduced.

We can make these statements more quantitative in the following way.
To first approximation, neglecting the contribution from
charm mis-identification,
the
overall efficiency of the $b$-tagging requirements in the resolved category will be
given by the following expression:
\be
\label{btaggingeff}
{\rm EFF}_{\rm b-tag}\simeq \sum_{j=0}^{4}n^{\rm (b-jet)}_j\cdot f_b^{j}\cdot f_l^{4-j} \, ,
\ee
with $n^{\rm (b-jet)}_j$ being the fraction of events satisfying all the selection
requirements,
where $j$ jets out of the leading four jets of the event
contain $b$ quarks (with $p_T^b\ge 15$
GeV).
Similar expressions can be derived for
the boosted and intermediate categories.

The naive expectation is that all events in the $4b$ sample have $n^{\rm (b-jet)}_4\simeq 1$
and $n^{\rm (b-jet)}_j\simeq 0$ for $j\ne 4$, while the events in the $2b2j$ sample
should have $n^{\rm (b-jet)}_2\simeq 1$ and zero otherwise.
This leads to a ratio of overall $b$-tagging selection efficiencies
\be
\label{eq:naive}
\frac{ {\rm EFF}_{\rm b-tag} \lc 2b2j \rc}{{\rm EFF}_{\rm b-tag} \lc 4b\rc}
  \simeq
 \lp \frac{f_l}{f_b}\rp^2 \simeq 1.5\cdot 10^{-4} \, .
\ee
However, after the parton shower, the above estimate is no longer accurate.
First of all, we will have a non-negligible fraction $n^{\rm (b-jet)}_j$
with $j=3,4$ also in the $2b2j$ sample, due to $b$-quark pair radiation
during the shower.
Secondly, not all events in the $4b$ sample will lead to four small-$R$ $b$-jets,
due to a combination of selection cuts and
parton shower effects.
%

\begin{table}[t]
  \centering
  \small
  \begin{tabular}{|c|c|c|c|c|c|c|c|}
    \hline
  \multicolumn{2}{|c|}{}   &  $n^{\rm (b-jet)}_0$  &  $n^{\rm (b-jet)}_1$  &  $n^{\rm (b-jet)}_2$  & $n^{\rm (b-jet)}_3$ &
    $n^{\rm (b-jet)}_4$ & ${\rm EFF}_{\rm b-tag}$ \\
    \hline
    \hline
    Signal  &  $hh\to 4b$  &   0.1\%    & 3\%     &  25\%     & 53\%     & 20\%      & 8.5\%  \\
    \hline
    \multirow{3}{*}{Background}  &  QCD $4b$  & 1\%      &  8\%    &   27\%   &  44\%     & 20\%       &  8.4\% \\
     &  QCD $2b2j$  &   9\%    & 42\%     &  49\%    & 1\%     &  0.1\%     & 0.04\%  \\
    &  QCD $4j$  &   96\%    &  3.5\%     & 0.5\%     &  0.01\%    & $3\cdot 10^{-4}$\%      &
    $2\cdot 10^{-4}$\%\\
    \hline
  \end{tabular}
  \caption{\small
    The relative fractions  $n^{\rm (b-jet)}_j$ of events for the resolved selection
    for which out of the four leading small-$R$ jets of the
    event, $j$ jets
    contain at least one $b$-quark with $p_T^b\ge 15$ GeV.
    This information is provided
    for the di-Higgs signal events and for the three QCD background samples.
    The last column indicates the overall 
    selection efficiency as defined in
    Eq.~(\ref{btaggingeff})
    \label{tab:btaggingcheck}
  }
  \end{table}

In Table~\ref{tab:btaggingcheck} we collect
the values of $n^{\rm (b-jet)}_j$ for the signal and the three QCD background samples.
We find that rather than the estimate Eq.~(\ref{eq:naive}),
the correct ratio of $b$-tagging selection efficiencies is instead
\be
\frac{{\rm EFF}_{\rm b-tag} \lc 2b2j\rc}{{\rm EFF}_{\rm b-tag} \lc 4b\rc}=
  \frac{0.04\%}{8.4\%} \simeq 5\cdot 10^{-3} \, .
  \ee
  This suppression factor is of the same order as
  the ratio of $4b$ to $2b2j$ cross-sections
  in the resolved category before $b$-tagging.
    This explains why the $2b2j$ contribution cannot be neglected as compared
    to the irreducible $4b$ component of the QCD background.
    A similar calculation from the numbers in Table~\ref{tab:btaggingcheck}
    shows
    that, on the other hand, the $4j$ component of the background
    can be neglected.

\section{Multivariate analysis}
\label{sec:mva}

At the end of the loose cut-based analysis,
by combining the three event topologies,
we obtain a signal significance of $S/\sqrt{B}\simeq 0.8~(1.4)$
with all backgrounds (only QCD $4b$) considered.
This section describes how this signal significance
 can be enhanced when the cut-based analysis
 is complemented by multivariate techniques.
 These are by now a mature tool
 in high-energy physics data analysis, opening
 new avenues to improve the performance
of many measurements and searches at high-energy colliders.
In particular, the classification of events into
signal and
background processes by means of MVAs is
commonly used in LHC
applications~\cite{Baldi:2014pta,Aaltonen:2012qt,
  Wardrope:2014kya,Chatrchyan:2013zna,Dall'Osso:2015aia,Kang:2015uoc}.

In this section, first we present the specific MVA that we use,
based on feed-forward multi-layer neural networks.
Then we introduce the input variables that are
used in the MVA, including the jet substructure
variables, and then present the signal significance obtained
by applying the MVA.
Then we assess the robustness of the MVA strategy in
the case of significant contamination from pileup.

\subsection{Deep artificial neural networks}

The specific type of  MVA that we use to
disentangle signal and background events is
a multi-layer feed-forward artificial neural network (ANN),
known as a {\it perceptron}.\footnote{This type of ANNs are the same
  as those used to parametrize Parton Distribution Functions
in the NNPDF global analyses~\cite{DelDebbio:2004qj,Ball:2008by,Ball:2011mu,Ball:2010de}.}
This family of ANNs are also known as {\it deep neural networks},
due to their multi-layered architecture.
The MVA inputs are a set of kinematic variables describing the
signal and background
events which satisfy the requirements of the
cut-based analysis.
The output of the trained ANNs also allows for the identification,
in a fully automated way,
of the most relevant variables in the discrimination between 
signal and background.

\begin{figure}[t]
  \begin{center}
      \vspace{-1cm}
  \includegraphics[width=0.90\textwidth]{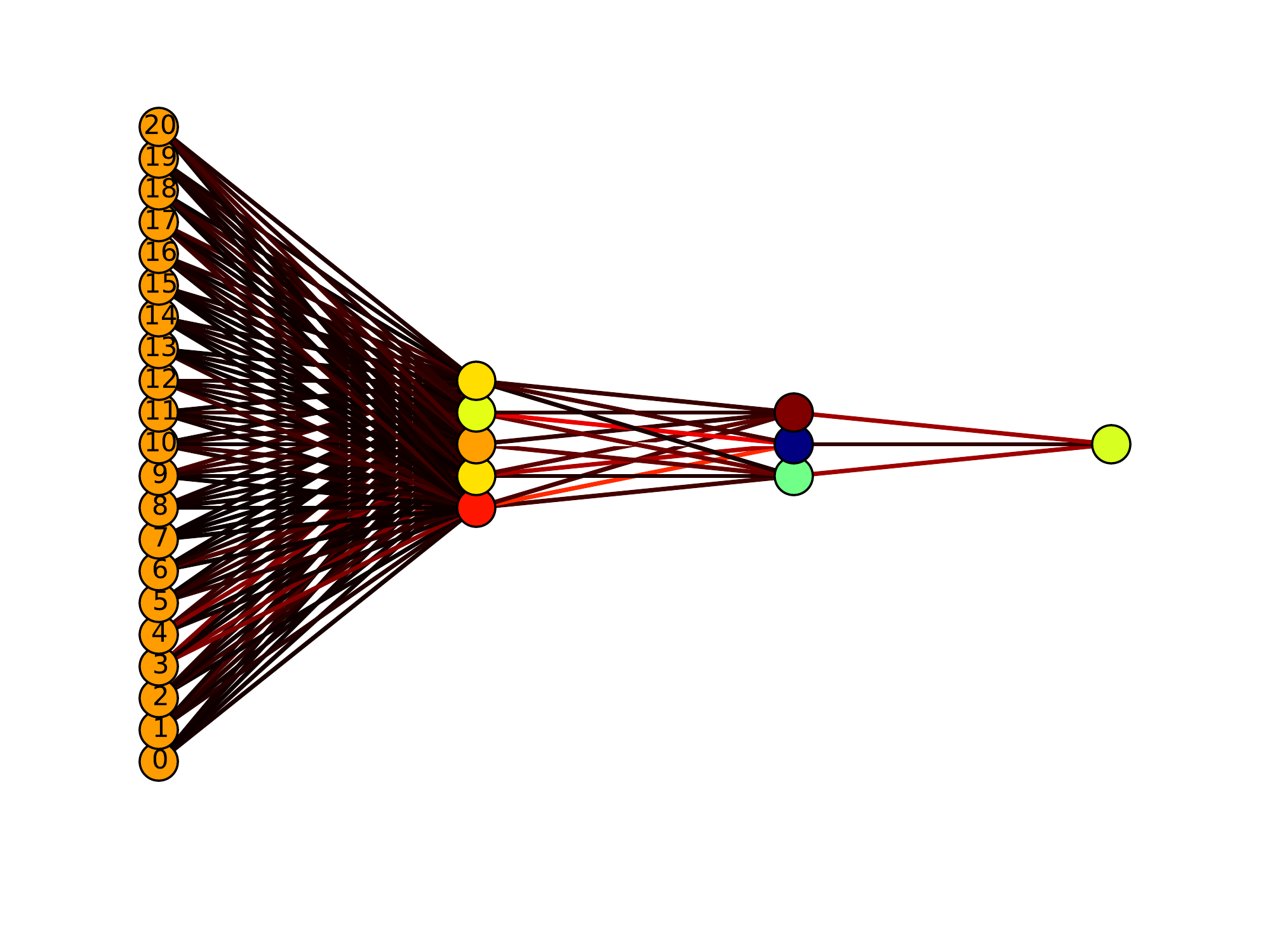}
  \vspace{-2cm}
  \caption{\small Schematic of the Artificial
    Neural Network (ANN)
    used for the analysis of the
    boosted
    category, with $N_{\rm var}=21$ input variables and thus
    the same number of neurons
  in the first layer.
  The color code in the neuron connections (the weights) is a heat map obtained
  at the end of the Genetic Algorithms training,
  with red indicating larger values and black indicating smaller values.
}
\label{fig:nnarch}
\end{center}
\end{figure}

In this work, the ANN that we use has the following architecture.
\be
\label{eq:nn1}
N_{\mathrm{var}}\times5\times3\times1 \, ,
\ee
where $N_{\mathrm{var}}$ represents the number of input variables for the MVA,
which is different in the resolved, intermediate, and boosted categories.
All neural-network layers use a sigmoid activation function, allowing
for a probabilistic
interpretation of the ANN output.
In Fig.~\ref{fig:nnarch} we show an illustrative
example of an ANN used in this work, corresponding 
to the case of the boosted category (thus $N_{\mathrm{var}}=21$, as we explain below).

The training of the ANN for the signal/background classification task
proceeds as follows.
Given a set of $N_{\mathrm{var}}$  kinematic variables $\{k\}_i$ associated with the event $i$, and a set of neural network weight
parameters $\{\omega\}$, we interpret the neural network output $y_i$
(the activation state of the
neuron in the last layer)
as the probability that the event $i$ originates from the signal process,
\be
y_i = P(y^\prime_i=1|\{k\}_i, \{\omega\} )\, ,
\ee
where $y_i^\prime$ represents the true classification of the event $i$, {\it i.e},
$y^\prime_i = 1$ for signal and $y^\prime_i = 0$ for background events.
With this interpretation, our general classification probability including background events is given by
\be
P(y_i^\prime|\{k\}_i, \{\omega\}) = y_i^{y^\prime_i}(1-y_i)^{1-y^\prime_i} \, ,
\ee
consequently we can define an error function $E(\{\omega\})$
to be minimized during the ANN training. In this case, the error function is
the cross-entropy function, defined as
 \bea
 &&E(\{\omega\}) \equiv -\log\left(\prod_i^{N_{\text{ev}}} P(y_i^\prime|\{k\}_i, \{\omega\})\right)\nonumber\\
 &&=
 \sum_i^{N_{\text{ev}}} \lc y^\prime_i\log{y_i} + (1-y^\prime_i)\log{(1-y_i)}\rc \, ,
 \label{cross-entropy}
 \eea
 where $N_{\text{ev}}$ is the number of
 Monte Carlo events that are used for the ANN training.
 The ANN is trained both on the signal and background MC events,
 so it is important to ensure that the input MC sample is large enough
 to avoid contamination from MC statistical fluctuations.
 
 The training of the neural networks therefore consists of the
 minimization of the cross-entropy error,
 Eq.~(\ref{cross-entropy}), which in this work is achieved using a
 Genetic Algorithm (GA).
 Genetic Algorithms~\cite{quevedo,tau,Abel:2014xta,Nesseris:2012tt} are
 non-deterministic
 minimization strategies suitable for the solution
 of complex optimization problems, for instance when a very large number
 of quasi-equivalent minima are present.
 GAs are inspired on natural selection processes
 that emulate biological evolution. 
 In our case, the GA training is performed for a very large 
 number of generations, $N_{\rm gen}=5\cdot 10^{4}$, to avoid the risk of
 under-training.
 We have verified that if a much larger number of generations
 are used, the results are unchanged.

 In addition,
 in order to avoid the possibility of over-fitting,
 we have used a cross-validation stopping
 criterion, in particular the same one as
 that used in the NNPDF3.0 analysis~\cite{Ball:2014uwa}.
 This cross-validation proceeds by dividing the input MC dataset into two disjoint sets,
 using one for training the ANN and the other for validation: the optimal
 stopping point is then given by the minimum of the error function
 Eq.~(\ref{cross-entropy}) to the validation sub-sample.
 This indicates the point where
the ANN begins to train upon  statistical fluctuations
in the input MC samples, rather than learning
the underlying (smooth) physical  distributions.
 
 \subsection{Input kinematic variables}
 \label{sec:input}

In this work we use different sets of
input variables for the three categories.
In the case of large-$R$ jets, we  exploit the available
information  on jet substructure.
For the three categories, boosted, intermediate and resolved,
the following common variables are used as input to the MVA:
\begin{itemize}
\item The transverse momenta of the leading and subleading Higgs, $p_{T,h_1}$ and $p_{T,h_2}$.
\item The transverse momentum of the reconstructed Higgs pair, $p_{T,hh}$.
\item The invariant masses of the leading and sub-leading Higgs candidates, $m_{h,1}$ and $m_{h,2}$.
\item The invariant mass of the reconstructed Higgs pair, $m_{hh}$.
\item The separation in the $\phi$--$\eta$ plane
  between the two Higgs candidates, $\Delta R_{hh}$.
  \item The separation in $\eta$  between the two Higgs candidates, $\Delta \eta_{hh}$.
\item The separation in $\phi$  between the two Higgs candidates, $\Delta \phi_{hh}$.
\end{itemize}
In addition, in the boosted category we use
  the transverse momenta of the leading, $p_{T,h_{1,1}}$ and $p_{T,h_{1,2}}$ and
  sub-leading, $p_{T,h_{2,1}}$ and $p_{T,h_{2,2}}$, Higgs candidate AKT03 subjets.
  In the resolved category instead,
  the corresponding variables are
  the transverse momenta $p_{T,i}$ of the four leading 
  $b$-tagged small-$R$ jets in the event.
  In the intermediate category, we use the
  transverse momenta of the subjets
  from the large-$R$ jet $p_{T,h_{1,1}}$ and $p_{T,h_{1,2}}$ and the
 transverse momenta $p_{T,i}$ of the two leading 
  $b$-tagged small-$R$ jets.
 Therefore, we have 13 variables which are common to the three categories.

 In the boosted and intermediate categories, we also include the jet substructure
 variables introduced in Sect.~\ref{sec:analysis} for the
 large-$R$ jets: the $k_t$ splitting scales
 $\sqrt{d_{12}}$, the ratio of 2-to-1 subjettiness $\tau_{12}$,
 and the ratios of energy correlation functions $C^{(\beta)}_2$ and
 $D_2^{(\beta)}$.
 This leads to
 a total of $N_{\mathrm{var}}=13,17$ and 21 variables for the
resolved, intermediate, and boosted categories, respectively.

Given that the MVA is able to identify the most discriminatory variables
in an automated way,
and to suppress those which have little effect, it is advantageous to
include a wide array of input variables.
This is one of the main advantages of ANNs in this context: 
their inherent redundancy means that 
adding additional information, even if carries very little weight,
should not degrade
the classification power of the MVA.

\subsection{MVA results}
\label{sec:signalsignificance}

\begin{figure}[t]
\begin{center}
\includegraphics[width=0.65\textwidth]{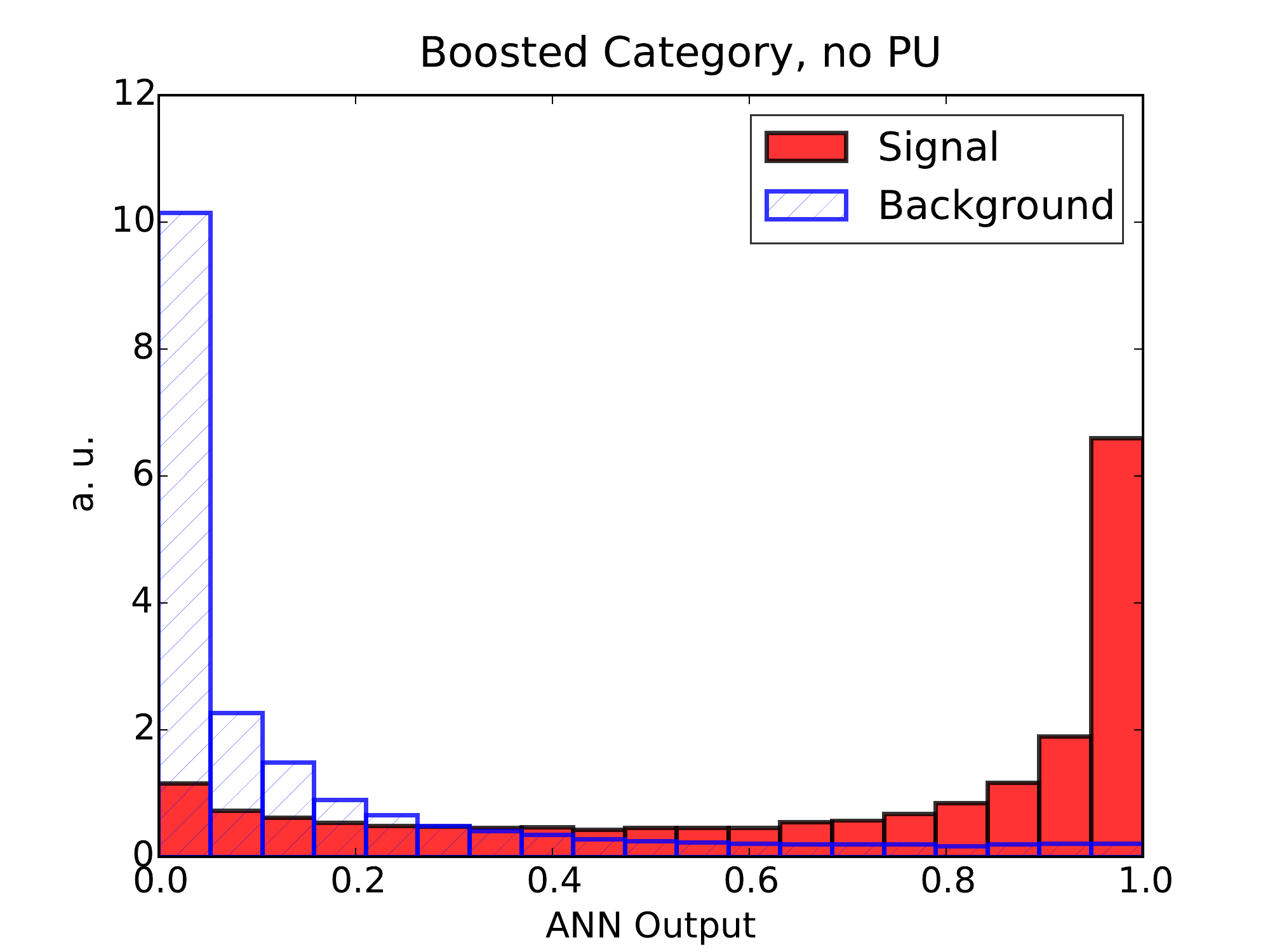}
\includegraphics[width=0.48\textwidth]{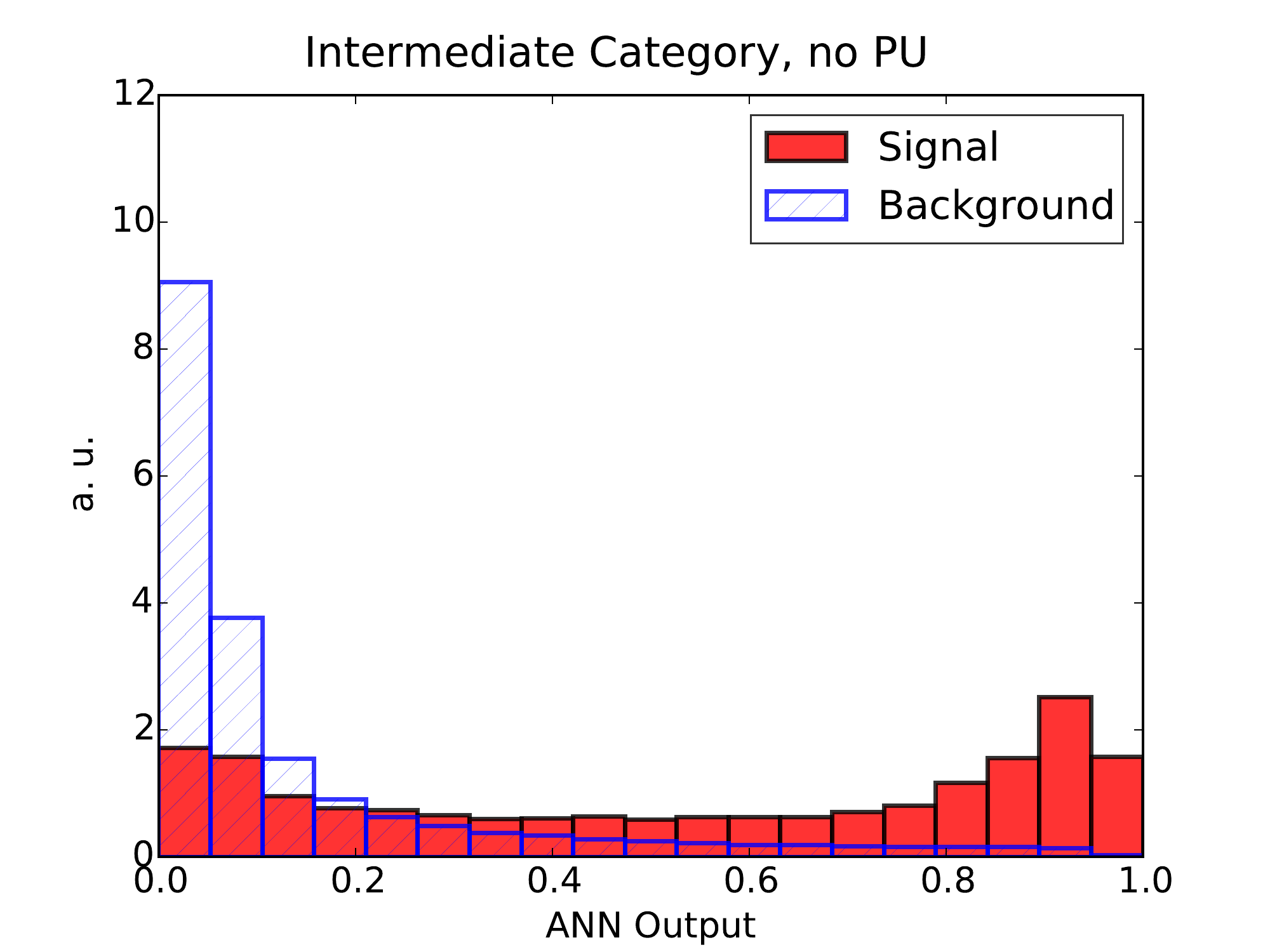}
\includegraphics[width=0.48\textwidth]{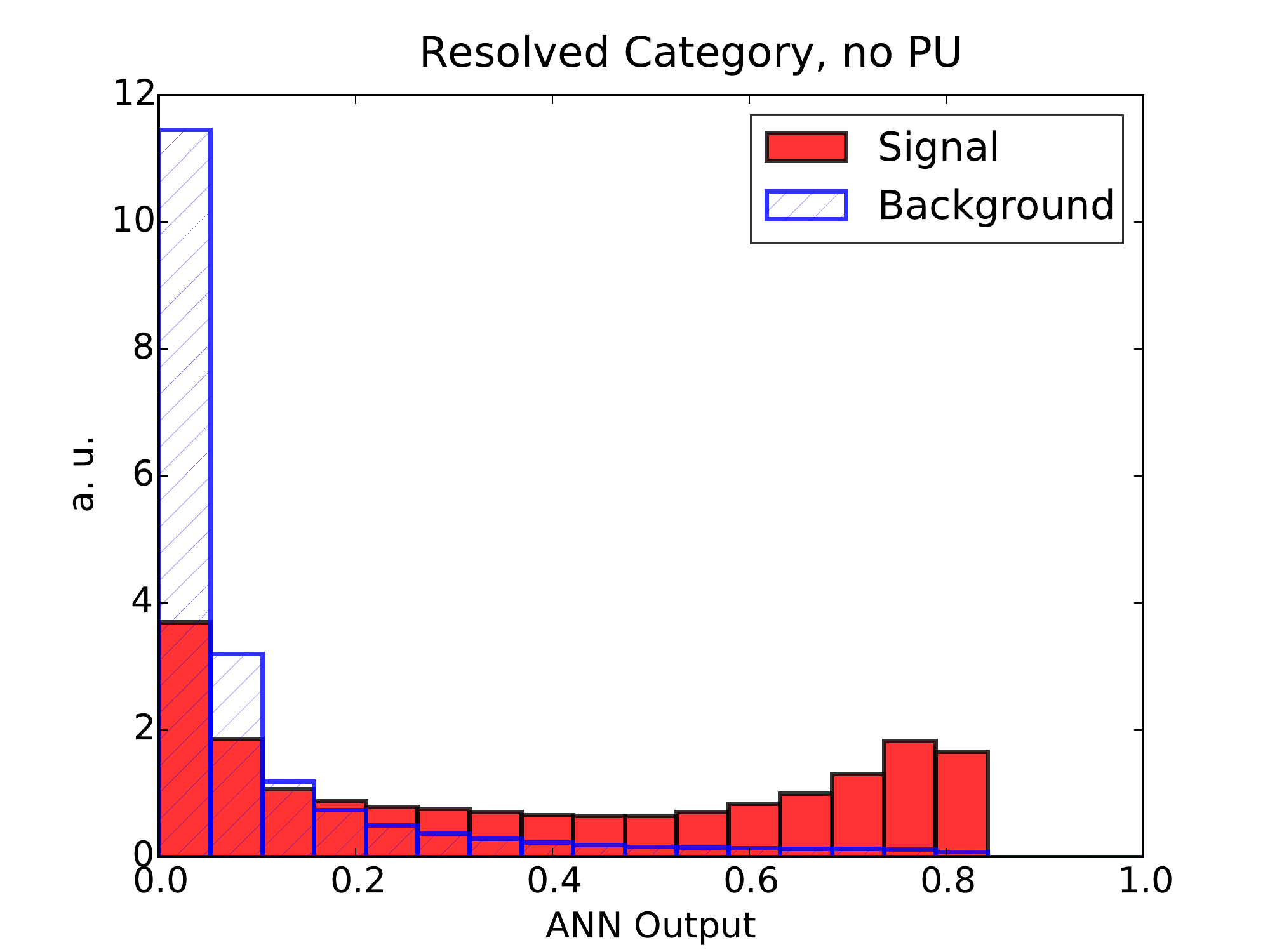}
\caption{\small The distributions, at the end of the
  GA training, 
  for the signal and background MC events in the three categories:
  boosted (upper plot), intermediate (lower left plot) and
  resolved (lower right plot), as a function of the ANN output.
}
\label{fig:nnresponse}
\end{center}
\end{figure}

We now present the results of the MVA, first without PU, and then
later including the effects of PU.
First of all, in Fig.~\ref{fig:nnresponse} we show the distribution of
the ANN output at the end of the GA minimization,
separately for the
boosted, intermediate and resolved categories.
All distributions are normalized so that their integral
  adds up to one.
The  separation between signal and background is achieved by introducing
a cut, $y_{\rm cut}$, on the ANN output, so that MC events with $y_i\ge
y_{\rm cut}$ are classified as signal events, and those with
 $y_i <
y_{\rm cut}$ as background events.
Therefore,
the more differentiated the distribution of the ANN output is
for signal and background events, the more efficient
the MVA discrimination will be.

From Fig.~\ref{fig:nnresponse} we see that in the boosted category the MVA can produce
a clear discrimination between signal and background, with the two distributions
forming peaks at their respective optimal limits.
This indicates that introducing a suitable cut
$y_{\rm cut}$
in the ANN output will substantially reduce the background,
while keeping a reasonable signal efficiency.
The performance of the MVA discrimination is similar,
although slightly worse, in the intermediate
and resolved categories.

\begin{figure}[t]
\begin{center}
  \includegraphics[width=0.49\textwidth]{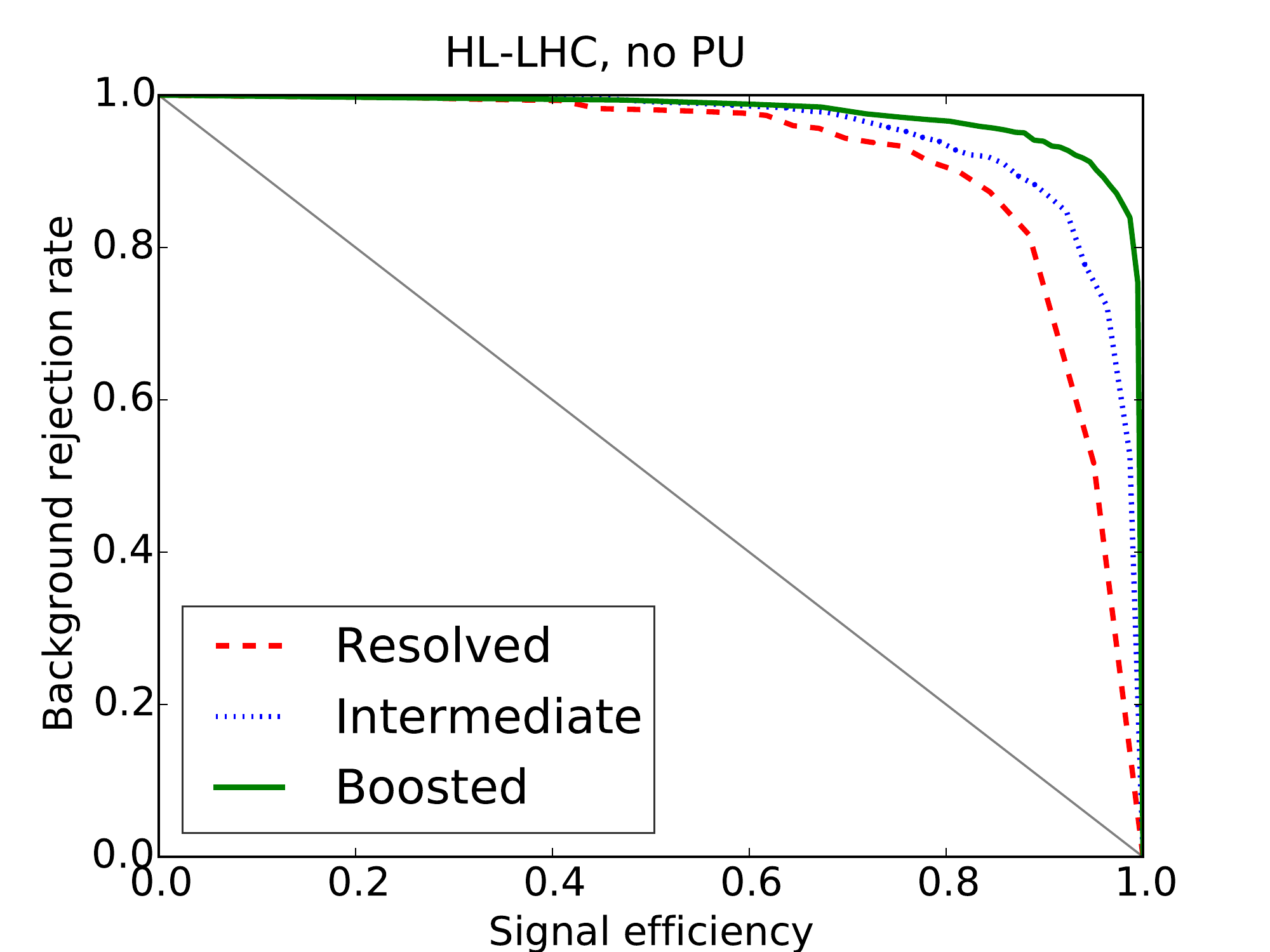}
  \includegraphics[width=0.49\textwidth]{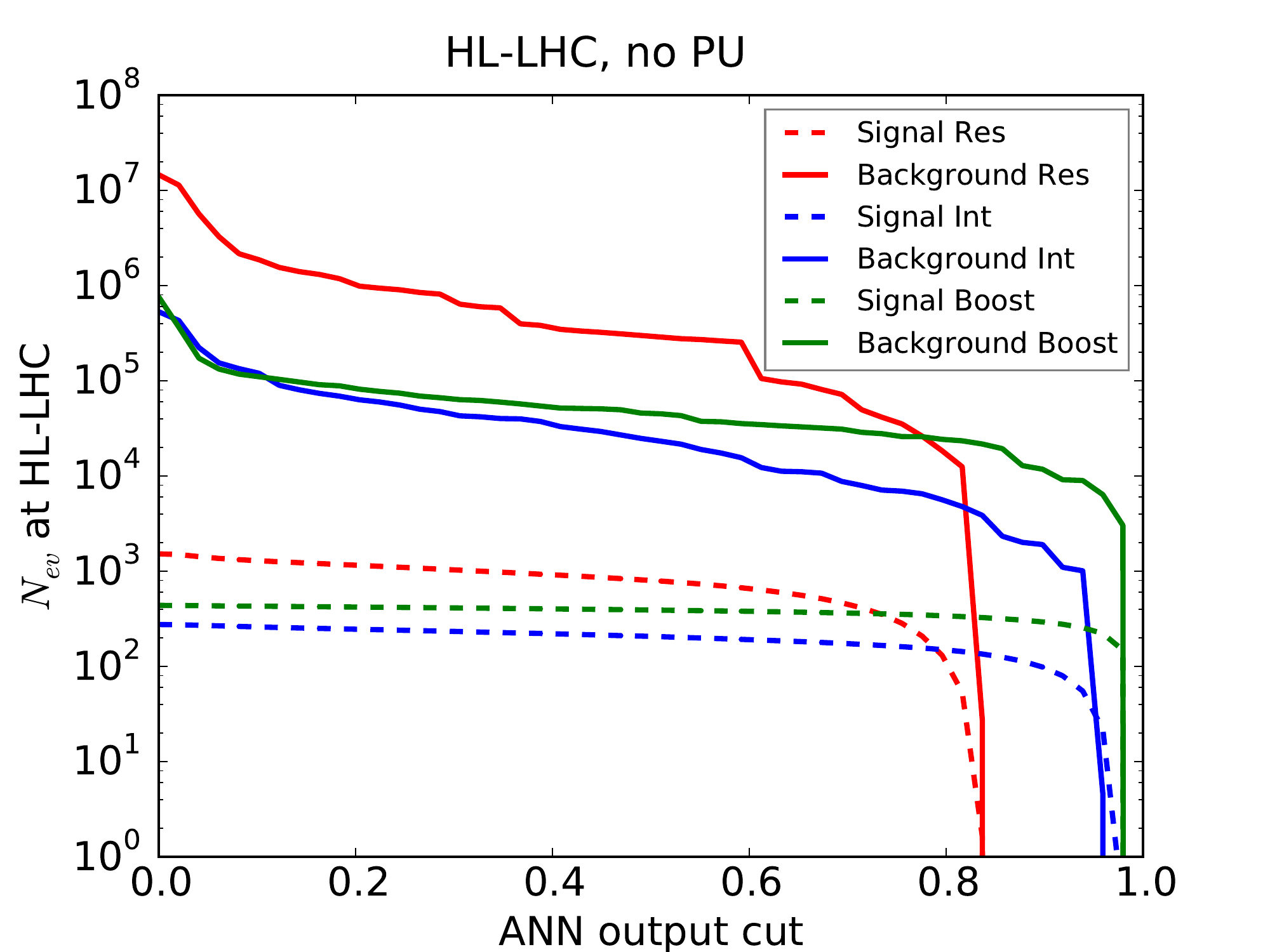}
\caption{\small Left: ROC curve for the background rejection rate as a function of the signal
  selection efficiency, as the cut $y_{\rm cut}$
  in the ANN output is varied.
  Right: Number of signal (dashed) and background (solid)
  events expected at the HL-LHC as a function of the $y_{\rm cut}$.
}
\label{fig:exampleroc}
\label{fig:nev2}
\end{center}
\end{figure}

The results for the signal selection efficiency and the 
background rejection rate as a function of the cut in the ANN output
$y_{\rm cut}$
define the so-called  Receiver-Operating Characteristic (ROC)
curve, shown in Fig.~\ref{fig:exampleroc}.
It is clear that we can achieve  high signal efficiency by using
a small value of $y_{\rm cut}$, but such a choice would be
affected by poor background
rejection.
Conversely, using a higher value of the cut will increase background rejection at the
cost of dropping signal efficiency.
As could already be inferred from the distribution of neural
networks output in Fig.~\ref{fig:nnresponse}, we find
that our MVA is reasonably efficient
in discriminating signal over background.
The performance is best in the case of the boosted category,
and then slightly worse in the resolved
and intermediate categories, consistent with the distributions of
the ANN outputs in
Fig.~\ref{fig:nnresponse}.
%

It is useful to estimate, for each value of
the cut in the ANN output $y_{\rm cut}$, how many
signal and background events are expected at the HL-LHC
with $\mathcal{L}=3$ ab$^{-1}$.
This comparison is shown in 
Fig.~\ref{fig:nev2}.
We observe that
in the boosted category, for a value $y_{\rm cut}\simeq 0.9$
we end up with around 300 signal events and $10^4$ background
events.
Similar results are obtained in the intermediate and resolved
categories: in the former we find 130 ($3\cdot 10^3$) signal (background)
events for $y_{\rm cut}\simeq 0.85$ (0.60), and in the latter
630 ($10^5$) signal (background) events for
$y_{\rm cut}\simeq 0.6$.
Therefore, the MVA achieves a
substantial background suppression
with only a
moderate reduction of signal efficiency.

A useful property of MVAs such as the one used in our
analysis
is that they can provide direct  physical insight about which of the
input variables contribute to the separation between
signal and background.
In the case of ANNs, this can be quantified by computing the sum
of the absolute values of all the weights connected to a given
input neuron $i$, that is
\be
\label{eq:totweight}
\omega^{\rm (tot)}_i \equiv \sum_{k=1}^{n^{(2)}} \Big|\omega^{(2)}_{ki}\Big| \, ,
\qquad i=1,\ldots,N_{\rm var} \, ,
\ee
with $\omega^{(2)}_{ki}$ the value of the weight connecting
the $k$-th neutron of the second layer with the $i$-th neuron of
the first (input) layer, and $n^{(2)}=5$ the number of
neurons in the second layer.
Those input variables with a larger value of $\omega^{\rm (tot)}_i$ will be those
that play a more significant role in enhancing the signal
discrimination using the MVA.
We note however
that the estimate provided
by Eq.~(\ref{eq:totweight}) is necessarily qualitative.

\begin{figure}[t]
  \begin{center}
    \includegraphics[width=0.49\textwidth]{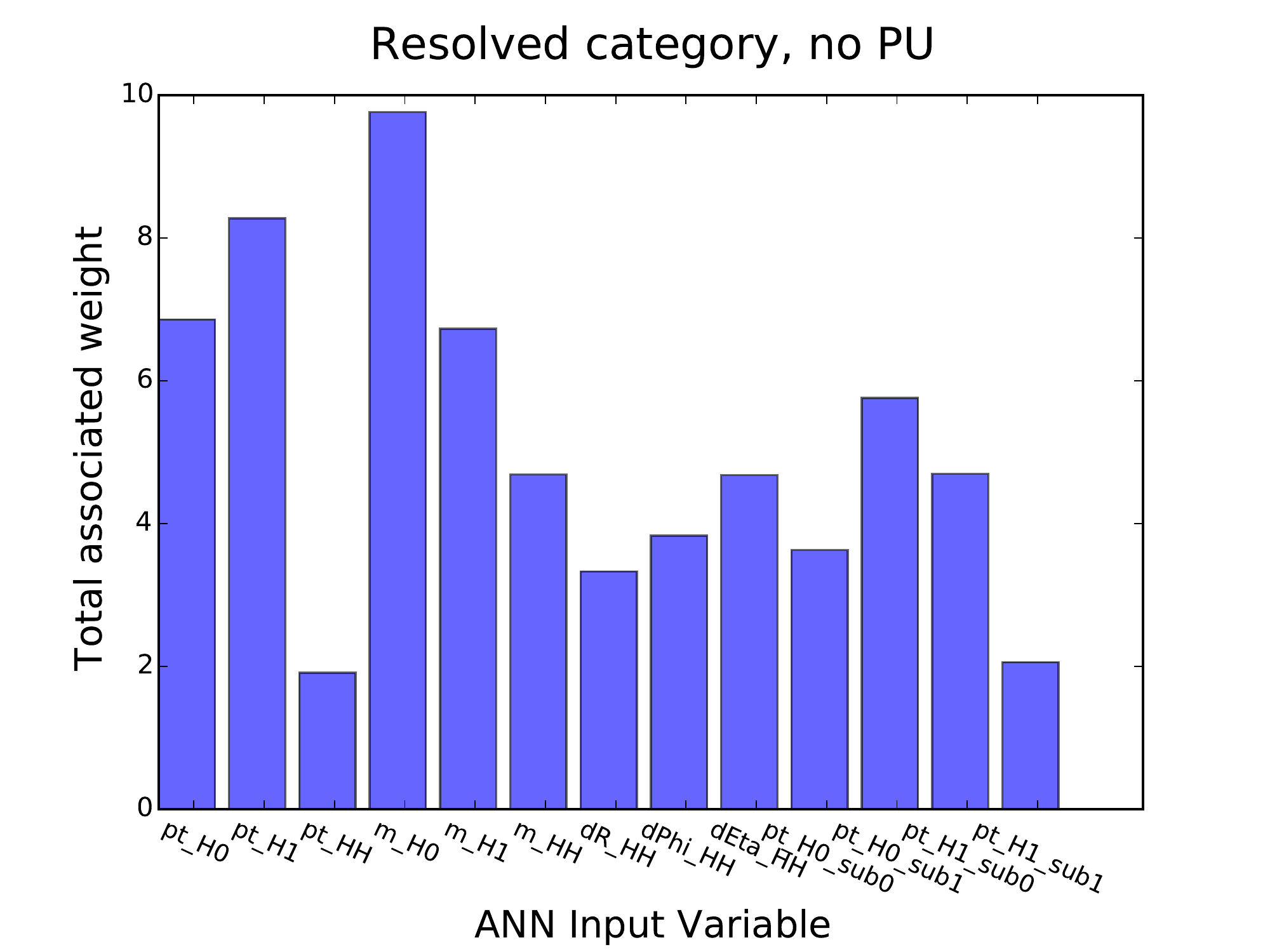}
\includegraphics[width=0.49\textwidth]{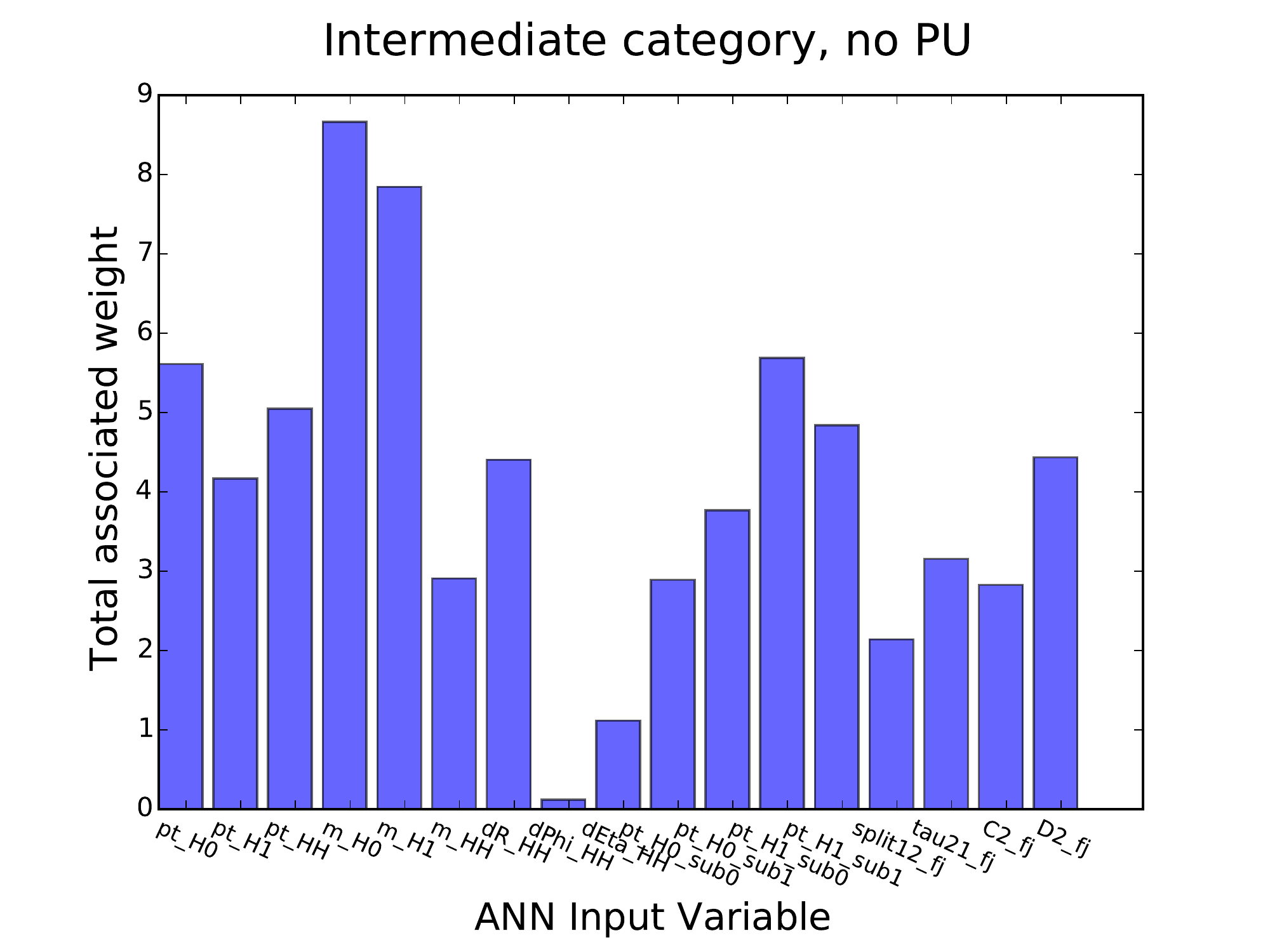}
\includegraphics[width=0.60\textwidth]{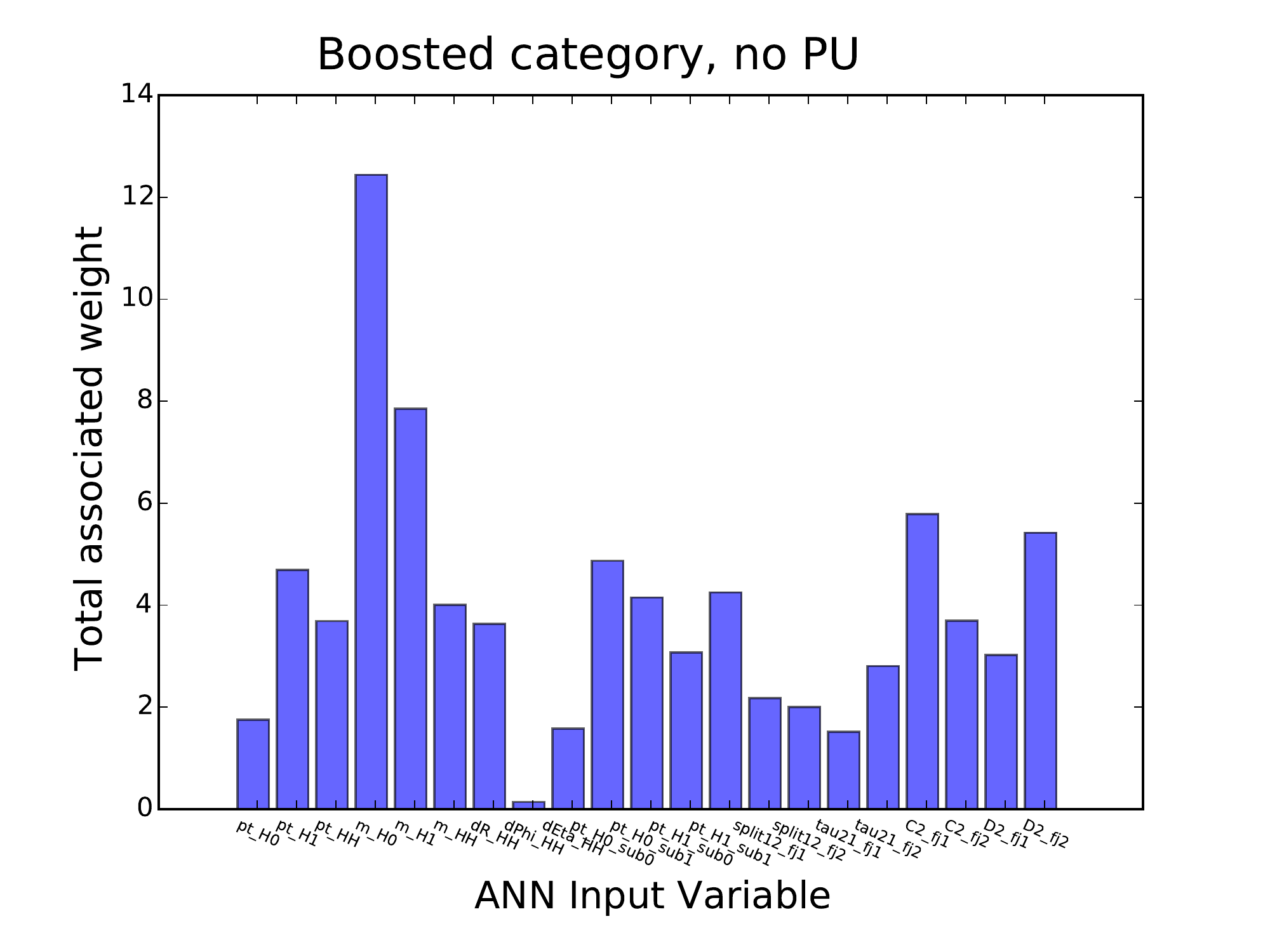}
\vspace{-0.5cm}
\caption{\small
Distribution of the total associated weight,
Eq.~(\ref{eq:totweight}) for each of the $N_{\rm var}$ input
variables of the resolved (upper left),  intermediate (upper right)
and boosted (lower plot)
categories.
}
\label{fig:nnweights}
\end{center}
\end{figure}

%
In Fig.~\ref{fig:nnweights} we show
the distribution of the total associated weight,
Eq.~(\ref{eq:totweight}) for each of the $N_{\rm var}$ input
variables of the three categories, using the
notation for the kinematic variables
as in Sect.~\ref{sec:input}.
In the 
resolved category, the variables that carry 
a higher discrimination power
are the transverse momentum of the two reconstructed Higgs candidates and
their invariant masses.
In the case of the boosted category, the invariant mass distribution
of the Higgs candidates is also the most discriminatory
variable, followed by the subjet $p_T$ distributions and
substructure variables such as $C_2^{(\beta)}$ and
$D_2^{(\beta)}$.

\begin{figure}[t]
\begin{center}
\includegraphics[width=0.48\textwidth]{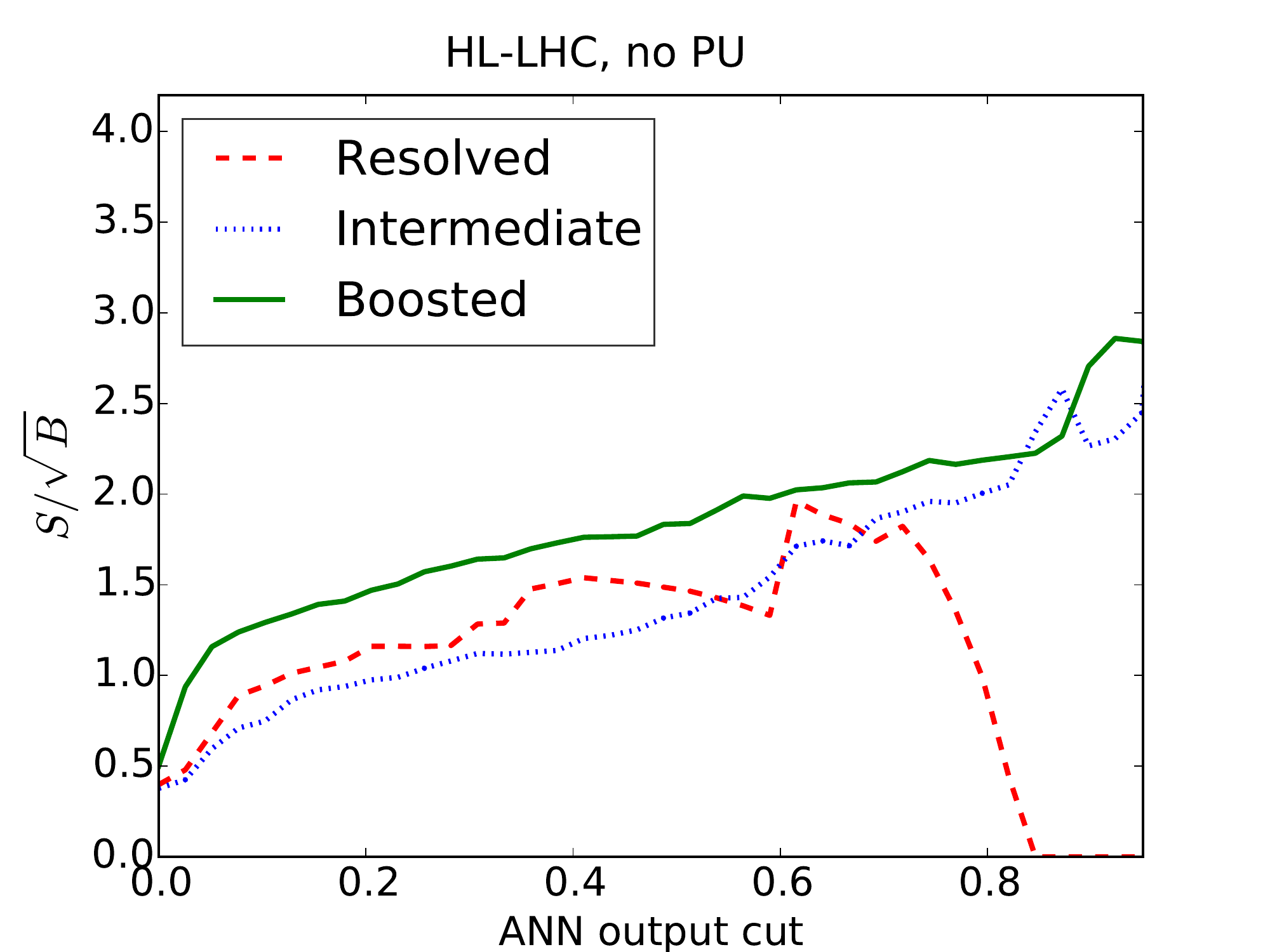}
\includegraphics[width=0.48\textwidth]{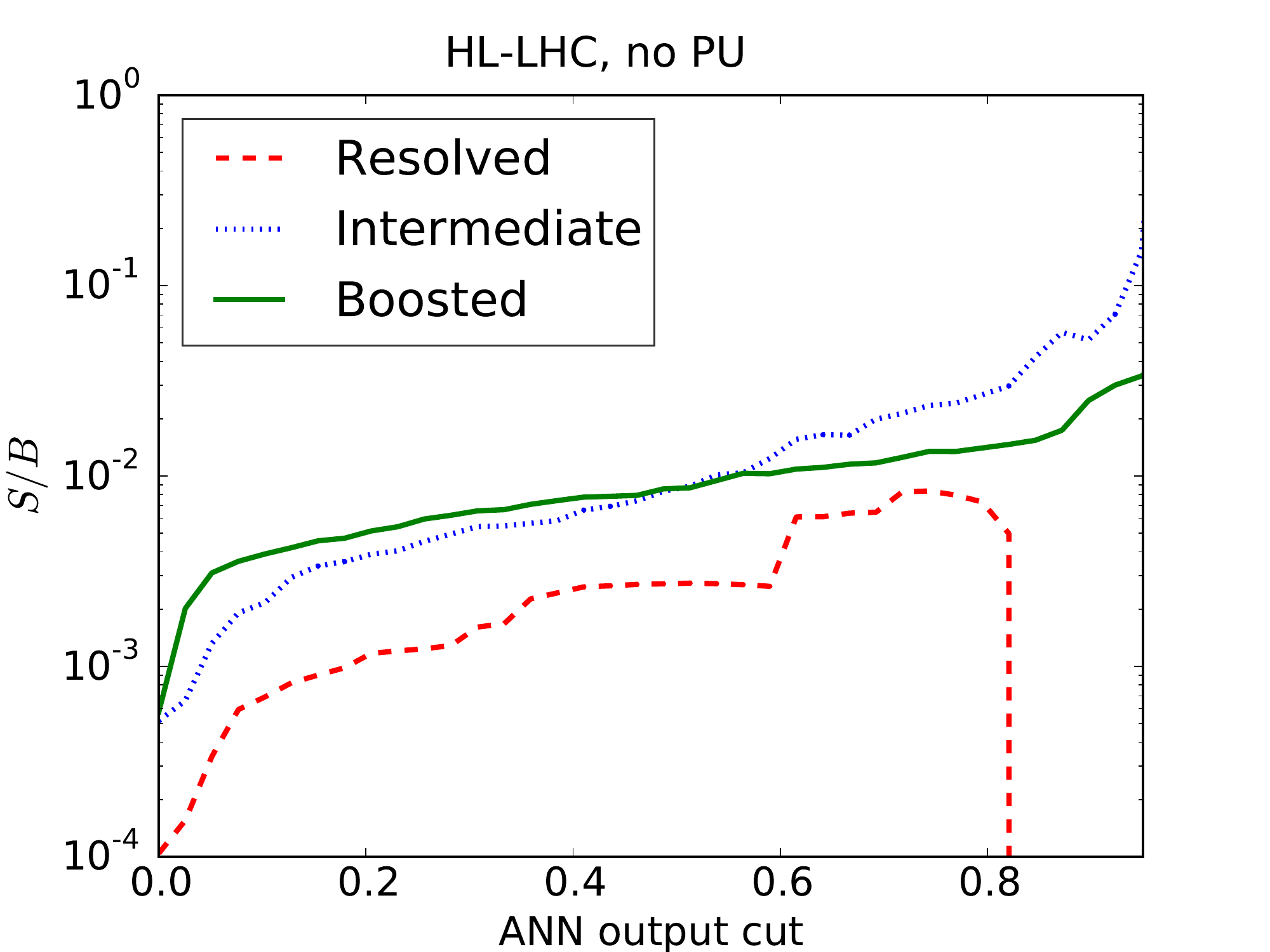}
\caption{\small
  The values of the signal significance, $S/\sqrt{B}$, and of the
  signal over background ratio, $S/B$, for the boosted, intermediate
  and resolved categories as a function of the cut
  $y_{\rm cut}$ in the ANN output.
  The $y_{\rm cut}=0$
  results are those at the end of the cut-based
  analysis.
}
\label{fig:sb_mva}
\end{center}
\end{figure}

The results for the signal significance $S/\sqrt{B}$ and
the signal over background ratio
$S/B$ as a function of $y_{\rm cut}$
for the three categories are given in 
Fig.~\ref{fig:sb_mva}.
The values 
for $y_{\rm cut}=0$ correspond to those at
the end of the loose cut-based analysis.
We observe how in the three
 categories there is a marked  improvement in signal
significance as compared to the pre-MVA results.
We also observe a substantial enhancement in $S/B$, arising
from the background suppression achieved by the MVA, reaching
values of 1\%, 6\% and 3.5\% in the resolved,
intermediate and boosted categories.
This improvement in $S/B$ is crucial to ensure the feasibility
of this measurement, since it allows systematic
uncertainties in the background determination to
be at most of a similar size.

The optimal value of the cut in the
ANN output, $y_{\rm cut}$,  can be determined from the maximisation of $S/\sqrt{B}$,
ensuring that the number of signal events $N_{\rm ev}$
expected at the HL-LHC does not become too low.
In  addition, we require
that the number of MC events used to define the signal
category (events with $y_i \ge y_{\rm cut}$)
is sufficiently large in order to avoid the biases and statistical
fluctuations associated to a small training sample.
In Table~\ref{table:cutflowMVA} we quote, for the optimal
value of $y_{\rm cut}$
in each category,
the number of signal and background events $N_{\rm ev}$ expected
at the HL-LHC, as well as $S/\sqrt{B}$ and $S/B$.
For completeness, we also include the corresponding
pre-MVA results.
%

\begin{table}[t]
  \centering
  \begin{tabular}{|c|l|c|c|c|c|}
    \hline
    \multicolumn{6}{|c|}{HL-LHC, no PU} \\
    \hline
    \hline
    Category  &   &  $N_{\rm ev}$ signal &  $N_{\rm ev}$ back  &  $S/\sqrt{B}$ & $S/B$ \\ 
    \hline
    \hline
    \multirow{2}{*}{Boosted} &  $y_{\rm cut}=0$  & 440 & $7.6\cdot 10^5$  & 0.5  & $6\cdot 10^{-4}$  \\
    &  $y_{\rm cut}=0.90$ & 290  & $1.2\cdot 10^4$   & 2.7  & 0.03 \\
    \hline
    \hline
    \multirow{2}{*}{Intermediate} &  $y_{\rm cut}=0$  & 280   & $5.3\cdot 10^5$
    & 0.4 & $5\cdot 10^{-4}$ \\
    &  $y_{\rm cut}=0.85$ & 130  & $3.1\cdot 10^3$  & 2.3 & 0.04\\
    \hline
    \hline
      \multirow{2}{*}{Resolved} &  $y_{\rm cut}=0$  & 1500 &  $1.5\cdot 10^{7}$ &  0.4 &$1\cdot 10^{-4}$ \\
    &  $y_{\rm cut}=0.60$ & 630  & $1.1\cdot 10^{5}$ & 1.9 & 0.01 \\
    \hline
      \end{tabular}
  \caption{\small Post-MVA results, for the optimal value of the
    ANN discriminant $y_{\rm cut}$ in the three categories, compared with the
    corresponding
    pre-MVA results ($y_{\rm cut}=0$).
    We quote the number of signal and
    background events expected for $\mathcal{L}=3$ ab$^{-1}$,
    the signal significance $S/\sqrt{B}$ and
    the signal over background ratio $S/B$.
    The pre-MVA results correspond to row C2 in
    Table~\ref{tab:cutflow_noPU_1}.
    \label{table:cutflowMVA}
  }
\end{table}

From Table~\ref{table:cutflowMVA} we see that
following the application of the MVA, 
the signal significance in the boosted category increases
from 0.5 to 2.7, with $S/B$ increasing from $0.06\%$ to $3\%$.
For the intermediate and resolved categories, $S/\sqrt{B}$
increases from 0.4 to 2.3 and 1.9 respectively, with
the signal over background ratio raising from
$0.05\%$ and $0.01\%$ to 4\% and 1\%.
Combining the three categories, taking into
account all background components, we obtain the overall signal
significance:
\be
\label{eq:soverb}
\lp \frac{S}{\sqrt{B}}\rp_{\rm tot} \simeq 4.0~(1.3) \, ,\quad
\mathcal{L}=3000~(300)\,{\rm fb}^{-1}\, ,
\ee
The signal significance for
$\mathcal{L}=3$ ab$^{-1}$
is thus
well above the threshold for the observation of Higgs
pair production.
However, given that the HL-LHC will be a high-PU environment,
which will affect the description of the various
kinematic distributions used as input to the MVA,
it is essential to quantify the robustness of these
results
in a realistic environment including the effects of
significant PU.


It should be emphasized that MVAs such as the ANNs used in this work can always be understood as
a combined set of correlated cuts.
Once the ANNs have been trained, it is possible to compare  kinematical distributions after and before the ANN  cut to verify its impact.
This information would allow in principle to perform a cut-based analysis, without the need of using ANNs,
and finding similar results.

\begin{figure}[t]
\begin{center}
\includegraphics[width=0.48\textwidth]{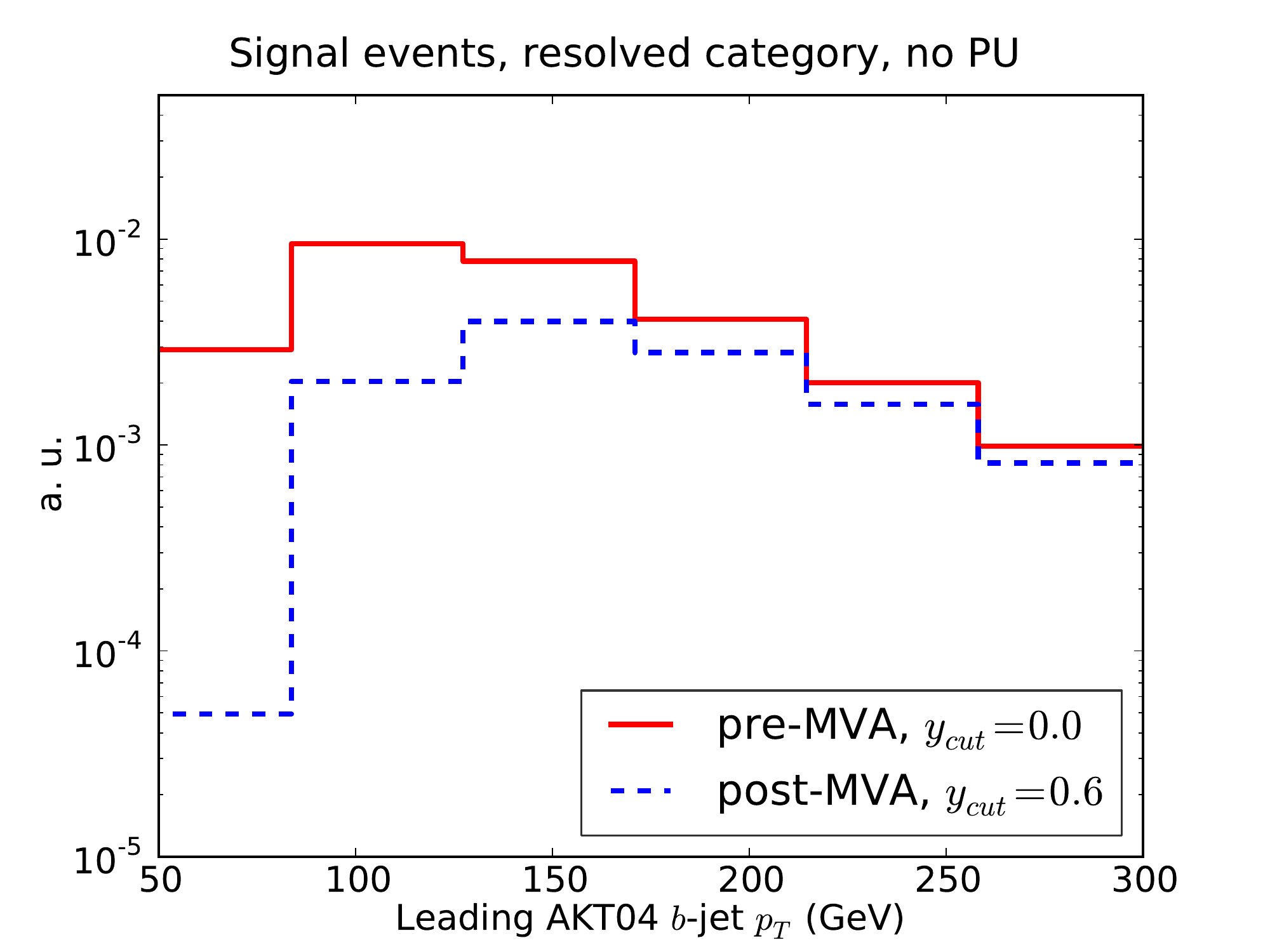}
\includegraphics[width=0.48\textwidth]{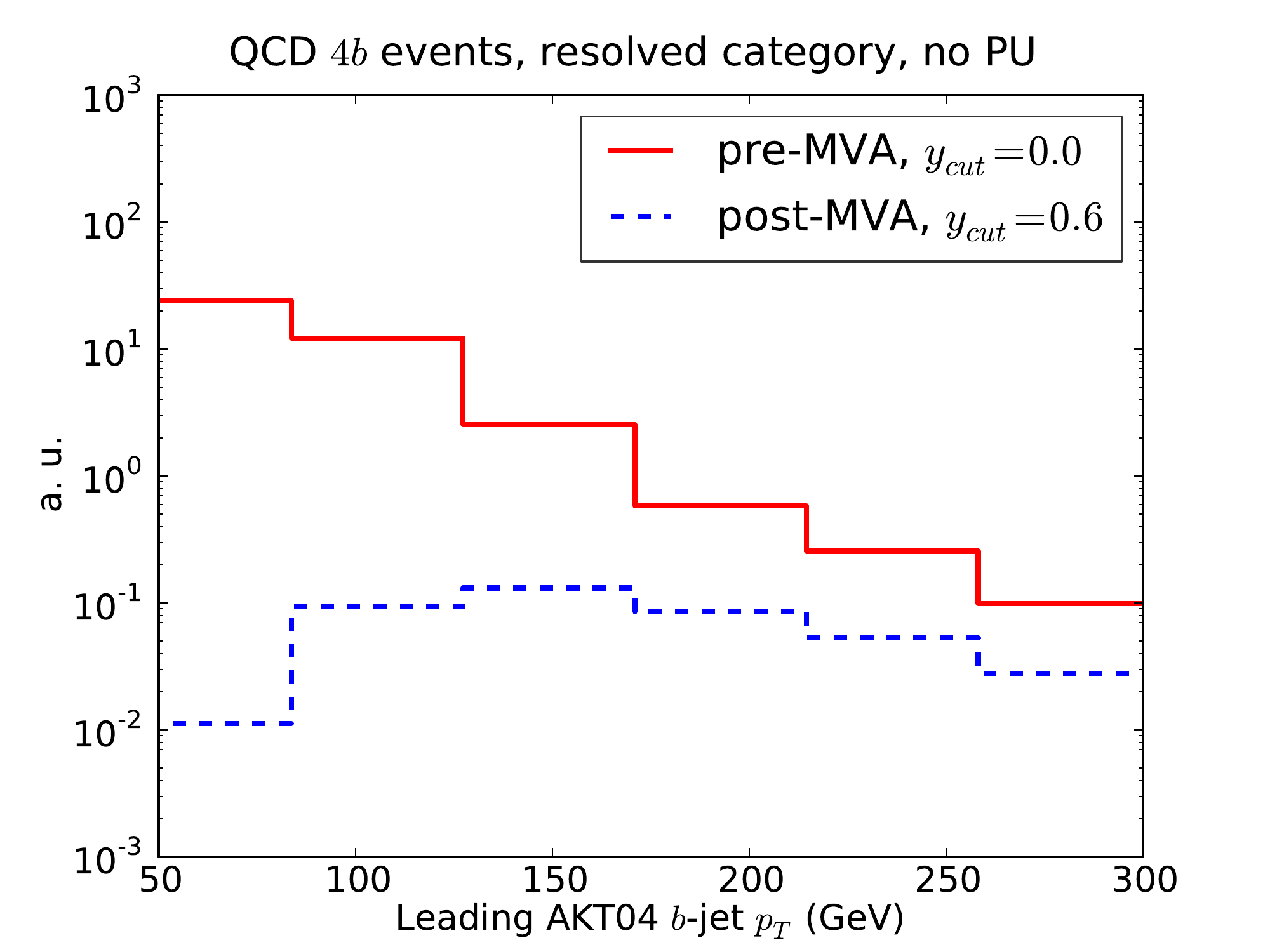}
\includegraphics[width=0.48\textwidth]{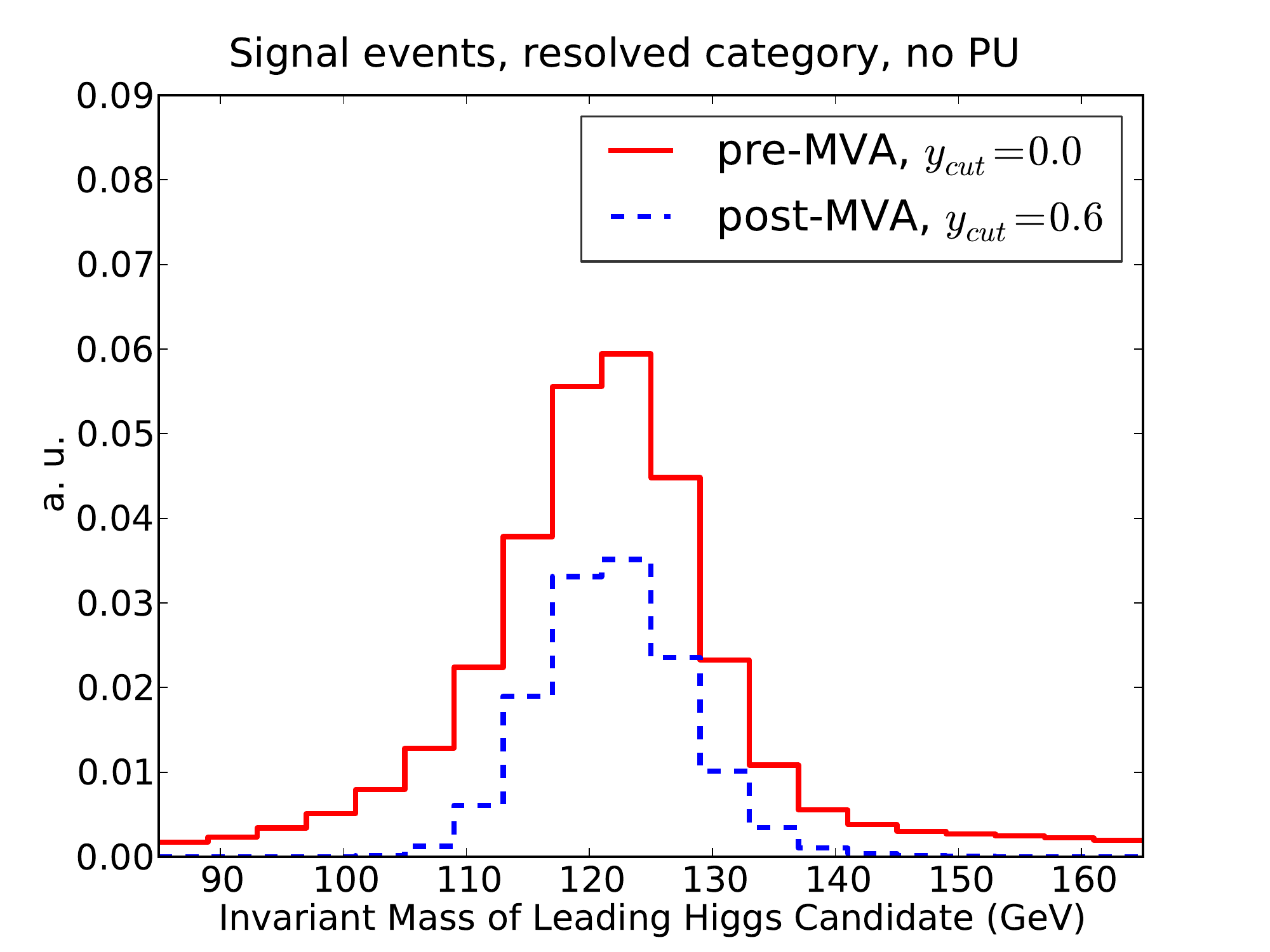}
\includegraphics[width=0.48\textwidth]{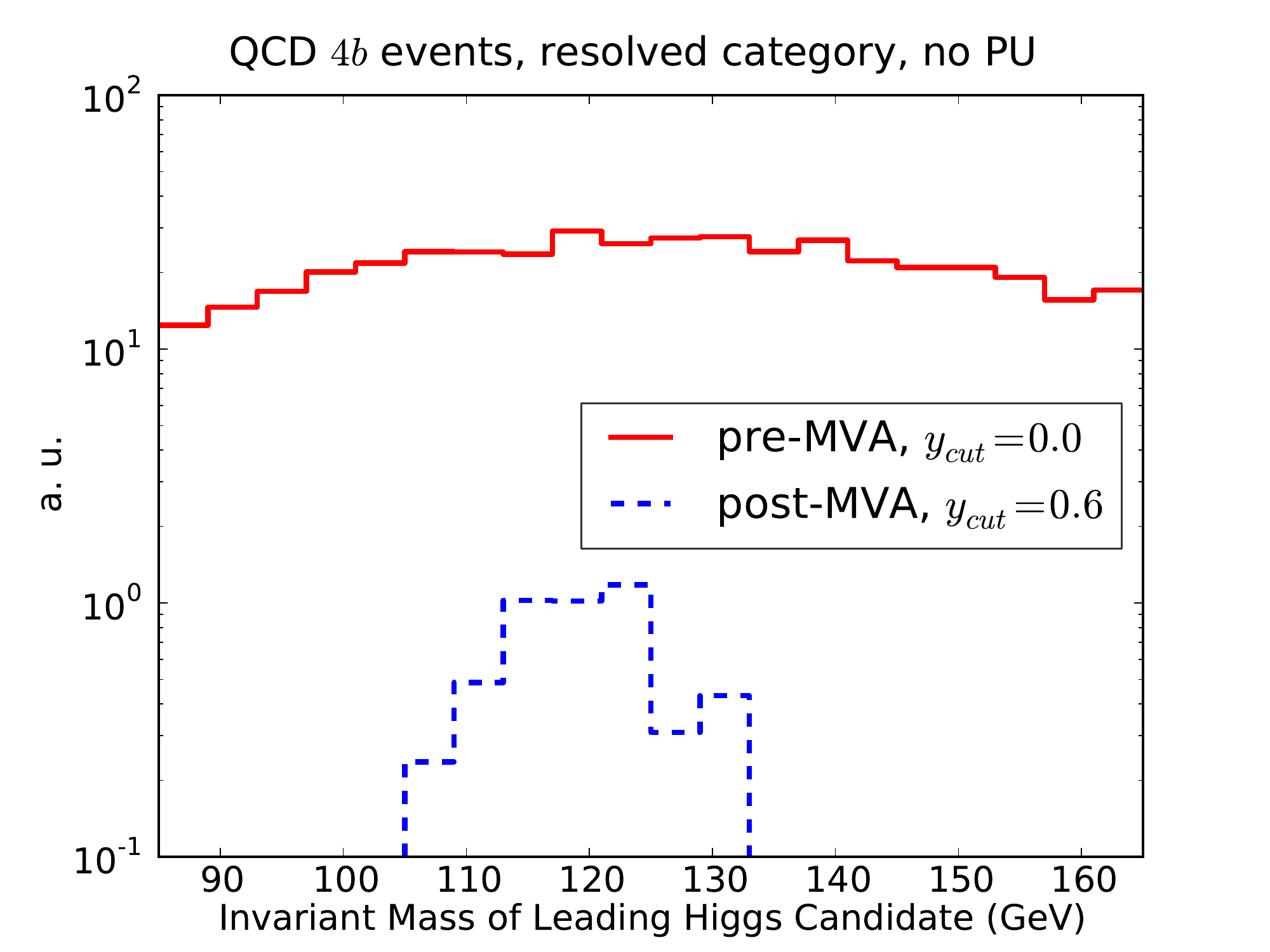}
\caption{\small
  The $p_T$ distribution of the leading AKT04 small-$R$ jets (upper plots)
  and
  the invariant mass of reconstructed Higgs candidates (lower plots) in the resolved
  category, comparing the pre-MVA results ($y_{\rm cut}=0$) with the post-MVA
  results ($y_{\rm cut}=0.60$) for signal (left) and background (right plot) events.
  In this case the distributions are not normalized, to better visualize the effects
  of the MVA cut.
}
\label{fig:pt_H0_sub0_res_noPU_ANNcut}
\end{center}
\end{figure}

To illustrate this point,
in Fig.~\ref{fig:pt_H0_sub0_res_noPU_ANNcut} we show
the $p_T$ distribution of the leading AKT04 small-$R$ jets
and the invariant mass of reconstructed Higgs candidates in the resolved
  category, comparing the pre-MVA results ($y_{\rm cut}=0$) with the post-MVA
  results ($y_{\rm cut}=0.60$) for signal and background  events.
  The distributions are not normalized, to better visualize the effect
  of the MVA cut.
  Unsurprisingly, the ANN cut effectively selects events which
  lead to similar kinematical distributions between signal
  and background events.
  In the case of the small-$R$ jets $p_T$ distribution, the
  ANN cuts favors the
  high-$p_T$ region, while for the invariant mass distribution
  only the region around the Higgs mass peak is selected for
  background events.

A particularly challenging aspect of our analysis is the modeling of the $2b2j$ and $4j$ background,
specially of the latter, which require extremely large MC samples.
In the analysis reported here, out of the original 3M $4j$ generated events, only around 100
survive the analysis cuts, and thus
these low statistics have associated a potentially large uncertainty
in the calculation of the post-MVA $4j$ cross-section.
On the other hand, since the $4j$ cross-sections
are always quite smaller than the sum of the $4b$ of the $2b2j$ components,
this low statistics should not modify qualitatively our conclusions above.
To verify explicitly this expectation, and obtain a more
robust estimate of the  background cross-section from mis-identified jets,
we have increased by a factor 10 the size of the $2b2j$ and $4j$
background samples, up to a total of 30M each.
Processing these events though our analysis, including retraining the MVA, we find
$(S/\sqrt{B})_{\rm tot}=3.9$, consistent with Eq.~(\ref{eq:soverb}), indicating that the low
statistics of the $4j$ background is not a limiting factor.

\subsection{Impact of PU in the MVA}

In this section we study how the MVA results are modified
when the analysis is performed including significant PU.
The loose cut-based analysis and the subsequent
MVA optimization have been performed using the same
settings as in the case without PU.
In Table~\ref{tab:cutflow_PU80_1}
we provide the pre-MVA cut flow in the case of PU80,
the corresponding version without PU being
Table~\ref{tab:cutflow_noPU_1}.
The interplay between the signal cross-sections and the various
background components is qualitatively unchanged as compared
to the no PU case.

\begin{table}[t]
  \centering
  \scriptsize
      \begin{tabular}{|l|cc|cccc|cc|cc|}
  \hline
\multicolumn{11}{|c|}{HL-LHC, Resolved category, PU+SK with $n_{\rm PU}=80$}\\
\hline
&  \multicolumn{6}{c|}{Cross-section [fb]} &  \multicolumn{2}{c|}{$S/B$}  &  \multicolumn{2}{c|}{$S/\sqrt{B}$}  \\
   &  $hh4b$ &  total bkg  &   $4b$    &  $2b2j$   &   $4j$    &
$t\bar{t}$ &
tot & $4b$ & tot & $4b$ \\
  \hline
  \hline
 C1a   &    $11$ &   $4.4 \cdot 10^8$   & $1.5 \cdot 10^5$ & $3.0 \cdot 10^7$ & $4.1 \cdot 10^8$ & $2.6 \cdot 10^5$     &  $ 2.4 \cdot 10^{-8}$    & $7.2 \cdot 10^{-5}$    &   0.03   & $1.5$\\
 C1b   &    $11$ &   $4.4 \cdot 10^8 $  & $1.5 \cdot 10^5$ & $3.0 \cdot 10^7$ & $4.1 \cdot 10^8$ & $2.6 \cdot 10^5$     &    $2.4 \cdot 10^{-8}$   & $7.2 \cdot 10^{-5}$    &   0.03   & $1.5$ \\
 C1c   &     $3$ &   $1.1 \cdot 10^8$   & $4.2 \cdot 10^4$ & $7.7 \cdot 10^6$ & $9.9 \cdot 10^7$ & $1.1 \cdot 10^5$     &   $2.8 \cdot 10^{-8}$    & $7.4 \cdot 10^{-5}$    &   0.02    & $0.8$ \\
 C2    &  $0.6$  &   $9.0 \cdot 10^3$   & $3.5 \cdot 10^3$ & $5.1 \cdot 10^3$ & $3.1 \cdot 10^2$ & $50$     &    $6.5 \cdot 10^{-5}$   & $1.7 \cdot 10^{-4}$    &   0.4                & $0.5$ \\
\hline
\end{tabular}
  $\,$ \\
  \vspace{0.5cm}
 \begin{tabular}{|l|cc|cccc|cc|cc|}
  \hline
  \multicolumn{11}{|c|}{HL-LHC, Intermediate category, PU+SK+Trim with $n_{\rm PU}=80$}\\
  \hline
&  \multicolumn{6}{c|}{Cross-section [fb]} &  \multicolumn{2}{c|}{$S/B$}  &  \multicolumn{2}{c|}{$S/\sqrt{B}$}  \\
   &  $hh4b$ &  total bkg  &   $4b$    &  $2b2j$   &   $4j$    &
$t\bar{t}$ &
tot & $4b$ & tot & $4b$ \\
  \hline
  \hline
 C1b     &  $2.7$  &   $8.1 \cdot 10^7$   & $2.1 \cdot 10^4$ & $5.2 \cdot 10^6$ & $7.6 \cdot 10^7$ & $3.0 \cdot 10^4$          &     $3.4 \cdot 10^{-8}$     & $1.3 \cdot 10^{-4}$        &   0.02    & $1.0 $ \\
 C1c     & $2.6 $  &   $6.2 \cdot 10^7 $  & $1.5 \cdot 10^4$ & $3.9 \cdot 10^6$ & $5.8 \cdot 10^7$ & $2.8 \cdot 10^4$          &     $4.1 \cdot 10^{-8}$     & $1.7 \cdot 10^{-4}$        &   0.02    & $1.1 $ \\
 C1d     & $0.5$   &   $2.8 \cdot 10^6$   & $7.9 \cdot 10^2$ & $1.9 \cdot 10^5$ & $2.7 \cdot 10^6$ & $6.5 \cdot 10^3$          &     $1.8 \cdot 10^{-7}$     & $6.2 \cdot 10^{-4}$        &   0.02    & $1.0 $ \\
 C2      & $0.09$  &   $2.6 \cdot 10^2$   & $47$             & $1.8 \cdot 10^2$ & $30$             & $2.2 $                   &      $3.4 \cdot 10^{-4}$     & $1.8 \cdot 10^{-3}$        &   0.3                & $0.7 $ \\
\hline
\end{tabular}
  $\,$ \\
 \vspace{0.5cm}
 \begin{tabular}{|l|cc|cccc|cc|cc|}
  \hline
  \multicolumn{11}{|c|}{HL-LHC, Boosted category, PU+SK+Trim with $n_{\rm PU}=80$}\\
   \hline
&  \multicolumn{6}{c|}{Cross-section [fb]} &  \multicolumn{2}{c|}{$S/B$}  &  \multicolumn{2}{c|}{$S/\sqrt{B}$}  \\
   &  $hh4b$ &  total bkg  &   $4b$    &  $2b2j$   &   $4j$    &
$t\bar{t}$ &
tot & $4b$ & tot & $4b$ \\
  \hline
  \hline
 C1a     &  $3.5 $  &   $4.1 \cdot 10^7$   & $1.0 \cdot 10^4$   & $2.7 \cdot 10^6$ & $3.8 \cdot 10^7$         & $2.0 \cdot 10^4 $    &   $8.6 \cdot 10^{-8}$    & $3.4 \cdot 10^{-4}$  &   0.03    & $1.9 $ \\
 C1b     & $2.5 $   &   $3.2 \cdot 10^7$   & $6.8 \cdot 10^3$   & $1.9 \cdot 10^6$ & $3.0 \cdot 10^7$         & $1.9 \cdot 10^4  $   &   $7.8 \cdot 10^{-8}$    & $3.6 \cdot 10^{-4}$  &   0.02  & $1.6 $ \\
 C1c     & $0.8 $   &   $2.2 \cdot 10^6$   & $5.4 \cdot 10^2$   & $1.4 \cdot 10^5$ & $2.0 \cdot 10^6$         & $4.8 \cdot 10^3   $  &   $3.8 \cdot 10^{-7}$    & $1.6 \cdot 10^{-3}$  &   0.03      & $2.0 $ \\
 C2      & $0.14$   &   $1.5 \cdot 10^2$   & $40$               & $86$             & $22$                     & $1.8 $               &   $ 9.0 \cdot 10^{-4}$   & $3.5 \cdot 10^{-3}$  &  0.6                & $1.2 $ \\
\hline
\end{tabular}

    \caption{\small Same as Table~\ref{tab:cutflow_noPU_1},
now for the case
    of PU80+SK+Trim.
 \label{tab:cutflow_PU80_1}}
\end{table}

In Table~\ref{table:cutflowMVA_PU} we compare the results
for the PU80+SK+Trim case between
  the pre-MVA loose cut-based analysis and
  the post-MVA results for the
  optimal values of the ANN output cut $y_{\rm cut}$.
  As in Table~\ref{table:cutflowMVA}, 
  we also quote 
   the number of signal and
    total background events expected
   for $\mathcal{L}=3$ ab$^{-1}$
    and the values of $S/\sqrt{B}$ and $S/B$.
We observe that the pre-MVA 
signal significance is close
to the results of the simulations
without PU for the three categories.
We now find values for $S/\sqrt{B}$ of 0.4, 0.3 and 0.6, in the resolved,
intermediate and boosted categories, respectively, to be compared
with the corresponding values without PU, namely 0.4, 0.4 and 0.5.
The number of selected
signal events in each category at the
end of the cut-based analysis is only mildly affected
by PU.
The slight pre-MVA improvement in $S/\sqrt{B}$ for the
boosted case arises from a reduction in the number
of background events that are classified in this category
as compared to the case without PU.

Once the MVA is applied, the signal significance in the 
resolved, intermediate and boosted
categories increases to 2.0, 1.9 and 1.5 respectively,
to be compared with the corresponding values
without PU, namely 1.9, 2.3 and 2.7.
Therefore, the post-MVA effect of PU on $S/\sqrt{B}$ is
a moderate degradation of the boosted and intermediate categories,
specially for the former,
while the resolved category is largely unchanged.\footnote{
  The impact of PU on the  separate significance of
  the three categories exhibits some
  dependence on the specific choice for $n_{\rm PU}$  and on the settings
  of the PU subtraction strategy.
  We find however that the
  overall signal significance from combining the three
  categories is similar in the $n_{\rm PU}=80$ and
  $n_{\rm PU}=150$ cases.
}
We also observe that, due
to the MVA, the
signal over background ratio is increased from 0.007\%, 0.03\% and
0.1\% up to 1\%, 3\% and 1\% in the resolved, intermediate
and boosted categories respectively.
This indicates that while this measurement is still highly challenging,
requiring a careful extraction of the QCD
background from the data, it should be within reach.

\begin{table}[t]
  \centering
  \begin{tabular}{|c|l|c|c|c|c|}
        \hline
     \multicolumn{6}{|c|}{HL-LHC, PU80+SK+Trim} \\
     \hline
         \hline
    Category  &   &  $N_{\rm ev}$ signal &  $N_{\rm ev}$ back  &  $S/\sqrt{B}$ & $S/B$ \\ 
    \hline
    \hline
    \multirow{2}{*}{Boosted} &  $y_{\rm cut}=0$  & 410   &  $4.5\cdot 10^5$ & 0.6   & $ 10^{-3}$  \\
    &  $y_{\rm cut}=0.8$ &  290  & $3.7\cdot 10^4$  & 1.5    & 0.01  \\
    \hline
    \hline
    \multirow{2}{*}{Intermediate} &  $y_{\rm cut}=0$  &  260  & $7.7\cdot 10^5$    & 0.3    &
     $3\cdot 10^{-4}$ \\
    &  $y_{\rm cut}=0.75$ & 140 & $5.6\cdot 10^3$  &  1.9   & 0.03 \\
    \hline
    \hline
    \multirow{2}{*}{Resolved} &  $y_{\rm cut}=0$  &  1800  & $2.7\cdot 10^7$
    & 0.4    &  $7\cdot 10^{-5}$  \\
    &  $y_{\rm cut}=0.60$ & 640  & $1.0\cdot 10^5$  &  2.0   & 0.01  \\
    \hline
      \end{tabular}
  \caption{\small Same as Table~\ref{table:cutflowMVA}, now for the case
    of PU80+SK+Trim.
        \label{table:cutflowMVA_PU}
  }
\end{table}

\begin{figure}[t]
  \begin{center}
    \includegraphics[width=0.49\textwidth]{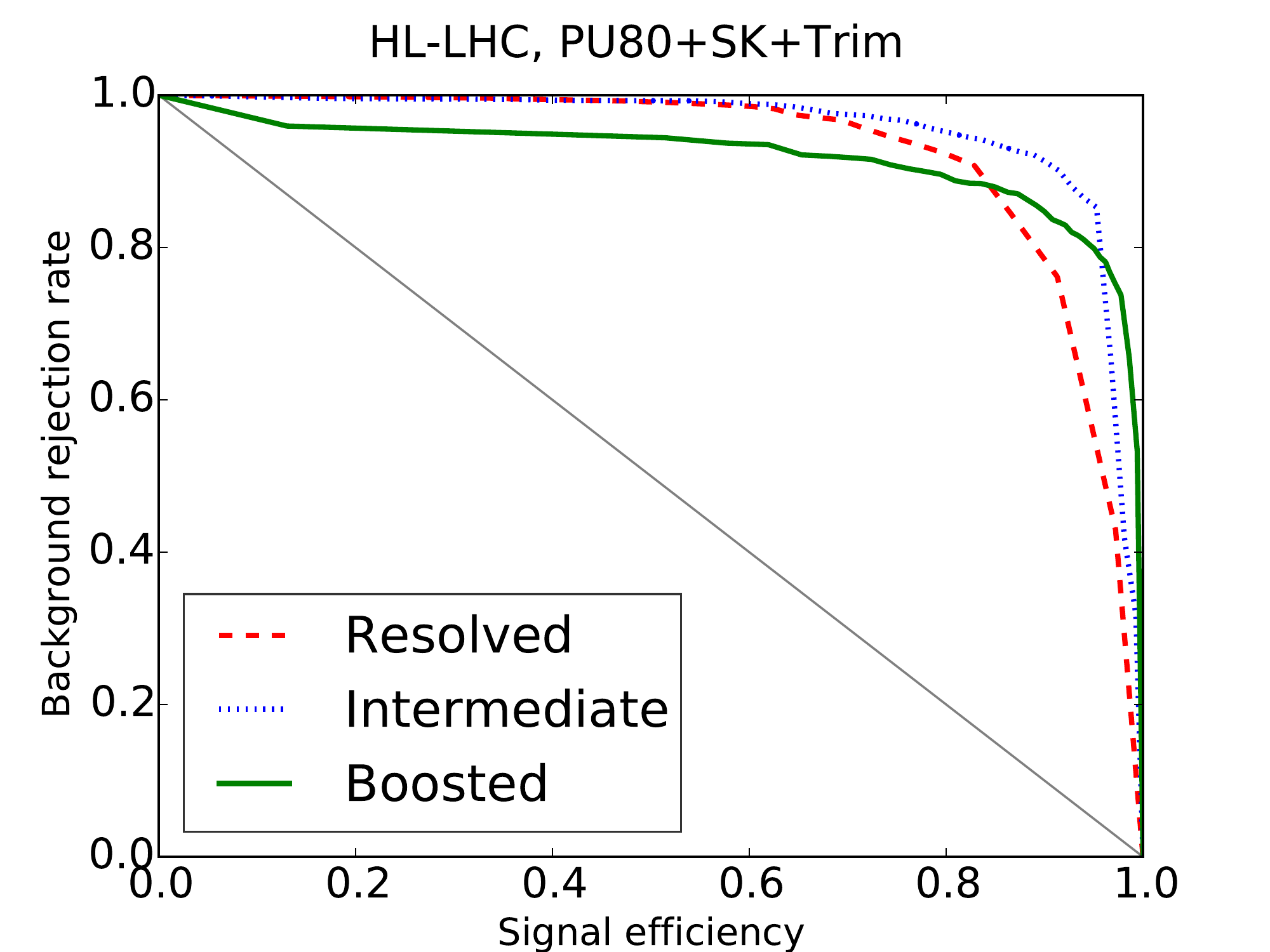}
\includegraphics[width=0.49\textwidth]{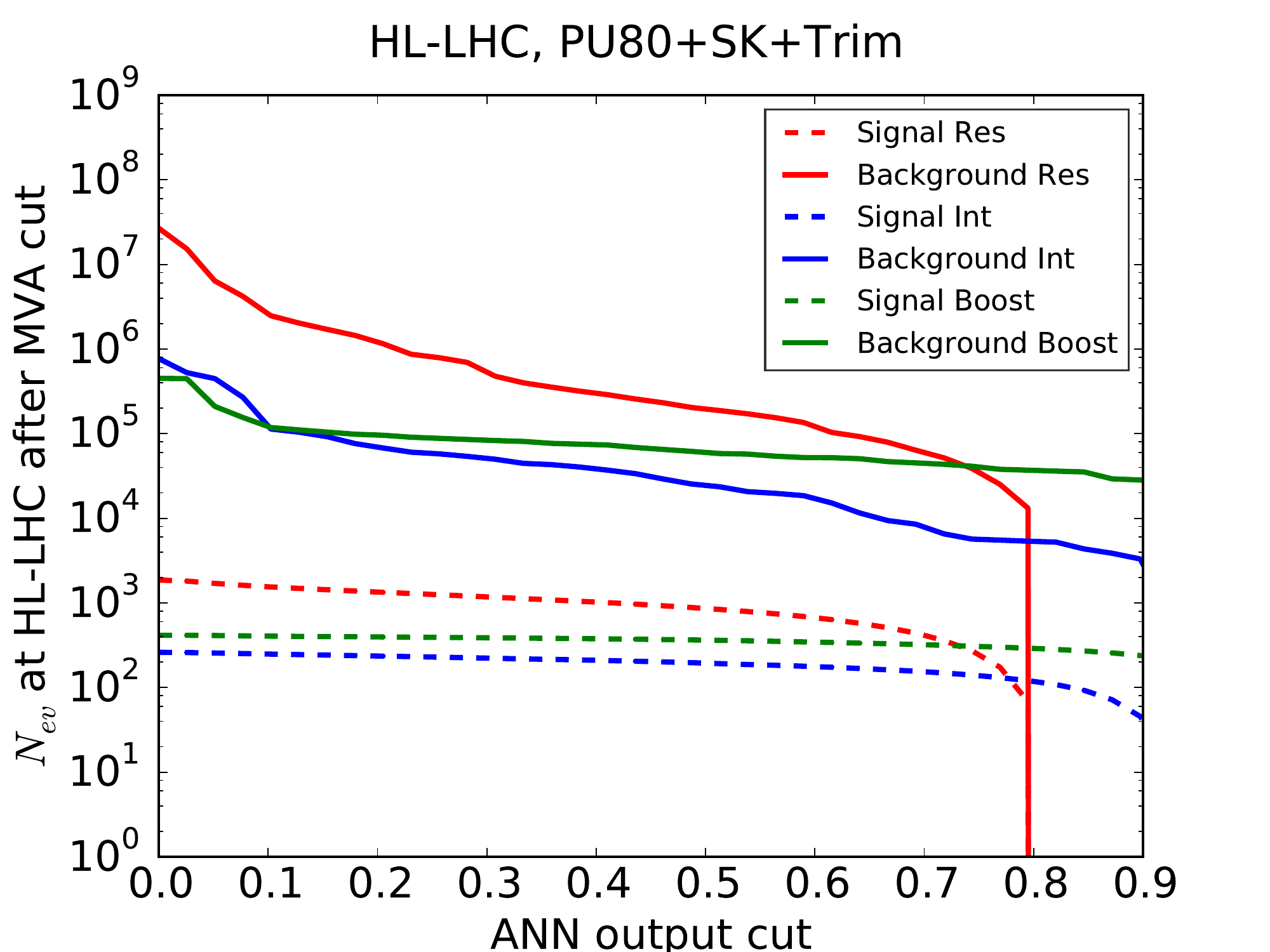}
\caption{\small Same as Fig.~\ref{fig:nev2}
 for
 the PU80+SK+Trim case.
\label{fig:nev2_PU}}
\end{center}
\end{figure}

\begin{figure}[t]
\begin{center}
\includegraphics[width=0.48\textwidth]{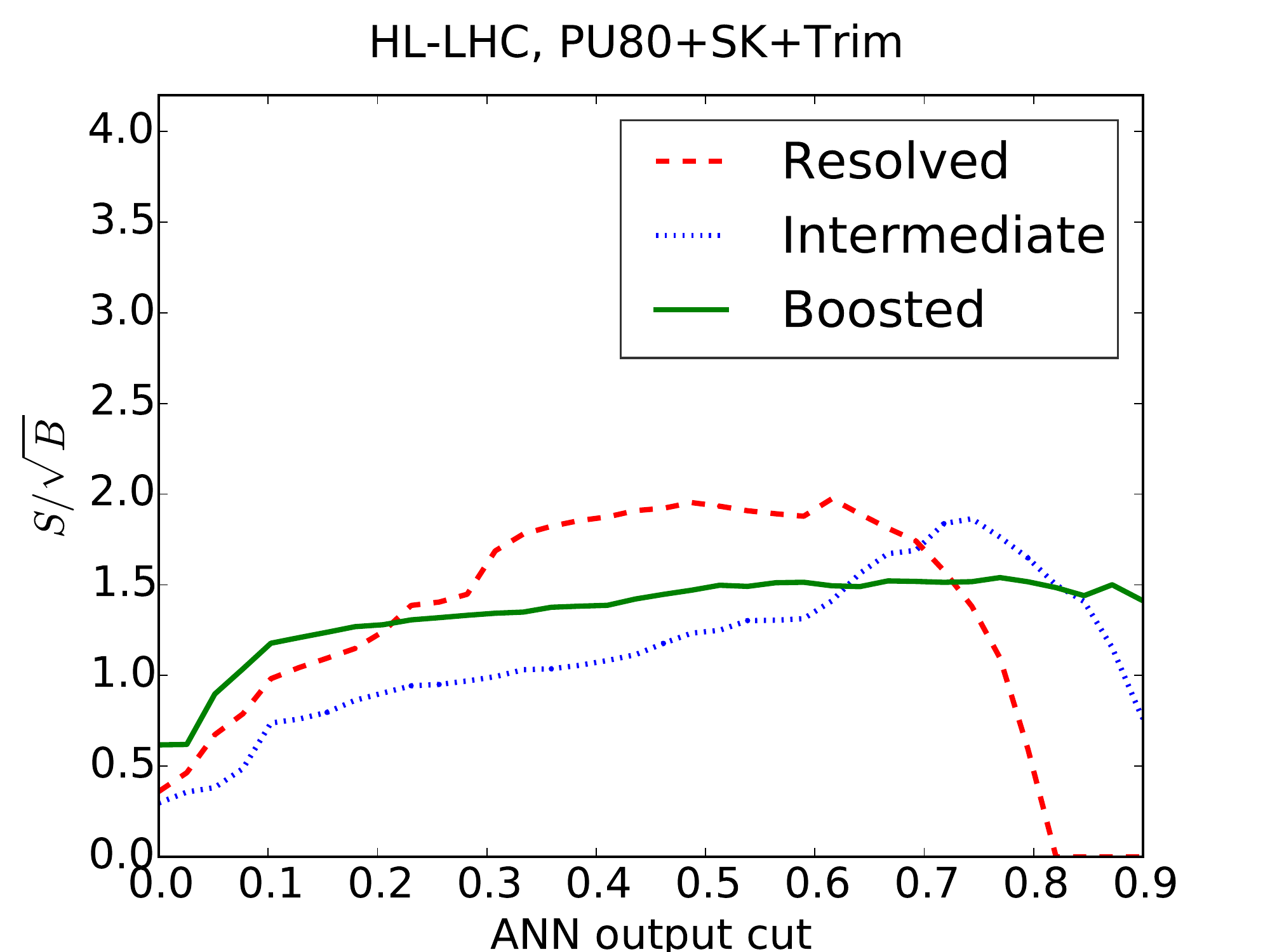}
\includegraphics[width=0.48\textwidth]{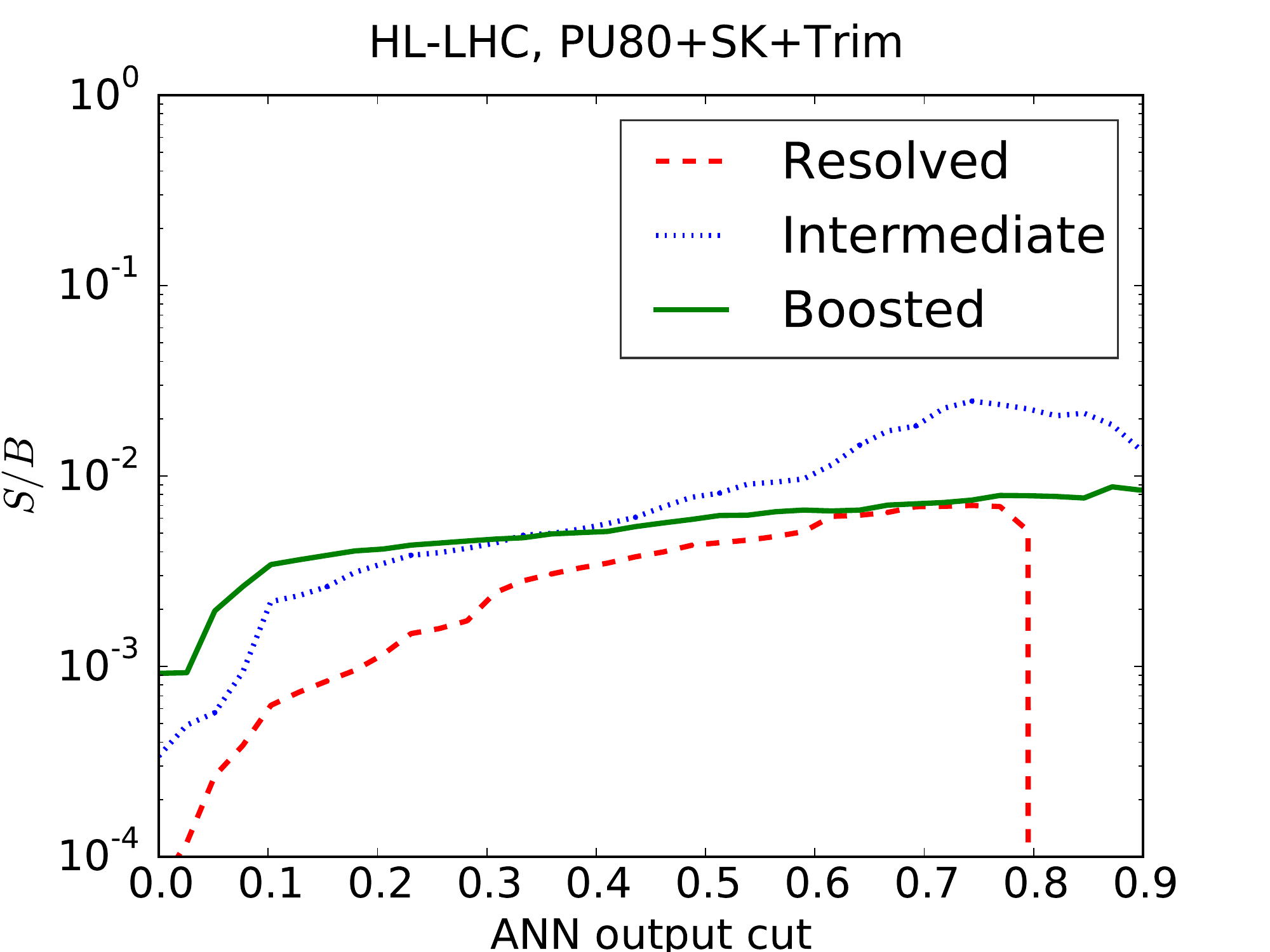}
\caption{\small 
  Same as Fig.~\ref{fig:sb_mva} for
  the PU80+SK+Trim case.
}
\label{fig:sb_mva_PU}
\end{center}
\end{figure}

In Fig.~\ref{fig:nev2_PU}
we show the number of signal and background events that
are expected for $\mathcal{L}=3$ ab$^{-1}$
as a function of
$y_{\rm cut}$, together with the corresponding ROC curve.
The slight degradation of the boosted category in the case
of PU can be seen by comparing with the corresponding
results without PU in Fig.~\ref{fig:nev2}.
In Fig.~\ref{fig:sb_mva_PU} we show the signal significance,
$S/\sqrt{B}$, and the signal over background ratio,
$S/B$, accounting now for the effects of PU.
The corresponding results in the case without PU were shown in
Fig.~\ref{fig:sb_mva}.
As can be seen, the MVA-driven enhancement remains robust in the
presence of PU, with $S/\sqrt{B}$ only moderately degraded.
Therefore, the qualitative conclusions drawn
in the case without PU also hold when the analysis
is performed in a high-PU environment.
Since no specific effort has been performed to
optimize PU subtraction, for instance by tuning the values
of the patch length $a$ in {\tt SoftKiller}
or the $p_T$ threshold during jet trimming,
we believe that
there should be still room for further improvement.

It is useful to quantify which of the MVA input variables
carry the highest discrimination power
in the case of PU, by means of
Eq.~(\ref{eq:totweight}),
and compare this with the corresponding
results without PU shown in Fig.~\ref{fig:nnweights}.
We have verified that 
the relative weight of the different input variables to the MVA
is mostly unchanged in the case of PU.
In the resolved category, the highest total associated weight is carried
by the Higgs candidates $p_T$ and invariant mass, as well
as by the $p_T$ of the individual small-$R$ jets.
For the boosted category, the highest weight is carried by the
Higgs invariant mass, followed by
the Higgs $p_T$, $m_{hh}$, the $p_T$ of the AKT03 subjets and the
substructure variables, with a similar weighting among them.

In Table~\ref{table:cutflowMVA_fakes} we provide
the post-MVA number of signal and background events  expected for
$\mathcal{L}=3$ ab$^{-1}$.
For the backgrounds, we quote
both
the total number, $N_{\rm ev}^{\rm tot}$,
and the  QCD $4b$ component only,
$N_{\rm ev}^{\rm 4b}$.
We quote results for the no PU and PU80+SK+Trim cases.
 We also quote in each case the corresponding values for the signal 
    significance and the signal over background ratio.
    Note that the MVA is always trained to the inclusive background sample,
    though differences
    in the kinematic distributions of the $4b$ and $2b2j$ processes are
    moderate,
    see Fig.~\ref{fig:histoBack}.
    From Table~\ref{table:cutflowMVA_fakes} one observes that
    all categories exhibit
    a marked improvement from eliminating the contamination
    from light and charm jet mis-identification.
    For instance, in the intermediate category,
    $S/\sqrt{B}$ increases from 2.3 to 3.3 (1.9 to 2.9)
    in the no PU (PU80) case, with similar improvements in the
    resolved and boosted categories.
    %

\begin{table}[t]
  \centering
  \scriptsize
  \begin{tabular}{|c|c|c|c|c||c|c||c|c|}
        \hline
        Category  &  &  signal  &  \multicolumn{2}{c||}{background}     &
        $S/\sqrt{B_{\rm tot}}$ & $S/\sqrt{B_{\rm 4b}}$  
        &  $S/B_{\rm tot}$ & $S/B_{\rm 4b}$\\
         &  &  $N_{\rm ev}$  &  $N_{\rm ev}^{\rm tot}$  &  $N_{\rm ev}^{\rm 4b}$   &
         &     &   &  \\
    \hline
    \hline
    \multirow{2}{*}{Boosted} &  no PU  & 290  & $1.2\cdot 10^4$  &  $8.0\cdot 10^3$    & 
      2.7 &  3.2  & 0.03 & 0.04 \\
      & PU80+SK+Trim & 290  &$3.7\cdot 10^4$ & $1.2\cdot 10^4$     &  1.5  & 2.7 &  0.01 & 0.02   \\
    \hline
    \hline
    \multirow{2}{*}{Intermediate} &  no PU   & 130   & $3.1\cdot 10^3$  & $1.5\cdot 10^3$    &
    2.3  & 3.3  &  0.04  & 0.08   \\
    & PU80+SK+Trim & 140 & $5.6\cdot 10^3$   & $2.4\cdot 10^3$   & 1.9  & 2.9  & 0.03 & 0.06  \\
    \hline
    \hline
    \multirow{2}{*}{Resolved} &   no PU  & 630 & $1.1\cdot 10^5$   & $5.8\cdot 10^4$
    & 1.9  & 2.7  & 0.01 & 0.01 \\
    & PU80+SK & 640  & $1.0\cdot 10^5$   & $7.0\cdot 10^4$   & 2.0 & 2.6  & 0.01 & 0.01 \\
    \hline
    \hline
    \multirow{2}{*}{\bf Combined} &   no PU  &  \multicolumn{3}{c||}{} 
    &  4.0 & 5.3  &   \multicolumn{2}{c|}{}  \\
    & PU80+SK+Trim &    \multicolumn{3}{c||}{}    & 3.1  & 4.7  &  \multicolumn{2}{c|}{}  \\
    \hline
      \end{tabular}
  \caption{\small Post-MVA number of signal and background events
    with  $\mathcal{L}=3$ ab$^{-1}$.
    For the backgrounds, both the total number, $N_{\rm ev}^{\rm tot}$,
    and the  $4b$ 
    component only, $N_{\rm ev}^{\rm 4b}$, are shown.
    Also provided are the values of the signal 
    significance and the signal over background ratio,
    both separated in categories and for their combination.
    We quote the results without PU and for PU80+SK+Trim.
         \label{table:cutflowMVA_fakes}
  }
\end{table}

In Table~\ref{table:cutflowMVA_fakes} we also provide
the results for $S/\sqrt{B}$ obtained by
combining the three categories.
Taking into
account all background components, we obtain for the case
of $n_{\rm PU}=80$
an overall signal
significance of
\be
\lp \frac{S}{\sqrt{B}}\rp_{\rm tot} \simeq 3.1~(1.0) \, ,\quad
\mathcal{L}=3000~(300)\,{\rm fb}^{-1}\, ,
\ee
indicating  that a measurement of
Higgs pair production in the $b\bar{b}b\bar{b}$ final state at the HL-LHC
should be 
above the threshold for observation, even when realistic PU conditions
are accounted for.
A similar signal significance is obtained in the case of
$n_{\rm PU}=150$.
Under the  assumption that
    the only relevant background would be the irreducible QCD $4b$ component,
    one obtains instead
    \be
\lp \frac{S}{\sqrt{B_{\rm 4b}}}\rp_{\rm tot} \simeq 4.7~(1.5) \, ,\quad
\mathcal{L}=3000~(300)\,{\rm fb}^{-1}\, .
\ee
Therefore, a measurement of Higgs pair production
in the $b\bar{b}b\bar{b}$ final state at the
HL-LHC  might be even above the threshold for discovery, provided
the effects due to mis-identification of light and charm jets as
$b$-jets can be reduced.

\section{Conclusions and outlook}
\label{sec:conclusions}

In this work we have presented a feasibility study for
 the measurement of Higgs pair production in the $b\bar{b}b\bar{b}$
final state at the LHC.
Our strategy is based on the combination of traditional
cut-based analysis with state-of-the-art multivariate techniques.
We take into account 
all relevant backgrounds, in particular
the irreducible $4b$
and the reducible 
$2b2j$ and $4j$ QCD multijets.
We have illustrated how the $2b2j$ component leads to
a contribution comparable to that of QCD $4b$ production,
due to a combination of  parton shower effects, $b$-quark 
pair radiation, and selection requirements.
We have also demonstrated the robustness of our analysis strategy
under the addition of significant PU.
In particular, we have explored two scenarios, $n_{\rm PU}=80$ and
$n_{\rm PU}=150$, and found a comparable overall signal significance
in the two cases.

Combining the contributions from the resolved,
intermediate and boosted categories, we find that, for
$\mathcal{L}=3$ ab$^{-1}$, the
signal significance for
the production of Higgs pairs turns out to be $S/\sqrt{B}\simeq 3$.
This indicates that, already from the $b\bar{b}b\bar{b}$
final state alone,
it should be possible to claim observation of Higgs pair production at
the HL-LHC.
Our study also suggests possible avenues that the LHC experiments
could explore to further improve this signal significance.
One handle would be to reduce the contribution from light and charm
jet mis-identification, ensuring that the irreducible $4b$ background 
dominates over the $2b2j$ component.
This would allow to enhance  $S/\sqrt{B}$ almost to the discovery
level, see Table~\ref{table:cutflowMVA_fakes}.
It would also be advantageous to improve the $b$-tagging efficiency, allowing
to achieve higher signal yields.
Another possibility would be to improve the mass resolution of the Higgs
reconstruction
in high-PU environments, and, more general,
to optimize the PU subtraction
strategy in order
to reduce the impact of PU in the modelling
of kinematic variables and the associated
degradation in the MVA discrimination.

Another challenging aspect of the measurement of Higgs pairs
in the $b\bar{b}b\bar{b}$ final state is achieving an efficient
triggering strategy.
In order to reduce the rate from background QCD processes sufficiently, while
being able
to access the relevant $p_T$ regimes, (multi-)jet triggers
using $b$-quark tagging information online for one or more jets are
likely to be
necessary.
The additional rejection provided by these triggers could
enable events to be selected efficiently, with four
jets down to $p_T=40$ GeV in the resolved category,
and boosted Higgs decays in large-$R$ jets down to jet transverse momenta of
$p_T=200$ GeV.
In addition,
good control of the multijet backgrounds and the
experimental systematics of the MVA inputs will be important to achieve
these sensitivities.

Our strategy relies on the modeling of the kinematic
distributions of signal and background events, since these provide
the inputs to the MVA discriminant.
In this respect, it would be important, having established the key
relevance of the $b\bar{b}b\bar{b}$ channel for the study of
Higgs pair production, to revisit and improve the
theoretical modeling of our signal and background simulation,
in particular using NLO calculations matched to
parton showers both for signal~\cite{Frederix:2014hta,Maierhofer:2013sha}
and for backgrounds~\cite{Alwall:2014hca,Gleisberg:2008ta}.

One important implication of this work is that it should
be possible to significantly
improve  the accuracy on the extraction of
the Higgs trilinear coupling $\lambda$ from
a measurement of the
$\sigma\lp hh\to b\bar{b}b\bar{b}\rp$ cross-section, as compared
to existing estimates.
A determination of $\lambda$ in our approach is however
rather
non-trivial, involving
 not only regenerating signal samples
 for a wide range of values of  $\lambda$, but also
 repeating the analysis
optimisation, including the MVA training, for each
of these values.
This study is left to a future
publication, where we will also
compare the precision from the $b\bar{b}b\bar{b}$ final state
with the corresponding  precision 
that has been reported from other final states such as
 $b\bar{b}\gamma\gamma$
and $b\bar{b}\tau\tau$.
It will also be  interesting to perform
this exercise for a 100 TeV hadron collider~\cite{Barr:2014sga,
  Azatov:2015oxa,Papaefstathiou:2015iba,
  Arkani-Hamed:2015vfh}.
While at 100 TeV the
signal yields would be increased, also the (gluon-driven) QCD
multijet background would grow strongly.
Revisiting
the present analysis, including the MVA optimization,
at 100 TeV would also allow us
to assess the accuracy of an extraction of the trilinear
coupling $\lambda$ from the $b\bar{b}b\bar{b}$ final state
at 100 TeV.

In this work we have considered only the SM production mechanism,
but many BSM scenarios predict deviations
in Higgs pair production, both at the level of total rates
and of
differential distributions.
In the absence of new explicit degrees of freedom,
deviations from the SM can be parametrized in
the EFT framework using higher-order
operators~\cite{Azatov:2015oxa,Goertz:2014qta}.
Therefore, we plan to study the constraints
on the coefficients of these effective
operators that can be obtained from measurements
of various kinematic distributions
in the $hh\to b\bar{b}b\bar{b}$ process.
Note that the higher rates of the $b\bar{b}b\bar{b}$ final state as compared to
other final states, such as
$b\bar{b}\gamma\gamma$, allow for better constraints upon operators
that modify the high-energy behavior
of the theory, for instance,
it would become possible
to access the tail of the $m_{hh}$ distribution.

As in the case of the extraction of the Higgs
trilinear coupling $\lambda$, such a study
would be a computationally intensive task, since
BSM dynamics will modify the shapes of the kinematic
distributions and thus in principle each point in the EFT parameter
space would require a re-optimization with a newly trained
MVA.
In order to explore efficiently the BSM parameters
without having to repeat the full analysis
for each point, modern statistical techniques
such as the Cluster Analysis method proposed
in Ref.~\cite{Dall'Osso:2015aia} might be helpful.

\bigskip
\bigskip
\begin{center}
\rule{5cm}{.1pt}
\end{center}
\bigskip
\bigskip

{\bf\noindent  Acknowledgments \\}
We thank F.~Bishara, R.~Contino, A.~Papaefstathiou and
G.~Salam for useful discussions on the topic
of Higgs pair production.
We thank E.~Vryonidou and M.~Zaro for
assistance with di-Higgs production
  in {\tt MadGraph5\_aMC@NLO}.

  \noindent
  The work of K.~B. is supported by a Rhodes Scholarship.
  D.~B., J.~F. and C.~I. are supported by the STFC.
  J.~R. and N.~H. are
supported by an European Research Council Starting Grant ``PDF4BSM".
  J.~R. is supported by an STFC Rutherford Fellowship and
  Grant ST/K005227/1 and ST/M003787/1.

\appendix

\section{Single Higgs backgrounds}
\label{app:singlehiggs}

As discussed in Sect.~\ref{mcgeneration},
in our analysis we neglect single Higgs production processes,
since they are much smaller than both the signal and the main
QCD multijet backgrounds.
To explicitly demonstrate this, we have generated LO samples
using  {\tt MadGraph5\_aMC@NLO}
for the
following single-Higgs processes:
  \begin{enumerate}
  \item $Z(\to b\bar{b})h(\to b\bar{b})$ (electroweak)
  \item $t\bar{t}h(\to b\bar{b})$
    \item $b\bar{b}h(\to b\bar{b})$ (QCD)
  \end{enumerate}
  For each processes, we have  generated 1M events, and in
 Table~\ref{HK}  we list resulting the
  LO and NLO cross-sections at the generation level.
  The subsequent decays and the
  corresponding branching fractions are not included in these cross-sections,
  since
  these are taken care by the {\tt Pythia8} parton shower.
  The values of these branching fractions 
  are listed in Table~\ref{HBF}, corresponding
  to the  most recent averages from the PDG.
  In the case of the $t\bar{t}h$ process, we
  consider only the fully hadronic decays
  of the top quark, since leptonic and semi-leptonic decays
  can be suppressed
  by means of a lepton veto.
  
  \begin{table}[h]
\begin{center}
\begin{tabular}{|c|c|c|c|}
\hline
Sample & LO & NLO & $K$-factor\\
\hline\hline
$Zh$ (13 TeV) & $6.5 \cdot 10^{-1}$ pb & $ 7.7 \cdot 10^{-1}$ pb & 1.19 \\
$t\bar{t}h$ (13 TeV) & $3.8 \cdot 10^{-1}$ pb & $4.6 \cdot 10^{-1}$ pb & 1.29 \\
$b\bar{b}h$ (13 TeV) &  $4.9 \cdot 10^{-1}$ pb & $6.1 \cdot 10^{-1}$ pb & 1.22 \\
\hline
\end{tabular}
\caption{\small LO and NLO cross-sections at the generation level for the single-Higgs background
  processes listed above, computed using {\tt MadGraph5\_aMC@NLO}.
  The subsequent decays and the corresponding branching fractions are not included in these generation-level cross-sections. \label{HK}
}
\end{center}
  \end{table}%

  \begin{table}[h]
\begin{center}
\begin{tabular}{|c|c|c|c|}
\hline
Sample & Decay & Branching Fraction\\
\hline\hline
$Zh$ & ($Z\to b\bar{b}$)($h\to b\bar{b}$) & 0.086 \\
$t\bar{t}h$ & $(W\to q\bar{q})^2$($h\to b\bar{b}$) & 0.26 \\
$b\bar{h}h$ & $h\to b\bar{b}$ & 0.57 \\
\hline
\end{tabular}
\caption{\small The values of the branching fractions applied to the single-Higgs
  background processes from Table~\ref{HK}, corresponding to
  the most updated PDG values. \label{HBF}}
\end{center}
  \end{table}%

  In Table~\ref{HBxsec}
  we show the signal and background cross-sections at the end of the cut-based analysis, before the MVA is applied,
  in the case without PU.
  We separate the results into the three exclusive categories used in our analysis.
  From this comparison, we see that as expected, at the end of the cut-based analysis, the single-Higgs
  backgrounds are smaller than the QCD multijet background by several orders of magnitude.
  In addition, we find that already at the end of the cut-based analysis the di-Higgs
  signal is also larger than all the single-Higgs backgrounds in all the selection categories.
  Since this discrimination can only be improved by the MVA, we
  conclude that neglecting single-Higgs backgrounds is a reasonable
  approximation.
  From Table~\ref{HBxsec} we also observe that in the resolved
  and intermediate categories $Zh\to b\bar{b}b\bar{b}$ is
  the dominant single-Higgs background, while $t\bar{t}h(\to b\bar{b})$ is
  instead the most important one in the boosted category.
  
  \begin{table}[h]
\begin{center}
\begin{tabular}{|c|c|c|c|c|}
\hline
& Sample &  \multicolumn{3}{c|}{Pre-MVA cross-section (fb)}\\
 & &  Boosted  & Intermediate & Resolved \\[0.1cm]
\hline\hline
Signal & $hh\to b\bar{b}b\bar{b}$ & $3.5\cdot 10^{-1}$  & $2.2\cdot 10^{-1}$ &  $1.2\cdot 10^{0}$ \\[0.1cm]
\hline
\multirow{4}{*}{Backgrounds} & QCD multijet &  $2.5\cdot 10^{+2}$ & $1.8\cdot 10^{+2}$ & $4.9\cdot 10^{+3}$ \\
&$Z(\to b\bar{b})h(\to b\bar{b})$ & $2.0\cdot 10^{-2}$ & $1.2\cdot 10^{-1}$ & $7.5\cdot 10^{-1}$ \\
&$t\bar{t}h(\to b\bar{b})$ & $5.1\cdot 10^{-2}$ & $6.3\cdot 10^{-3}$ & $4.0\cdot 10^{-1}$ \\
&$b\bar{b}h(\to b\bar{b})$ & $2.3\cdot 10^{-3}$ & $5.5\cdot 10^{-3}$ & $2.6\cdot 10^{-1}$\\
\hline
\end{tabular}
\end{center}
\caption{\small \label{HBxsec} Signal and background cross-sections at the end of the cut-based analysis
  (before the MVA is applied), in the case without PU.
  We separate the results into the three exclusive categories used in our analysis.
}
  \end{table}%


\providecommand{\href}[2]{#2}\begingroup\raggedright\endgroup

\end{document}